\documentclass[a4paper,fleqn,usenatbib]{mn2e}

\usepackage[T1]{fontenc}
\usepackage{ae,aecompl}

\usepackage{graphicx}	
\usepackage{amsmath}	
\usepackage{amssymb}	
\usepackage{natbib}
\usepackage{xspace}
\usepackage{rotate} 
\usepackage{subfigure}
\usepackage{float}
\usepackage[normalem]{ulem}
\usepackage{wrapfig}
\usepackage{longtable}

\newcommand{\xte}{\textsl{RXTE}\xspace}
\newcommand{\rxte}{\textsl{RXTE}\xspace}
\newcommand{\integral}{\textsl{INTEGRAL}\xspace}
\def\myfrac#1#2{\left(\frac{#1}{#2}\right)}
\def\eqn#1{~(\ref{#1})}
\def\Eq#1{Eq.~(\ref{#1})}

\title[Flattening of the Cyclotron Line/Luminosity Relation in GX 304$-$1]{Discovery and Modeling of a Flattening of the Positive Cyclotron Line/Luminosity Relation in GX 304$-$1 with \xte}

\author[R. E. Rothschild et al.]{
\parbox{\textwidth}{\raggedright Richard E. Rothschild,$^{1}$\thanks{E-mail:rrothschild@ucsd.edu}
Matthias K\"uhnel,$^{3}$
Katja Pottschmidt,$^{2}$
Paul Hemphill,$^{1}$
Konstantin Postnov,$^{5,6}$
Mikhail Gornostaev,$^{5,6}$
Nikolai Shakura,$^{5}$
Felix F\"urst,$^{7}$
J\"orn Wilms,$^{3}$
R\"udiger Staubert,$^{4}$
and
Dmitry Klochkov$^{4}$
}
\\
\\
$^{1}$Center for Astrophysics and Space Sciences, University of California, San Diego, 9500 Gilman Dr., La Jolla, CA
  92093-0424, USA\\
$^{2}$CRESST, Department of Physics, and Center for Space Science and
Technology, UMBC, Baltimore, MD 21250, USA, and \\
NASA Goddard Space Flight Center, Code 661, Greenbelt, MD 20771, USA\\
$^{3}$Dr. Karl-Remeis-Sternwarte and ECAP, Sternwartstr. 7, 96049 Bamberg, Germany\\
$^{4}$Institut f\"ur Astronomie und Astrophysik, Universit\"at T\"ubingen, Sand 1, 72076 T\"ubingen, Germany\\
$^{5}$Sternberg Astronomical Institute, Moscow State University, Universtetskij pr. 13, 119234 Moscow, Russia\\
$^{6}$Faculty of Physics, Moscow State University, Leninskie Gory 1, 119991 Moscow, Russia\\
$^{7}$Cahill Center for Astronomy and Astrophysics, California Institute of Technology, MC 290-17, 
1200 E. California Blvd, Pasadena, \\
CA 91125, USA
}
\date{Accepted XXX. Received YYY; in original form ZZZ}

\pubyear{2016}

\begin{document}
\label{firstpage}
\pagerange{\pageref{firstpage}--\pageref{lastpage}}
\maketitle

\begin{abstract}
The \xte observed four outbursts of the accreting X-ray binary transient source, GX 304$-$1 in 2010 and 2011.
We present results of detailed 3$-$100 keV spectral analysis of 69 separate observations, and report 
a positive correlation between cyclotron line parameters, as well as other spectral parameters,
with power law flux. The cyclotron line energy, width and depth versus flux, and thus luminosity, correlations
show a flattening of the relationships with increasing luminosity, which are well described by 
quasi-spherical or disk accretion that yield the surface magnetic field to be $\sim$60 keV. 
Since HEXTE cluster A was fixed aligned with the PCA field of view and 
cluster B was fixed viewing a background region 1.5 degrees off of the source direction during 
these observations near the end of the \xte mission, the cluster A background 
was estimated from cluster B events using HEXTEBACKEST. This made possible the detection of the 
$\sim$55 keV cyclotron line and an accurate measurement of the continuum. Correlations of all spectral 
parameters with the primary 2$-$10 keV power law flux reveal it to be the primary driver of the spectral shape. 
The accretion is found to be in the collisionless shock braking regime. 
\end{abstract}

\begin{keywords}
pulsars: individual (GX 304-1) -- X-rays: binaries -- stars: neutron -- magnetic fields -- X-rays individual (GX 304-1)
\end{keywords}



\section{INTRODUCTION}
The study of neutron star magnetic fields in accreting X-ray pulsars has progressed significantly over the past few 
decades through the observations of cyclotron resonance scattering features (CRSFs), or cyclotron lines. 
Beginning with the discovery in 1976 of such a feature in Her~X$-$1 \citep{Trumper78}, we now have identified 
about two dozen accreting X-ray pulsars that exhibit cyclotron line 
features\footnote{http://www.sternwarte.uni-erlangen.de/wiki/doku.php?id=cyclo:start}. 
The fundamental line energies range from 10 to 55 keV, implying magnetic field strengths from about 1 to 5 TG. 
Recent work to model the accretion column emission from a physics-based point of view is based upon the accreted 
material passing through a radiative, radiation dominated shock and forming a thermal mound just above the surface 
at the magnetic poles, as first proposed by  \citet{Davidson73}. Conditions in the infalling supersonic material  
are dominated by either radiation pressure at high luminosities or Coulomb interactions at lower luminosities before 
settling on the neutron star surface \citep[e.g., ][ and references therein]
{Becker12,Postnov15b}. At the lowest luminosities no shock is formed and the material flows unabated 
until reaching the mound of material piled up on the magnetic poles. At high luminosities -- defined as above the critical 
luminosity where radiation pressure dominates over gas pressure \citep{Mushtukov15a}-- an increase in flux 
causes the shock, and thus 
the scattering region, to rise and sample lower magnetic field strengths, giving rise to a negative correlation 
of the cyclotron line energy with luminosity. Physically, the structure of accretion column starts changing with decreasing mass accretion rate when the photon diffusion time across the optically thick column becomes comparable to the matter settling time from the radiative shock height, and generally can be different in different sources. First estimates 
\citep[e.g. ][]{Basko76} shows it to be around $10^{37}$~erg s$^{-1}$ if the height of the radiative shock above the neutron star surface is comparable to the accretion column radius. With further decrease in the mass accretion rate onto the neutron star magnetic poles, the accretion flow decelerates most likely due to plasma instabilities leading to the formation of a collisionless shock, as numerical calculations performed at $\dot M<10^16$~g s$^{s}$ \citep[e.g. ][]{BykovKrassilshchikov04} suggest. The intermediate regime (i.e. between the radiative shock at high accretion rates and collisionless shock) is the most difficult to treatment, and still is to be explored numerically with taking into account the relevant complicated microphysics. In the collisionless shock regime, the height of the the scattering region decreases with increasing mass accretion rate thus producing a positive correlation of the cyclotron line with luminosity.
\citet{Nishimura14} reproduces the same correlations
with the line forming region being that between the top of the thermal mound and a height equal to twice the accretion 
column radius, both of which rise as the luminosity increases. \citet{Poutanen13} have asserted a reflection model for 
the cyclotron line formation in which the shocked infalling matter generates X-rays that illuminate the atmosphere of the 
neutron star. In this case, increased accretion, and thus increased luminosity, increases the height of the X-ray emitting 
region and thus increases the area of the neutron star that is illuminated. This increased area contains lower values of 
the dipole magnetic field and thus the resulting cyclotron line has a lower value.

To date six accreting X-ray pulsars are known to have correlations of the 
fundamental cyclotron line energy with luminosity: one with a negative correlation, V~0332+53 
\citep{Tsygankov06, Klochkov11}, and five with a positive correlation, Her~X$-$1 \citep{Staubert07,Klochkov11,Staubert14,Staubert16}, GX~304$-$1 \citep{Klochkov12}, A~0535+26 \citep{Klochkov11},
the first harmonic of Vela~X$-$1 \citep{Fuerst14}, and Cep~X$-$4 \citep{Fuerst15}.
Note: 4U0115+63 is no longer deemed to have a correlation of the cyclotron line energy with luminosity
\citep{Mueller13,Boldin13}. \citet{LaParola16} have recently published results from analysis of 
\textit{Swift}/BAT observations of Vela X-1 where they find a positive correlation of the first harmonic 
cyclotron line energy with luminosity, and in addition, find a flattening of the correlation with increasing luminosity.
Other spectral components, such as the power law index \citep[e.g., ][]{Malacaria15,Postnov15b} and iron line flux, 
have been seen to vary with accretion rate as expressed by the X-ray flux.

GX 304$-$1 was first detected in a balloon flight \citep{McClintock71} and later by the \textsl{Uhuru} satellite 
as 2U1258-61 \citep{Giacconi72}. It is an accreting neutron star exhibiting a teraGauss magnetic field in a 
high mass X-ray binary system with its companion B2Vne star, 
V850 Cen \citep{Mason78,Reig97}. The system has an orbital period of 132.1885$\pm$0.022 days  
\citep{Sugizaki15}, a pulse period of $\sim$272 seconds \citep{McClintock77}, 
and a distance of 2.4$\pm$0.5 kpc \citep{Parkes80}. After a nearly three decade period of quiescence, GX 304$-$1 
emerged in 2008 \citep{Manousakis08}, and began a series of regularly spaced outbursts in late 2009 
\citep[see fig. 1 of][]{Yamamoto11}. 
A cyclotron resonance scattering feature at $\sim$54 keV was discovered during the 2010 August outburst  
\citep{Yamamoto11}, and a possible positive correlation with flux was suggested. This has been confirmed with 
recent \integral results by \citet{Klochkov12}, who found the line varying between 
$\sim$48\,keV and $\sim$55\,keV, and by \citet{Malacaria15} who found the range to be 50 to 59 keV with newer
\integral calibrations. Four outbursts in 2010 and 2011 were observed by \xte until its demise in 2012 January.

In this work we present an analysis of \xte data of the outbursts in 2010 March/April, 2010 August, 
2010 December/2011 January, and 2011 May, which represent 72 separate observations, of which 69 were 
analyzed in detail. 
From this we determine the variations of various spectral 
components with respect to unabsorbed power law flux, with which all are correlated. We present 
the Observations and Data Reduction in Section 2, Data Analysis in Section 3, Results in 
Section 4, and Discussion in Section 5, and present our conclusions in Section 6. In Appendix A we give the 
background and analysis that is 
the basis for the cluster A background estimation tool, HEXTEBACKEST. In Appendix B we give the tables of 
best-fit spectral parameters and plot them versus unabsorbed power law flux. Also in Appendix B we 
present representative contour plots of the cyclotron line parameters versus various spectral components.
In Appendix C we discuss 
tests of the HEXTE background estimation and plot 
the systematic normalization constants. 

\section{OBSERVATIONS AND DATA REDUCTION}

\subsection{Observations}
The \textit{Rossi X-ray Timing Exlorer} (\rxte)  observed GX 304$-$1 72 different times over its operational lifetime from 1996 to 2012, with three outbursts 
(2010 August, 2010 December, and 2011 May), numbering 69 observations, covered extensively. The outburst in 2010 March/April 
outburst had only 3 observations, and they are included to show consistency with the other outbursts. 
Three of the observations had less livetime than the GX~304-1 pulse period (Table~\ref{tab:rxte_all} numbers 
10, 52, and 62), and they were not included in subsequent analyses. Table~\ref{tab:rxte_all} gives the dates, 
ObsIds, livetimes, and rates for both the Proportional Counter Array \citep[PCA; ][]{Jahoda06} Proportional Counter 
Unit 2 (PCU2) and for the High Energy X-ray Timing Experiment \citep[HEXTE; ][]{Rothschild98} Cluster~A. Rates for 
PCU2 and HEXTE Cluster~A are background subtracted.  The sequential numbering of the individual observations 
is for identification in subsequent tables.

\subsection{Data Reduction}
PCA data were restricted to the 3$-$60 keV range of the top xenon layer of PCU2 due to the extensive calibration of 
this detector \citep{Jahoda06} that did not experience high voltage break down during the mission and thus were 
included in all 
PCA observations. The observational data were filtered to accept only observations with elevation above the 
Earth's limb of greater than 10$^\circ$, observation times more than 30 minutes from the start of the previous 
South Atlantic Anomaly passage, and electron rate below 0.5, instead of the nominal 0.1, due to the high X-ray 
flux adding counts to the electron detection portions of the proportional counter. The HEXTE data utilized the 
PCU2 filter criteria, were restricted to the 20$-$100 keV range, and data from both clusters were included in the analyses. 
The PCU2 background was estimated using PCABACKEST, and the PCU2 response was generated for the 
specific observation day using PCARSP. Due to rocking mechanism failures in the latter stages of the \xte 
mission, HEXTE cluster~A was continuously pointed on-source (after 2006 October 20), and cluster~B was continuously pointed 1.5$^\circ$ 
off-source (after 2009 December 12)  to collect background data for all observations\footnote{see http://heasarc.gsfc.nasa.gov/docs/xte/whatsnew/big.html for details of HEXTE rocking.}. The background spectrum for cluster~A was then 
generated from that of cluster~B using HEXTEBACKEST, as discussed in subsequent subsections and Appendix A. 
The cluster~A spectral response was generated using HEXTERSP, which did not vary during the mission due to 
HEXTE's automatic gain control.

\begin{figure}
\includegraphics[width=3.0in]{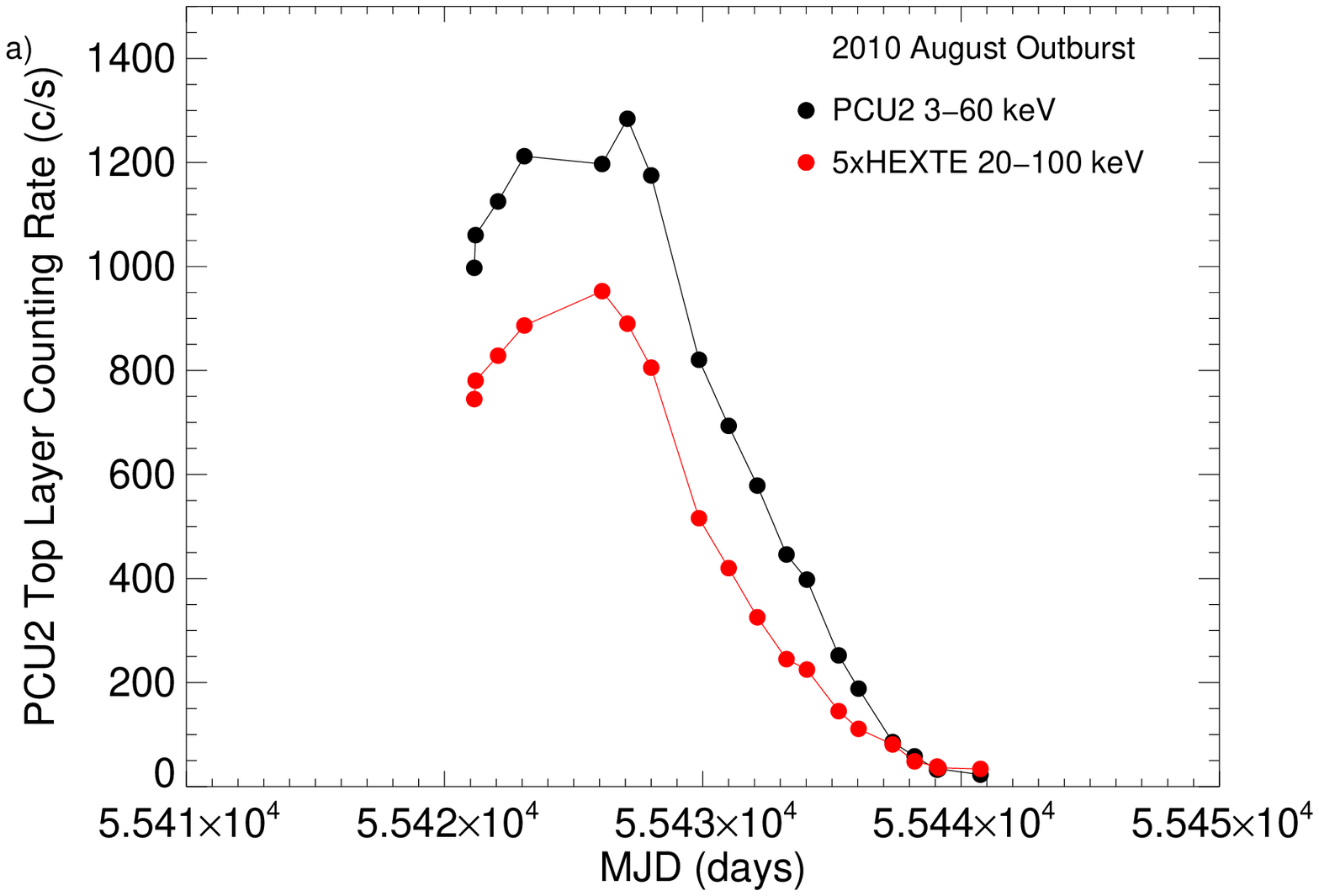}\\
\includegraphics[width=3.0in]{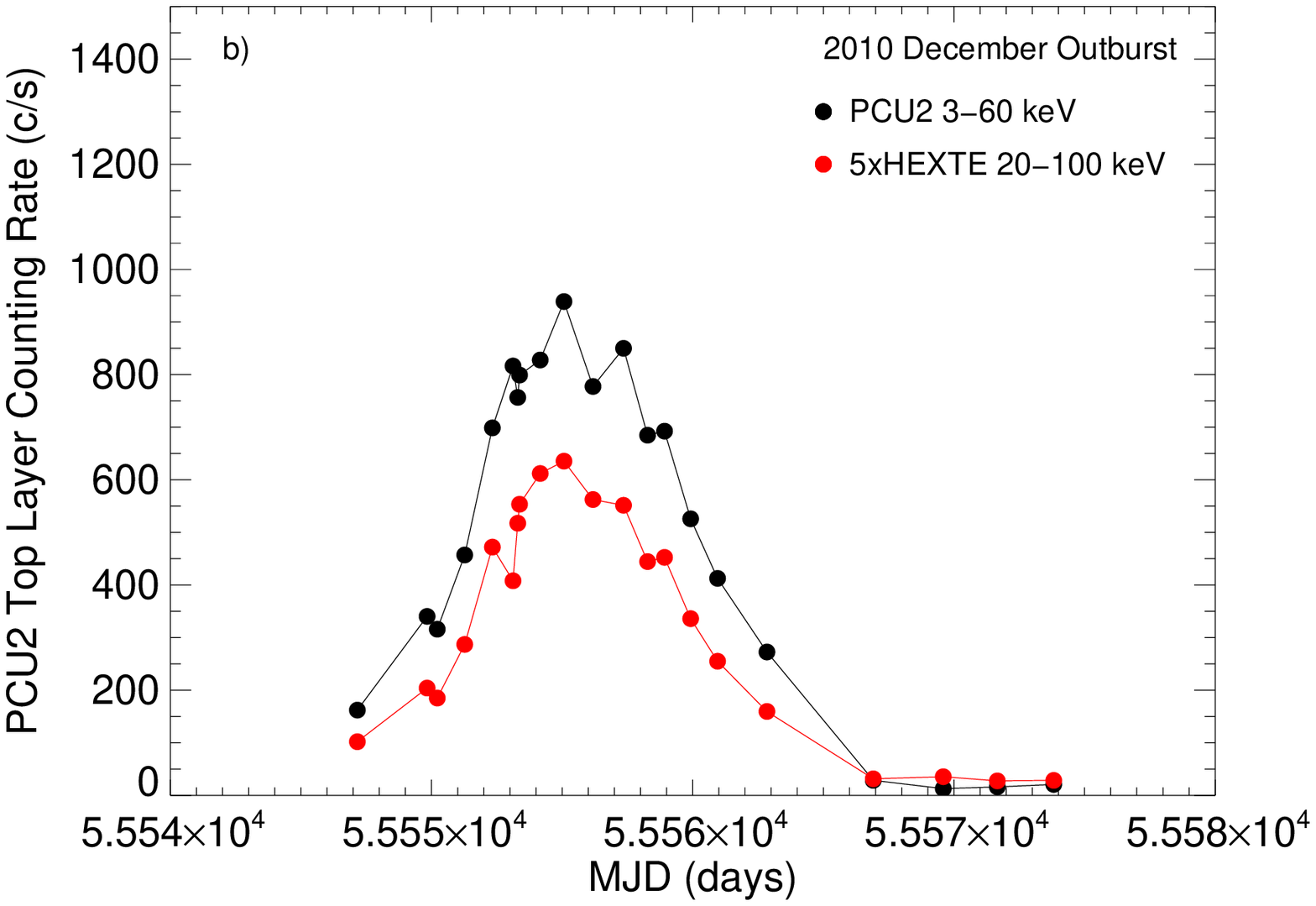}\\
\includegraphics[width=3.0in]{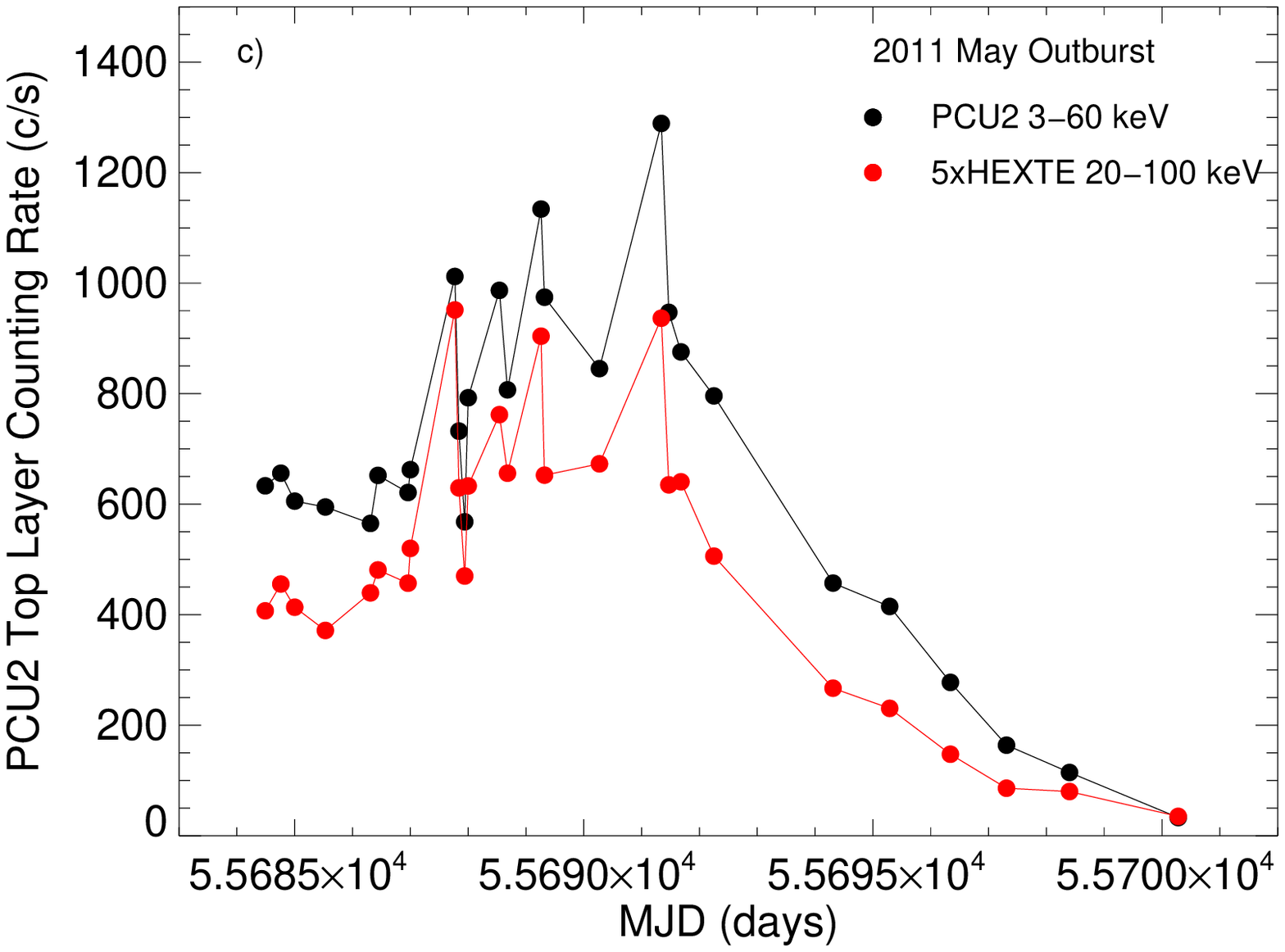}\\
\caption{The PCU2 top layer 3$-$60 keV background subtracted, counting rates and the HEXTE cluster~A 
20$-$100 keV background subtracted, counting rates as a function of the observation date in Modified Julian Days. 
The 2010 August outburst is seen in panel a), the 2010 December outburst in panel b), and the 2011 May outburst 
in panel c). The PCU2 data are in black and the HEXTE data are in red. The HEXTE data have been multiplied by five.
\label{fig:pca_hexte_rates}}
\end{figure}

The 3$-$60 keV PCU2, top layer, background subtracted, counting rates and the 20$-$100 keV HEXTE cluster~A, 
background subtracted, counting rates for each of the three observing epochs are shown in panels a), b), and c) 
in Fig.~\ref{fig:pca_hexte_rates}. The HEXTE rates are multiplied by five in order to visually compare them with 
those of the PCU2. The 2010 August epoch 
observations cover from just before the maximum through decay to the beginning of  a low state. 

\clearpage
\onecolumn
\begin{table}
\caption{\rxte Observations of GX 304$-$1}
\scriptsize
\label{tab:rxte_all}
\begin{tabular}{llllrrrr}
\# & Date & ObsID & MJD$^a$ & PCA Lvt$^b$ & PCA Rate$^c$ & HEXTE Lvt$^b$ & HEXTE Rate$^d$\\ 
\hline
1 &2010 Mar 27          & 95417-01-01-00   & 55282.34 & 2880 & 108.5$\pm$0.2   & 1620 &    13.7$\pm$0.3\\
2 & 2010 Mar 27         & 95417-01-01-01   & 55282.61 & 2192 & 114.1$\pm$0.2   & 1480 &    12.1$\pm$0.3\\
3 & 2010 Apr 6            & 95417-01-02-00   & 55292.68 & 3296 & 195.8$\pm$0.3   & 2275 &    69.8$\pm$0.3\\
\hline
4 &2010 Aug 13          & 95417-01-03-03   & 55421.15 & 2304 &  997.4$\pm$0.7  & 1399 & 149.0$\pm$0.4 \\
5 &2010 Aug 13          & 95417-01-03-00   & 55421.20 & 3712 & 1060.2$\pm$0.5 & 2300 & 156.1$\pm$0.3 \\
6 &2010 Aug 14          & 95417-01-03-01   & 55422.07 & 5408 & 1125.0$\pm$0.5 & 1480 & 165.7$\pm$0.4 \\
7 &2010 Aug 15          & 95417-01-03-02   & 55423.09 & 6096 & 1212.4$\pm$0.4 & 1961 & 177.3$\pm$0.4 \\
8 &2010 Aug 18          & 95417-01-04-00   & 55426.10 & 3328 & 1197.0$\pm$0.6 &   184  & 190.5$\pm$1.1\\
9 &2010 Aug 19          & 95417-01-04-01   & 55427.08 & 3216 & 1289.0$\pm$0.6 & 1966  & 178.0$\pm$0.4 \\
10 &2010 Aug 19        & 95417-01-04-02   & 55427.99 &      64 &  1470.0$\pm$4.8 &     33 & 186.3$\pm$2.9\\
11 &2010 Aug 20          & 95417-01-05-00   & 55428.00 & 3120 & 1175.0$\pm$0.6 & 1922  & 161.1$\pm$0.4 \\
12 &2010 Aug 21          & 95417-01-05-01   & 55429.85 &   992 &    820.6$\pm$0.9 &    638 & 103.2$\pm$0.6 \\
13 &2010 Aug 23          & 95417-01-05-02   & 55431.00 & 2016 &    693.5$\pm$0.6 &  1159 &   84.0$\pm$0.4 \\
14 &2010 Aug 24          & 95417-01-05-03   & 55432.11 & 3408 &    578.7$\pm$0.4 &  2072 &   65.1$\pm$0.3 \\
15 &2010 Aug 25          & 95417-01-05-04   & 55433.24 & 1184 &    446.2$\pm$0.6 &    870 &   49.0$\pm$0.4 \\
16 & 2010 Aug 26          & 95417-01-05-05   & 55434.03 & 1328 &    397.9$\pm$0.6 &    770 &   45.0$\pm$0.4 \\
17 &2010 Aug 27          & 95417-01-06-00   & 55435.26 & 1696 &    252.2$\pm$0.4 &   1234 &  29.0$\pm$0.3 \\
18 &2010 Aug 28          & 95417-01-06-01   & 55436.03 & 1984 &    188.3$\pm$0.3 &   1105 &  22.2$\pm$0.3  \\
19 &2010 Aug 29          & 95417-01-06-02   & 55437.35 & 1568 &      85.6$\pm$0.3 &    1108 &  16.2$\pm$0.4 \\
20 &2010 Aug 30          & 95417-01-06-03   & 55438.20 & 2336 &      58.2$\pm$0.2 &    1648 &    9.7$\pm$0.3 \\
21 &2010 Aug 31          & 95417-01-06-04   & 55439.07 & 1440 &      30.8$\pm$0.2 &      817 &    8.1$\pm$0.3 \\
22 &2010 Aug 31          & 95417-01-06-06   & 55439.13 & 1344 &      34.1$\pm$0.2 &      828 &    7.2$\pm$0.3 \\
23 &2010 Sep 1     & 95417-01-06-05   & 55440.75 &    832 &     22.9$\pm$0.2 &     543 &     6.8$\pm$0.4 \\
\hline 
24 &2010 Dec 17    & 95417-01-07-00   & 55547.16 & 16400&  162.1$\pm$0.1  & 10680&   20.4$\pm$0.1\\
25 &2010 Dec 19    & 95417-01-07-01   & 55549.83 &  2944 &   340.3$\pm$0.4  &  1793 &   40.8$\pm$0.3 \\
26 &2010 Dec 20    & 95417-01-07-02   & 55550.22 & 12210&  315.9$\pm$0.2  &   7418 &   37.0$\pm$0.1 \\
27 &2010 Dec 21    & 95417-01-07-03   & 55551.27 &  7744 &  457.1$\pm$0.2  &   4651 &   57.4$\pm$0.2 \\
28 &2010 Dec 22    & 95417-01-07-04   & 55552.33 &  2848 &  698.7$\pm$0.5  &   1738 &   94.4$\pm$0.3 \\
29 &2010 Dec 23    & 95417-01-07-05   & 55553.12 &  8880 &  816.4$\pm$0.3  &   5625 &   81.6$\pm$0.3  \\
30 &2010 Dec 23    & 95417-01-07-06   & 55553.30 &  3664 &  756.6$\pm$0.5  &    2181 & 103.5$\pm$0.3 \\
31 &2010 Dec 23    & 95417-01-07-07   & 55553.37 &  3200 &  799.2$\pm$0.5  &    1807 & 110.7$\pm$0.3 \\
32 &2010 Dec 24    & 95417-01-08-00   & 55554.16 &  3408 &  827.8$\pm$0.5  &    2156 & 122.4$\pm$0.3\\
33 &2010 Dec 25    & 95417-01-08-01   & 55555.07 &  3520 &  939.0$\pm$0.5  &       216 & 127.1$\pm$0.9\\
34 &2010 Dec 26    & 95417-01-08-02   & 55556.18 &  3344 &  775.5$\pm$0.5  &     2016 & 112.5$\pm$0.3\\
35 &2010 Dec 27    & 95417-01-08-03   & 55557.35 &    768 &  850.1$\pm$1.1  &       407 &  110.3$\pm$0.8\\
36 &2010 Dec 28    & 95417-01-08-04   & 55558.27 &  2736 &  684.8$\pm$0.5  &    1592 &  88.9$\pm$0.3 \\
37 &2010 Dec 28    & 95417-01-08-05   & 55558.92 &  5760 & 692.5$\pm$0.4   &    3812 &  90.5$\pm$0.2 \\
38 &2010 Dec 29    & 95417-01-08-06   & 55559.92 &  5136 &  525.8$\pm$0.3  &    3295 &  67.2$\pm$0.2 \\
39 &2010 Dec 30    & 95417-01-08-07   & 55560.95 &  3344 &  412.6$\pm$0.4  &    2199 &  51.0$\pm$0.3 \\
40 &2011 Jan 1       & 96369-01-01-00   & 55562.80 &  9939 & 272.6$\pm$0.2  &    6471  &  31.9$\pm$0.1\\
41 &2011 Jan 5       & 96369-01-01-01   & 55566.91 & 2524 &  28.5$\pm$0.1     & 1673    &     6.3$\pm$0.2 \\
42 &2011 Jan 8       & 96369-01-02-00   & 55569.59 & 1744 &  12.8$\pm$0.1     & 1105    &     7.1$\pm$0.3 \\
43 &2011 Jan 10     & 96369-01-02-01   & 55571.66 & 2544 & 16.1$\pm$0.1     & 1619     &     5.5$\pm$0.2 \\
44 &2011 Jan 12     & 96369-01-02-02   & 55573.82 & 2528 & 20.8$\pm$0.1     & 1496    & 5.7$\pm$0.2  \\
\hline
45 &2011 May 3                  & 96369-01-03-00   & 55684.49 & 1280 & 633.2$\pm$0.7 &  866 &  81.4$\pm$0.4 \\
46 &2011 May 3                  & 96369-01-03-01   & 55684.76 &  960 & 656.1$\pm$0.8  &  660 &  91.1$\pm$0.6  \\
47 &2011 May 4                  & 96369-01-03-02   & 55685.00 & 1168 & 605.5$\pm$0.7 &  774 & 82.7$\pm$0.5 \\
48 &2011 May 4                  & 96369-01-04-00   & 55685.53 & 1984 & 594.9$\pm$0.6 & 1387 &  74.3$\pm$0.3\\
49 &2011 May 5                  & 96369-01-05-00   & 55686.31 & 3584 & 565.2$\pm$0.4 & 2056 &  87.9$\pm$0.3 \\
50 &2011 May 5                  & 96369-01-05-01   & 55686.44 & 6272 & 652.1$\pm$0.3 & 4313 &  96.2$\pm$0.2 \\
51 &2011 May 5                  & 96369-01-05-02   & 55686.96 & 1136 & 621.1$\pm$0.8 &   749  &  91.4$\pm$0.5 \\
52 &2011 May 6                  & 96369-02-01-00   & 55687.00 &      32 & 475.7$\pm$4.0 &     13  & 65.4$\pm$3.4\\
53 &2011 May 6                  & 96369-02-01-000 & 55687.00 & 17730& 662.4$\pm$0.2 & 9977 & 104.0$\pm$0.1\\
54 &2011 May 6                  & 96369-02-01-02   & 55787.77 & 1056 & 1012.0$\pm$1.0&  714 & 190.3$\pm$0.7\\
55 &2011 May 6                  & 96369-02-01-03  & 55687.84 &  768 &  732.2$\pm$1.0 &  514 & 125.9$\pm$0.7 \\
56 &2011 May 6                  & 96369-02-01-04   & 55687.94 & 1104 & 568.1$\pm$0.07 &  738 &  94.0$\pm$0.5 \\
57 &2011 May 7                  & 96369-02-01-01G  & 55688.00 & 18300 & 792.6$\pm$0.2 & 10000 & 126.6$\pm$0.1 \\
58 &2011 May 7                  & 96369-02-01-05   & 55688.54 & 3072 & 986.9$\pm$0.6 & 830 & 152.4$\pm$0.5  \\
59 &2011 May 7                  & 96369-02-01-06   & 55698.68 & 1344 & 806.9$\pm$0.8 &  919 & 131.2$\pm$0.5  \\
60 &2011 May 8                  & 96369-01-06-00   & 55689.26 & 2064 & 1134.0$\pm$0.7 & 1131 & 180.8$\pm$0.5 \\
61 &2011 May 8                  & 96369-01-06-01   & 55689.32 & 2912 & 974.5$\pm$0.6 & 1643 & 130.5$\pm$0.3  \\
62 &2011 May 9                   & 96369-01-06-02   & 55690.27 &   96 & 845.3$\pm$3.0 &   51 & 134.6$\pm$2.1 \\
63 &2011 May 10                 & 96369-01-06-03   & 55691.34 &  656 & 1289.0$\pm$1.4 &  433 & 187.3$\pm$0.8  \\
64 &2011 May 10                 & 96369-01-06-04  & 55691.47 & 1344 & 947.1$\pm$0.8 &  962 & 127.0$\pm$0.5  \\
65 &2011 May 10                 & 96369-01-07-00   & 55691.68 & 1728 & 875.5$\pm$0.7 & 1170 & 128.1$\pm$0.5  \\
66 &2011 May 11                 & 96369-01-07-01   & 55692.25 & 4076 & 795.9$\pm$0.4 & 2372 & 101.2$\pm$0.3 \\
67 &2011 May 13                 & 96369-01-08-00   & 55694.31 & 7056 & 457.2$\pm$0.3 & 4703 &  53.4$\pm$0.2 \\
68 &2011 May 14                 & 96369-01-08-01   & 55695.29 & 1136 & 414.9$\pm$0.6 &  603 &  46.1$\pm$0.5 \\
69 &2011 May 15                 & 96369-01-09-00   & 55696.34 & 3760 & 277.5$\pm$0.3 & 2544 &   29.5$\pm$0.2 \\
70 &2011 May 16                 & 96369-01-09-01   & 55617.31 &  832 & 163.9$\pm$0.5 &  496 &  17.2$\pm$0.4  \\
71 &2011 May 17                  & 96369-01-10-00   & 55698.40 & 3104 & 114.5$\pm$0.2 & 2165 &   16.0$\pm$0.2  \\
72 &2011 May 19                 & 96369-01-10-01   & 55700.28 &  480 & 32.9$\pm$0.3 & 345 &  7.0$\pm$0.5 \\
\hline
\end{tabular}
\\
$^a$ Start time of the observation\\
$^b$ Livetime in seconds\\
$^c$ 3$-$60 keV count rate in c/s\\
$^d$ 20$-$100 keV count rate in c/s\\
\end{table}
\clearpage
\normalsize
\twocolumn

 The 2010 December epoch covers a full outburst from just after the start to well into the low state, but not reaching 
the peak intensities of the other two epochs. 
\xte began observing the 2011 May epoch after it was well underway, similarly to that of the 2010 August epoch, 
and followed it to the low state. While all three light curves show similar decreases from peak values to a low state,
the third epoch shows substantial counting rate variability approaching and at the peak of the outburst. As shown below, the majority of this variability is due to large variations in column density. 

Systematics of 0.5\% ($<$15 keV) and 1\% (15$-$60 keV) were added to the 
PCU2 data for observations 5, 7, 8, 26, 29, 50, 53, 57, 60, and 61 
to reduce the chi-square to an acceptable range for interpretation of parameter uncertainties. Addition of similar 
systematic errors to the other PCU2 data would have resulted in unreasonably low chi-square values in the 
spectral fitting. Otherwise, no systematic uncertainties were added to PCU2 data. No systematic uncertainties 
were added to the HEXTE data. In addition no spectral binning of either PCU2 or HEXTE-A data was used.

\section{DATA ANALYSIS}
For each ObsID, the spectral histograms of PCU2 covering 3$-$60 keV and HEXTE cluster~A covering 
20$-$100 keV were simultaneously fit using ISIS 1.6.2-30 \citep{Houck00}, and verified with 
XSPEC 12.8.2 \citep{Arnaud96}. For this analysis, two spectral 
models were utilized. The \texttt{cutoffpl} model approximated the continuum with a power law times an 
exponential to form a continuously steepening continuum, plus a blackbody 
(\texttt{CUTOFFPL + BBODY}), and the \texttt{highecut} model used a power law that abruptly changes 
to an exponentially falling continuum at a break energy
(\texttt{POWERLAW  x HIGHECUT}). Both models included low energy photoelectric absorption with 
interstellar  abundances
(\texttt{TBnew})\footnote{This is a revised version of the absorption model \texttt{TBABS} of \citet{Wilms00}}. 
The abundances of \citet{Wilms00} and cross sections of \citet{Verner95} were used in the analysis. The 
continuum was further modified by a Gaussian shaped cyclotron resonance scattering feature, or cyclotron line,
 (\texttt{GAUABS}) for those observations when the depth was measured, or had a lower limit, at 
 greater than 90\% confidence level. In addition, narrow ($\sigma$=10 eV), Gaussian line components were 
 added fixed at 6.40 and 7.06 keV representing iron K$\alpha$ and K$\beta$ emission with the K$\beta$ flux 
 set to 13\% of the K$\alpha$ flux.  

As presented in Appendix A, HEXTEBACKEST is based upon the channel by channel comparison of cluster~A 
and cluster~B background data for all of the observations in 2009 that included South Atlantic Anomaly passages. 
As such, the correlation parameters in each spectral bin are an average. Fig.~\ref{fig:bkgex} gives an idea of 
the spread in the data for two spectral channels. For any given observation, the correction factors will not give 
a cluster~A background prediction that exactly expresses the background that would have been observed by 
cluster~A, if it were rocking. Additionally, as the mission progressed from 2009, the satellite experienced lower 
and lower altitudes with the attendant increased magnetic rigidity and lesser South Atlantic Anomaly fluxes. 
This resulted in a somewhat lower background in the instruments. Consequently, four narrow  Gaussians with 
fixed energies at 30.17, 39.04, 53.0, and 66.64 keV, representing  corrections to 
the HEXTEBACKEST estimated fluxes of the four major HEXTE background lines were included in the modeling 
(see Appendix A for a description of HEXTEBACKEST and Appendix B for a presentation of the systematic lines 
versus 2$-$10 keV flux). The four energies were determined by averaging the individual fitted values during 
preliminary spectral analyses. The 30 keV and 67 keV lines are the strongest in the HEXTE background. 
While the lines at 39 keV and 53 keV are of lesser strength, they may affect the measurement of the energy 
of the known cyclotron line at $\sim$50-55 keV \citep{Yamamoto11}, and were thus included in the fitting procedure. 

The `10 keV feature', which is seen in fits to accreting pulsar spectra 
\citep[e.g., ][]{Coburn02},  was modeled by a negative Gaussian at 10.5 keV, when its inclusion reduced chi-square 
by 10 or more. No clear correlation was seen with respect to the detection of the 10 keV feature 
and power law flux.  A systematic feature in the PCU2 fits occurs at about 3.88 keV in some of 
the observations, and it was modeled as a fixed energy, negative, narrow Gaussian, if the fitted depth was 
inconsistent with zero at the 90\% confidence level.

The HEXTE model included the above mentioned parameters plus
a constant representing the fractional difference in the response collecting area with respect to PCU2. The 
HEXTE constant was generally near 0.88, and was included in the variables of a given fitting procedure. 
Calculation of the PCU2 dead time showed that the deadtime correction was only a few percent at the highest 
PCU2 counting rates, and thus, no PCU2 deadtime correction was made. The HEXTE deadtime was calculated 
as an integral part of the data preparation using HEXTEDEAD. Since HEXTEDEAD is based upon average rates from 
two upper level discriminator rates \citep{Rothschild98}, any individual observation may deviate from the average. 
Thus, to compensate for the few percent uncertainty of the PCA background and HEXTE 
background and deadtime models, the 
background subtractions were optimized with multiplicative parameters (RECOR in XSPEC and CORBACK in ISIS)
during the fitting process. All uncertainties are expressed as 90\% confidence.

The XSPEC model forms were:

\noindent F(E)=Recor*Const*TBnew*(Powerlaw*Highecut*Gauabs + Gauss(Fe$_{K\alpha}$) + Gauss(Fe$_{K\beta}$)) + Sys

\noindent or

\noindent F(E)=Recor*Const*TBnew*(Cutoffpl*Gauabs + Bbody + Gauss(Fe$_{K\alpha}$) + Gauss(Fe$_{K\beta}$)) + Sys

\noindent where

\noindent Sys = Gauss(3.88 keV) + Gauss(10.5 keV) + Gauss (30.17 keV) + Gauss(66.37 keV) + Gauss (39.04 keV) + Gauss (53.00 keV)

The best fit continuum parameters for all observations using the \texttt{highecut} and \texttt{cutoffpl} 
models are given in Appendix B as Tables~\ref{tab:best_fit_highecut_cont} and \ref{tab:best_fit_cutoffpl_cont}.  
The best fit spectral line parameters are given in Tables~\ref{tab:best_fit_highecut_lines} and 
\ref{tab:best_fit_cutoffpl_lines}. Plots of the various continuum parameters versus unabsorbed power law 
(\texttt{highecut}) or unabsorbed power law times exponential (\texttt{cutoffpl}) fluxes can be found in 
Appendix B and plots of the \texttt{recor} parameter and the HEXTE constant can be found in Appendix C.
For those fittings where the search for the depth of the cyclotron line reached zero, no values for the 
cyclotron line parameters were reported and only double dashes are in Tables B1 and B2. For those fittings where
a lower limit on the depth was found but not an upper limit, lower limits are given and values for the cyclotron line energy and width are given. Otherwise, both high and low limits are given.
Examples of correlations between the fitted cyclotron line parameters and background lines at 53 keV and 66 keV, 
as well as versus the cutoff energy and folding energy of the continuum, are displayed in Appendix B for high and 
low flux observations \#9 (12$\times10^{-9}$ ergs cm$^{-2}$ s$^{-1}$) and \#39 (4.7$\times10^{-9}$ ergs cm$^{-2}$ s$^{-1}$) . In addition the correlation between the folding and cutoff energies is shown 
for those examples. At the lower flux levels, the correlation contours are somewhat bimodal and that the more 
significant maximum occurred for the higher value of the cyclotron line parameter.

\begin{figure}
\center
\includegraphics[width=3.1in]{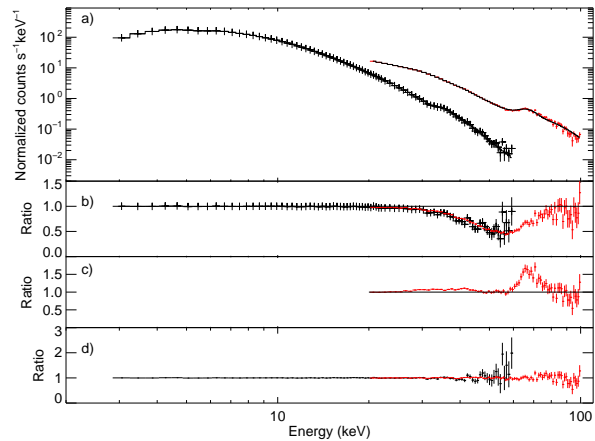}
\caption{\textbf{a)}: The PCU2 (black) and HEXTE-A (red) counts histogram for ObsID 95417-01-04-01 
(2010 August 19) plotted versus energy. The best-fit model is the solid histogram 
in black. \textbf{b)}:  The residuals of the best-fit model with the depth of the cyclotron line set to zero. The residuals are expressed as the Ratio of the data to the model. \textbf{c)}: The residuals to the best-fit HEXTE model with the four 
additional HEXTE-A background line fluxes set to zero. \textbf{d)}: The residuals to the best-fit model with all 
parameters set to their best-fit values. The solid black line in the three Ratio residuals denotes a value of 1.\label{fig:bright_spectrum}}
\end{figure}

As an example, the fit to ObsID 95417-01-04-01 is shown in Figure~\ref{fig:bright_spectrum}. The effects of 
excluding a cyclotron line component (panel b) and excluding the four HEXTE-A background lines (panel c) 
are shown as the ratio of the data to the model. Panel d) gives the ratio when all parameters are at their best-fit values. 
The reduced chi-square for this fit was 1.09 for 151 degrees of freedom. Note that the cyclotron line is clearly seen in the 
high energy portion of the PCU2 data (panel b), thus supporting the background estimation technique for HEXTE 
cluster~A.

\section{RESULTS}
The two spectral models employed in the analysis generally lead to qualitatively similar results. From hereon 
throughout the rest of the paper, the \texttt{highecut} model results will be the subject of the discussion
for two reasons. First,  it has one parameter less than the \texttt{cutoffpl} model, and secondly, the continuum 
parameters do not influence each other to the degree that they do in the \texttt{cutoffpl} model, where the black 
body flux and the photon index are strongly correlated. This results in the parameters using the \texttt{highecut} 
model being better determined, such as the power law index and 2$-$10 keV continuum flux.  The stable 
behavior of the column density at lower fluxes in the \texttt{highecut} model is preferred over the strong correlation 
with flux seen when modeling with the \texttt{cutoffpl} model. Section 4.3 gives a short discussion of the 
\texttt{cutoffpl} model fitting.

\subsection{Peak Phase Zero Offsets}
\begin{figure}
\includegraphics[width=3.2in]{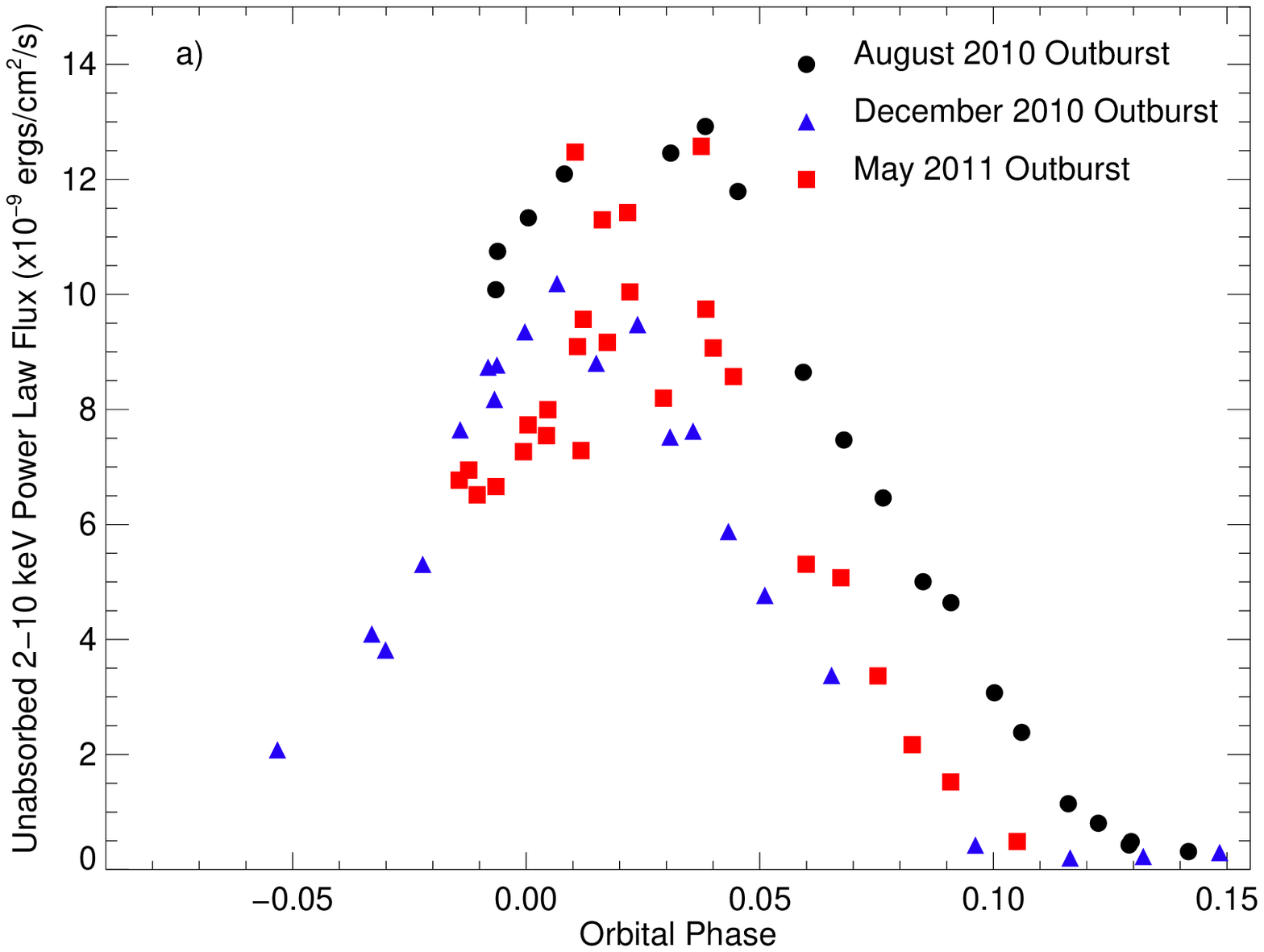}
\includegraphics[width=3.2in]{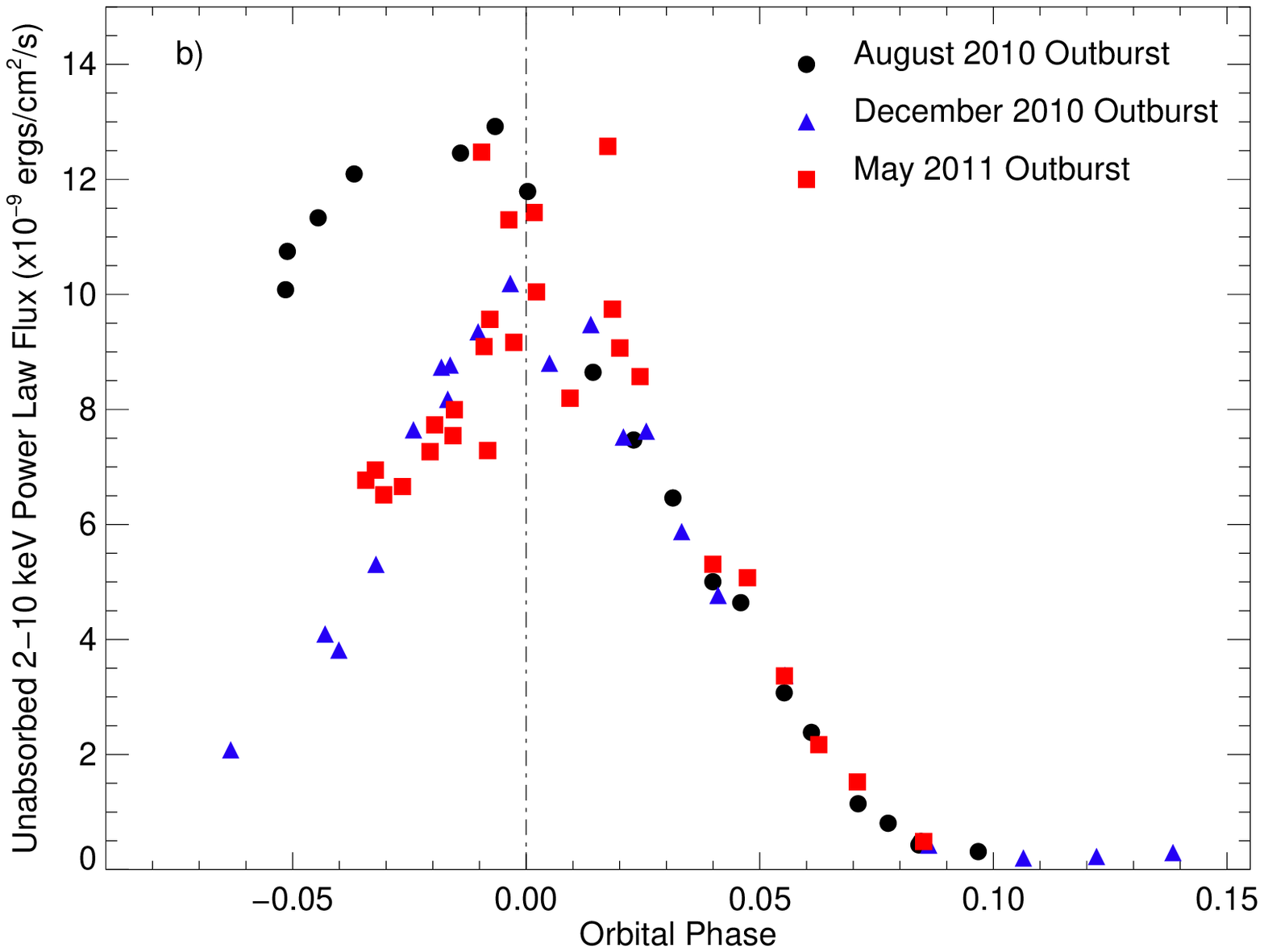}
\caption{\textbf{a}: The unobscured power law 2$-$10 keV power law flux plotted versus orbital phase for the three 
outbursts in 2010 August, 2010 December, and 2011 May of GX 304$-$1, as observed by \xte. \textbf{b}: The same data 
but with the orbital phases shifted by $-$0.045, $-$0.010, and $-$0.020, respectively, to match the 2010 August 
and 2011 May data to the falling portion of the 2010 December outburst. The 90\% uncertainties are generally less 
than the size of the data points.\label{fig:three_outbursts}}
\end{figure}

The orbital period of 132.1885$\pm$0.022 days \citep{Sugizaki15}, 
and T$_0$=MJD 55554.75, determined from \textit{MAXI} observations,
were used to generate the respective orbital phases for each observation. 
The three sets of observations (now versus fitted 2$-$10 keV power law flux) have quite similar outburst 
decay profiles (Fig~\ref{fig:three_outbursts}-a) with 
rise to peak flux and then decay to the lowest fluxes. By shifting the overall orbital phases slightly, the decay 
portions of the profiles align well (Fig.~\ref{fig:three_outbursts}-b). The amounts of the peak epoch phase shifts 
were determined by first centering the midpoint of the peak of the 2010 December data on phase zero, since 
that outburst showed a relatively complete rise and fall of the flux. Then the remaining 
two data sets were shifted to align their falling portions to that of the 2010 December data. The resulting  
phase shifts are $-$0.045 for 2010 August, $-$0.010 for 2010 December, and $-$0.020 for 2011 May. These phase 
shifts amount to 5.9, 1.3, and 2.6 days earlier than the orbital period derived from the \textsl{MAXI} data would 
have suggested. This is consistent with the residual offsets from the orbital model in fig.~2 of \citet{Sugizaki15} 
for these three outbursts covered by \xte. This reveals that the shapes of the outbursts are quite similar once 
the flux drops below $\sim$10$\times 10^{-9}$ erg cm$^{-2}$ s$^{-1}$. The rising portions of the 2010 December 
and 2011 May outbursts also appear consistent with each other below $\sim$10$\times 10^{-9}$ erg cm$^{-2}$ 
s$^{-1}$. The first four 2010 August observations (black filled circles) may indicate that the 2010 August outburst 
exhibited an outburst with wider extent than the others, or was indicative of flaring during the rising portion of the 
outburst.  The four 2011 May data points (red filled squares) above the common outburst trend are indicative 
of flaring near the peak of the 2011 May outburst. The three 2010 March/April points are not included here, 
since a phase shift could not be determined from so few points.

The flaring activity seen on the rising portion of the 2011 May outburst in Fig.~\ref{fig:pca_hexte_rates}, is absent in 
Fig.~\ref{fig:three_outbursts} and is attributable to the variation in column density affecting the PCU2 counting rate
(see Fig.~\ref{fig:highecut_continuum_flux}a). 
Individual points do remain above the overall outburst trend in  Fig.~\ref{fig:three_outbursts}, which may be considered 
flaring to some extent. Such flaring may be similar to the flaring activity seen on the rising portion of the 2005 
August/September outburst of A0535$+$26 \citep{Postnov08,Caballero08}, and attributed to a low mode 
magnetospheric instability. These GX 304$-$1 data, however, do not show a significant change in the cyclotron line 
energy for any of the high flux points, whereas the A0535$+$26 data did, and other than the four earliest 2010 
August outburst points, the points above the trend are at the maximum of the outbursts, and not on the rising edge, 
as in the 2005 August/September flares \citep{Postnov08,Caballero08}.

\subsection{Variations with Power law Flux}
\label{sec:variations}
Fig.~\ref{fig:highecut_continuum_flux} reveals that the \texttt{highecut} spectral parameters from the four 
outbursts have the same variations with power law flux and essentially the same values at any given flux 
level. Thus the accretion process for matter onto the neutron star was the same for all four outbursts.

A complete discussion of the column density variations is presented in K\"uhnel et al. (in preparation) where a large 
($\times$3) 
column density enhancement event is detected in the 2011 May outburst (red points in 
Fig.~\ref{fig:highecut_continuum_flux}a) and a smaller ($\times$0.5) enhancement 
is seen in the 2010 December data (blue points in Fig.~\ref{fig:highecut_continuum_flux}a). These values associated 
with the large and small increases in column density are significant outliers from the overall trend of decreasing 
column density with increasing power law flux above a few 10$^{-9}$ ergs cm$^{-2}$ s$^{-1}$ and a constant 
value below that flux value.  

The power law index has a strong positive correlation with power law flux (Fig.~\ref{fig:highecut_continuum_flux}b).  
The four early 2010 August points noted earlier are now indistinguishable from the overall correlation with flux, 
which supports the contention that the flux is the primary driver of the continuum spectral 
shape. The continuum cut-off break energy (Fig.~\ref{fig:highecut_continuum_flux}c) exhibits two distinct levels in 
the \texttt{highecut} model: $\sim$7.8 keV and $\sim$5.0$-$6.5 keV. 
The sharp transition from high to lower cut-off break energies appears at
$\sim$6.5$\times 10^{-9}$ ergs cm$^{-2}$ s$^{-1}$, or (4.5$\pm$0.9)$\times 10^{36}$ ergs s$^{-1}$ for a 
distance of 2.4$\pm$0.5 kpc \citep{Parkes80}. The continuum folding energy shows an overall trend of 
decreasing energy with increasing power law flux (Fig.~\ref{fig:highecut_continuum_flux}d).

The cyclotron line energy ($E_\mathrm{cyc}$; Fig.~\ref{fig:highecut_continuum_flux}g) is found to range from 
50 to 60 keV with an ever increasing value with power law flux in agreement with \citet{Klochkov11}.  
The widths ($W_\mathrm{cyc}$; Fig.~\ref{fig:highecut_continuum_flux}h) vary with power law flux from 4 to 12 keV, and the depths
($\tau_l$; Fig.~\ref{fig:highecut_continuum_flux}i) range from $\sim$1.1 down 
to $\sim$0.4, beyond which the depth is not significantly detected.  For the cyclotron line energy and width, 
a positive correlation is clearly seen, while that for the depth or strength is less clear. 

The iron line flux (Fig.~\ref{fig:highecut_continuum_flux}f) shows a relatively smooth increase with flux.
The iron line equivalent width variation with power law flux (Fig.~\ref{fig:highecut_continuum_flux}e)
was somewhat constant versus 
flux with large scatter between 2 and 4$\times 10^{-9}$ ergs cm$^{-2}$ s$^{-1}$ and at fluxes in excess of 
10$\times 10^{-9}$ ergs cm$^{-2}$ s$^{-1}$. 

\subsection{\texttt{Cutoffpl} Fits} \label{sec:cutoffpl} 
Fig.~\ref{fig:cutoffpl_continuum_flux} shows the variation of spectral parameters 
with cutoff power law flux. Due to the shape of the cutoff power law and the blackbody component, the shape of the 
continuum is somewhat different than that of a straight power law. Therefore the values of the column density and 
power law index are slightly different than those from the \texttt{highecut} model. The column density still drops with 
increasing cutoff power law flux above $\sim 3 \times 10^{-9}$ ergs cm$^{-2}$ s$^{-1}$ and the two column density 
enhancements are still above the trend. Where the \texttt{highecut} column density values leveled off at a value of 
$\sim 7 \times 10^{22}$ cm$^{-2}$, those for \texttt{cutoffpl} drop to $\sim 3  \times 10^{22}$ cm$^{-2}$ below 
$\sim 1 \times 10^{-9}$ ergs cm$^{-2}$ s$^{-1}$. Similarly for the power law index, while \texttt{highecut} values 
have a linear series of values over the entire power law flux range, the \texttt{cutoffpl} values exhibit an abrupt 
change from the 
linear trend of the index at $\sim 1 \times 10^{-9}$ ergs cm$^{-2}$ s$^{-1}$ to that of a constant value of $\sim$0.75 
with large uncertainties. The blackbody temperature is constant at $\sim$1.1 keV from the lowest cutoff power law 
fluxes to $\sim 1 \times 10^{-9}$ ergs cm$^{-2}$ s$^{-1}$, beyond which it rises linearly with flux to $\sim$2.7 keV. At 
ever-increasing cutoff power law flux, the trend is to decrease somewhat, albeit with large uncertainties. The 
cyclotron line parameters and the iron line fluxes variations are quite similar to those found in the \texttt{highecut} 
modeling.

\subsection{Color/Intensity Diagrams}
\begin{figure}
\includegraphics[width=2.8in]{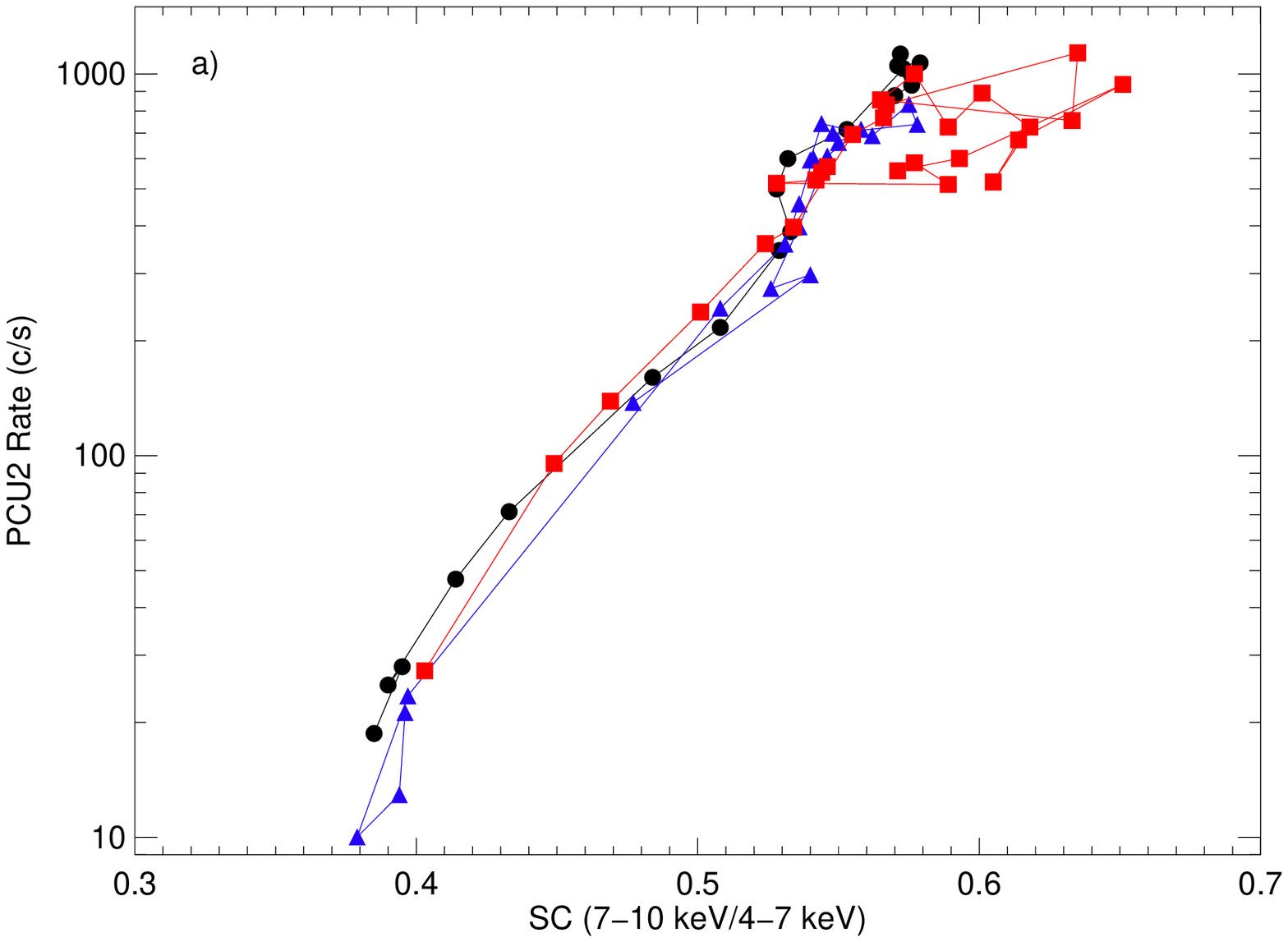}
\includegraphics[width=2.8in]{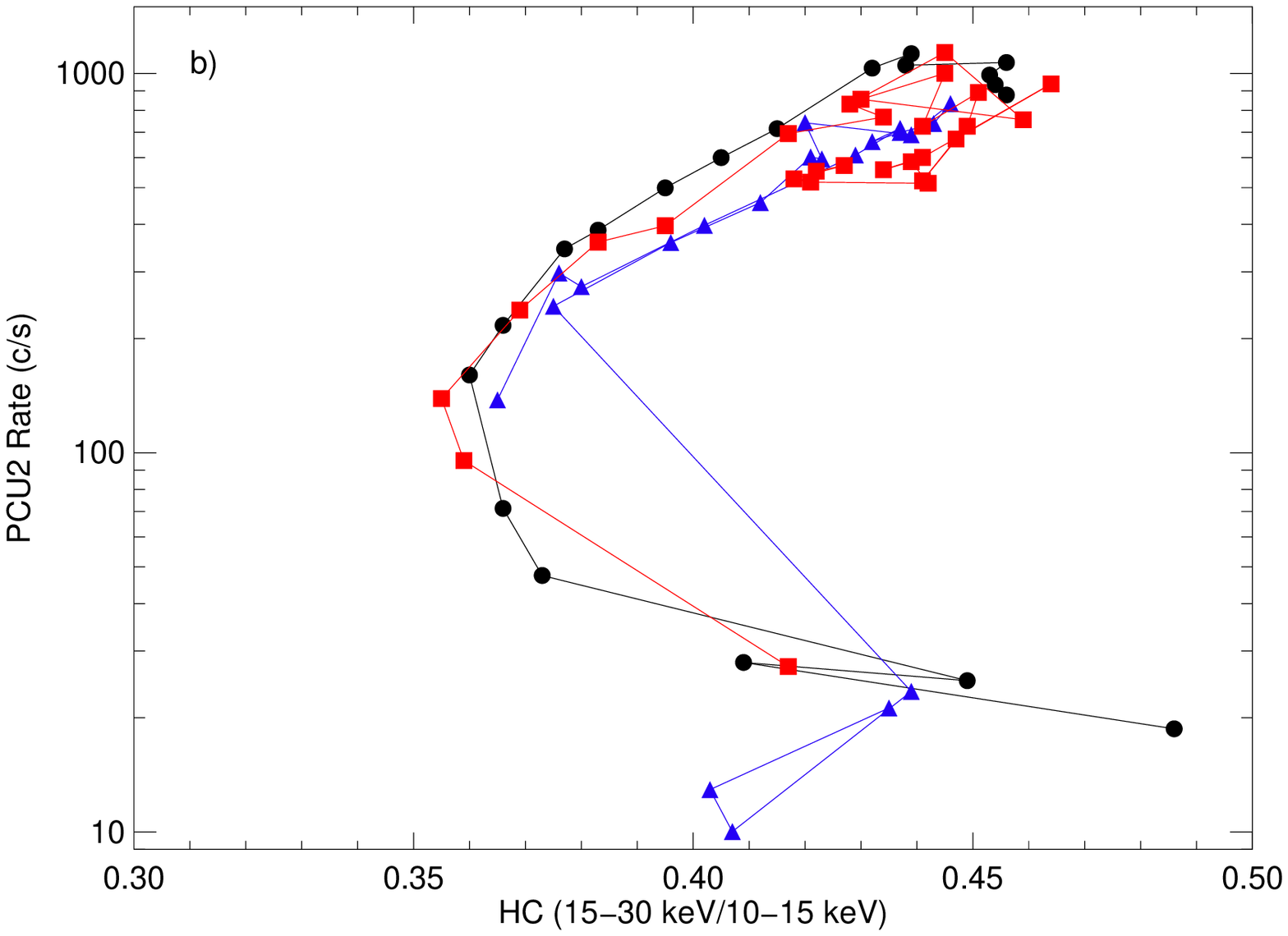}\\
\caption{\textbf{a}: Soft color (7$-$10 keV/4$-$7 keV) plotted versus PCU2 4$-$30 keV counting rate for the three 
outbursts observed, following the 
prescription of \citet{Reig13}. \textbf{b}: Hard color (15$-$30 keV/10$-$15 keV) plotted versus PCU2 4$-$30 keV 
counting rate.The colors for the 
three outbursts are the same as Fig~\ref{fig:three_outbursts}.\label{fig:sc_flux}}
\end{figure}

We have created the GX~304$-$1 soft (SC) and hard color (HC) versus intensity diagrams following the prescription 
of \citet{Reig13} with the intensity being the PCU2 4$-$30 keV count rate, the soft color being the ratio of the PCU2 
4$-$7 keV to 7$-$10 keV count rates, and the hard color being the ratio of 15$-$30 keV to 10$-$15 keV rates. 
Fig~\ref{fig:sc_flux}a shows the SC versus intensity and Fig~\ref{fig:sc_flux}b the HC versus intensity. Both show 
increases in the color indices with increasing intensity, as expected for a hardening of the power law flux with intensity.
For the SC/intensity diagram, the 2010 August and 2010 
December outbursts follow the same track throughout their observations. The 2011 May outburst also follows the 
same track except for the period of time when the large column density enhancement was present. The larger 
column density values reduce the 4$-$7 keV fluxes and therefore raise the value of the soft color ratios. The hard 
color/intensity plot shows overlapping tracks for the three outbursts without the large deviations at 
higher intensity seen in the soft color plot, except for two of the the four last observations in 2010 December. 
The general trend of a reduction in soft and hard color indices throughout the outbursts can be attributed to the steepening 
of the power law component as the power law flux decreased, and the reversal of the hard color diagram below 
$\sim$100 counts per second may be attributable to the hardening of the spectrum at low fluxes as expressed in 
the spectral fitting by the increased values of E$_{fold}$. All together 
the soft and hard color/intensity diagrams imply that the accretion onto the neutron star was nearly identical in all 
three outbursts.

\subsection{Variance-Weighted Averages}\label{sec:variance}
Since we have demonstrated the nearly identical spectral performance in the four outbursts through the 
continuum parameters' variations versus source flux and through the overlapping color intensity 
diagrams, we have performed a variance-weighted average of the power law index, cyclotron line energy and width,
the iron line flux, and its equivalent width in six flux 
bins of width 2$\times10^{-9}$ ergs cm$^{-2}$ s$^{-1}$ from zero to 12$\times10^{-9}$ ergs cm$^{-2}$ s$^{-1}$ 
in order to reduce the scatter in parameter values and reduce uncertainties. The residual flux and cyclotron 
line depth values are given without the lowest flux bin since at most only lower limits were achieved.

The average CRSF energy, $\langle E_{\mathrm{cyc},i}\rangle$, in a certain flux bin, $i$, was found by minimizing the $\chi_i^2$ defined as
\begin{equation}
\chi_i^2 = \sum_k \frac{(E_\mathrm{cyc,k} - \langle E_{\mathrm{cyc},i}\rangle)^2}{S(\sigma_\mathrm{cyc,k}^+, \sigma_\mathrm{cyc,k}^-)^2} 
\end{equation}
$\quad\text{with}\quad S=\begin{cases}
\sigma_\mathrm{cyc,k}^+ \quad\text{for~} E_\mathrm{cyc,k} - \langle E_{\mathrm{cyc},i}\rangle \le 0,\\
\sigma_\mathrm{cyc,k}^- \quad\text{for~} E_\mathrm{cyc,k} - \langle E_{\mathrm{cyc},i}\rangle > 0,\\
\end{cases}$\\

\noindent with the CRSF energy, $E_\mathrm{cyc,k}$, of each observation $k$ falling into the flux bin $i$, 
and the upper or the lower uncertainty, $\sigma_\mathrm{cyc,k}^+$ and $ \sigma_\mathrm{cyc,k}^-$, of the 
CRSF energy. The average CRSF width ($W_\mathrm{cyc,k}$), depth ($\tau_\mathrm{cyc,k}$), and residual 
flux ($r=F_\ell(E_\mathrm{cyc})/F_\mathrm{c}(E_\mathrm{cyc})$) in each flux bin was found in the same way.
The residual flux is related to the line 'optical depth',  $\tau_\mathrm{\ell}$, as $r=e^{-\tau_\mathrm{\ell}}$.
Note that in case of symmetric uncertainties, i.e., $\sigma_\mathrm{cyc,k}^+$ = $ \sigma_\mathrm{cyc,k}^-$, 
the average CRSF parameter value obtained is equivalent to the mean value weighted by the corresponding 
uncertainties \citep[see, e.g.,][]{bevington}. 

The results are plotted in Fig.~\ref{fig:ecyc_index} where 
the cyclotron line parameters are in the lefthand panels and the power law index, iron line flux and equivalent width 
are in the righthand panels. All, except the residual flux, show positive correlations with flux, with the cyclotron line parameters gradually flattening with increasing flux. In Section \ref{sec:cyclo_fits} below, we show successful fits 
to the cyclotron line parameters with both disk accretion and quasi-spherical accretion models.

\subsection{Comparison to Previous Observations}
\citet{Yamamoto11} presented spectral analyses of \xte observations of the first two thirds of the 2010 August 
outburst, plus that of a \textit{Suzaku} observation on 2010 August 13 after the second \xte observation. 
Their analysis differed from 
that of the present work by only covering the 3$-$20 keV band in PCU2, using no extra Gaussians to augment 
\texttt{hextebackest}, ignoring the HEXTE band from 61$-$71 keV, and normalizing the PCU2 to HEXTE spectra 
by assuming no HEXTE flux above 150 keV. In addition, a different spectral model for the continuum, \texttt{NPEX}, 
was used. Nevertheless, they discovered the cyclotron line and concluded that the line had a positive correlation 
with overall flux or it had a bi-modal distribution. Cyclotron line energies ranged from 49$-$54 keV, albeit with large 
uncertainties on those values from lower luminosities. \citet{Klochkov12} used 6 INTEGRAL observations 
covering the 2012 January outburst to confirm a positive correlation of the cyclotron line energy with flux 
employing the \texttt{cutoffpl} spectral model. The range of INTEGRAL cyclotron line energies was 48$-$ 55 keV. 
In the present work we have detected the cyclotron line in 54 of 69 observations, 
with individual energies ranging from 49 keV to 59 keV. 
\citet{Jaisawal16} recently presented results from two \textit{Suzaku} observations, one of which occurred at the 
time of the \xte observations (4 \& 5) on 2010 August 13. Their use of the \texttt{NPEX} and \texttt{CYCLABS} 
models for spectral fitting does not allow comparison to the present results due to the differing assumptions of 
spectral shapes. They did report, however, that the higher cyclotron line energies did occur for the 
brighter observation, as one would expect from the positive correlation with luminosity.

\section{DISCUSSION}
The present work covers three outbursts of GX~304$-$1 with twenty or more observations per outburst 
over a range of luminosities. The detailed modeling and corrections to the PCU2 background via the 
\texttt{RECOR} function and to the HEXTE background utilizing additional flux from the four prominent background 
lines in addition to \texttt{HEXTEBACKEST} plus \texttt{RECOR} has resulted in best-fit spectral parameters from 
spectra covering 3$-$100 keV with significant overlap in the 20$-$60 keV band, which allows for 
confirming the lower energy portions the HEXTE background subtraction.

\subsection{Scaling Laws of CRSF Properties}
\label{sec:cyclo_fits}

The correlations of the CRSF properties with flux during outbursts of GX 304-1 suggest that the mass accretion 
rate onto the neutron star poles is the driver of the CRSF changes. The CRSF formation is a very complicated 
problem that can be solved only numerically by taking into account the dynamics of the accretion flow near the 
neutron star surface coupled with the radiation in the strong magnetic field. Qualitatively, however, it is clear that 
at low accretion rates, when the radiation field is not very strong, the braking of the flow is mediated by Coulomb 
interactions in the accreting plasma (e.g. \cite{Nelson93}), while at high accretion rates the flow is decelerated 
mostly due to interactions with photons \citep{Davidson73}. The transition between these two extreme cases 
occurs gradually around some critical luminosity $\sim 10^{37}$~erg s$^{-1}$, which depends on the geometry 
of the flow and the structure of magnetic field near the neutron star surface and may be different in different 
sources (see \cite{Basko76}, and more recent calculations in \cite{Becker12}, \cite{Mushtukov15a}). At low 
luminosities, the CRSF energy in some sources (e.g. Her X-1) was found to positively correlate with X-ray flux, 
and in the simplest interpretation this can be due to a closer location of the effective site of CRSF formation 
with respect to the neutron star surface, where the magnetic field is stronger, with increasing mass accretion 
rate \citep{Staubert07}. Clearly, with increasing X-ray luminosity, transition to the radiation-dominated regime 
occurs, where the effective height of accretion column gets higher, and hence the CRSF energy is expected 
to decrease with increasing X-ray flux, as indeed observed in some bright transient X-ray pulsars 
(e.g. V0332+53, \cite{Tsygankov06}). \citet{LaParola16} make similar assumptions in the fitting of the Vela X-1 
first harmonic positive correlation of cyclotron line energy with luminosity.

Here we suggest a possible interpretation of the observed correlations in GX~304-1, assuming that the source, 
even at the highest X-ray flux in the outburst, is indeed well below the critical luminosity \citep{Becker12}, which 
implies it remains in the regime where the radiation effects are subdominant 
in braking the accretion flow. This will enable us to use the results of detailed calculations of the (effectively 
one-dimensional in this case) plasma flow above the neutron star surface. In this way we will obtain simple 
formulae that can be used to fit the observed correlations of the CRSF energy, $E_\mathrm{cyc}$, its width, 
$W$, the line residual flux, $r$, and its related line optical depth $\tau_\ell$ with changing X-ray flux
(see Table \ref{t:exponents}).

\subsubsection{Physical setup}

In the GX~304$-$1 case, the accretion flow 
decelerates in a collisionless shock \citep{LangerRappaport82}. The height 
of the collisionless shock above the neutron star surface, $H_\mathrm{s}$, is governed by energy exchange 
between protons (which tap most of the post-shock energy) and electrons, and the cooling
of electrons and ions via bremsstrahlung and cyclotron losses; photons participate in the post-shock dynamics 
of the flow via resonant and non-resonant scattering 
on electrons in the strong magnetic field, but their density is insufficient to produce a radiation-dominated shock  
\citep{BykovKrassilshchikov04}.  
With increasing mass accretion rate, $H_\mathrm{s}$ decreases because the plasma density increases, 
and the line formation region within the cyclotron resonant layer downstream of the shock gets closer to the 
neutron star surface. The CRSF formation is governed by the resonance electron scattering of thermal 
photons produced at the base of the accretion mound where most of the free-fall energy is released. 
Thus, the scaling with mass accretion rate appears for the line centroid energy, its width, residual flux, and depth.

A photon with energy $\hbar\omega$ experiences resonant scattering on an electron at the fundamental cyclotron
resonance frequency $\omega_\mathrm{cyc}$ in the magnetic field $B$, and
$E_\mathrm{cyc}=\hbar\omega_\mathrm{cyc}=\hbar eB/(m_\mathrm{e}c)$, where $e$ is the electron charge, 
$m_\mathrm{e}$ is the electron mass, and $c$ is the speed of light. Therefore, in the plasma above the neutron 
star surface, for each photon of energy $E$ there should be the cyclotron resonance scattering
radius, $r_\mathrm{res}=R_\mathrm{NS}((\hbar eB_\mathrm{NS}/m_\mathrm{e}c)/E)^{1/3}$, 
due to inhomogeneity of the dipole magnetic field, $B = B_\mathrm{NS}(R_\mathrm{NS}/r)^3$, where 
$R_\mathrm{NS}$ is the neutron star radius  and $B_\mathrm{NS}$ is the surface magnetic field at the magnetic 
pole \citep{Zheleznyakov96}. 

\clearpage
\onecolumn
\begin{figure}
\includegraphics[width=6.6in]{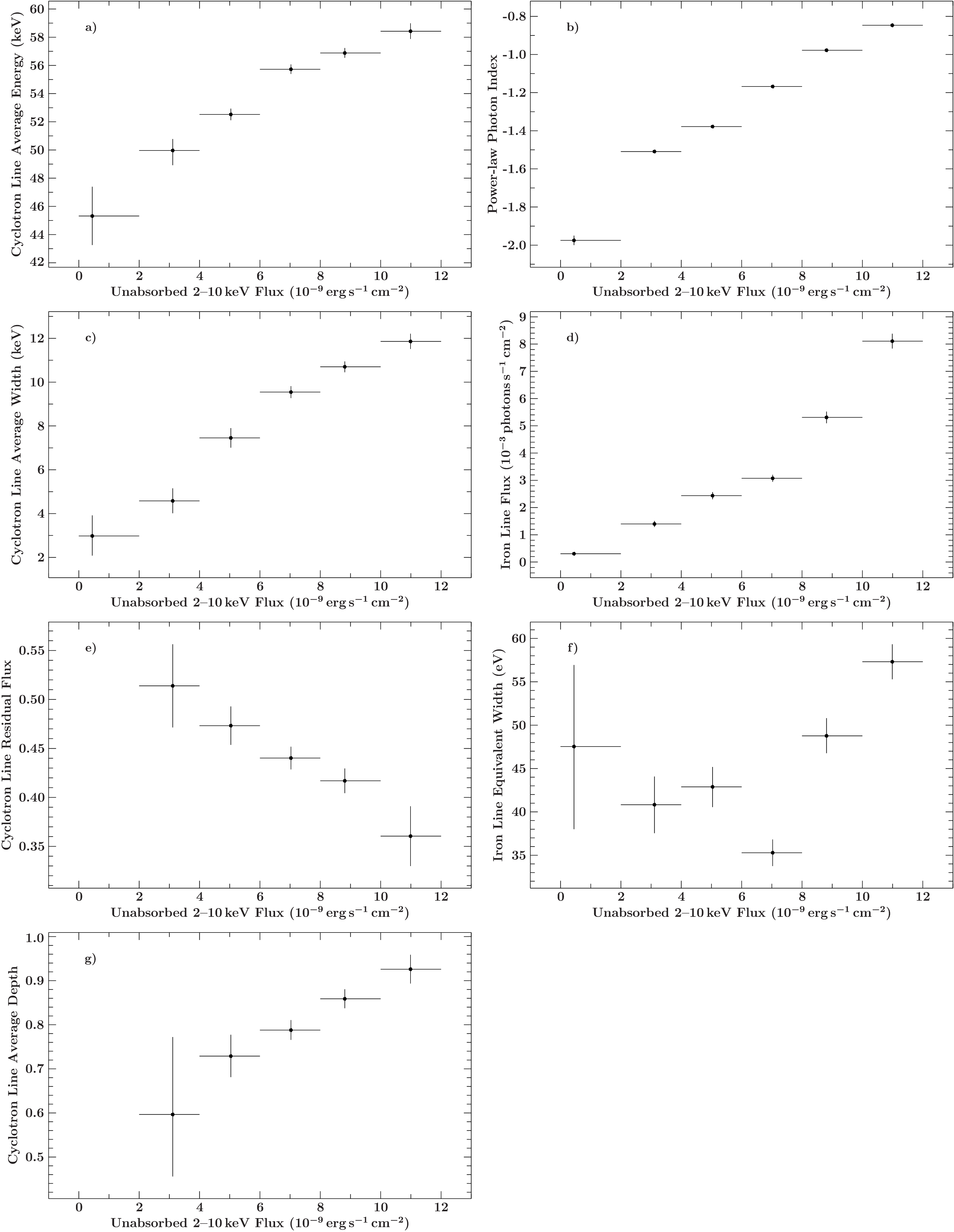}
\caption{The variance weighed average cyclotron line energies (a), power law indices (b), cyclotron 
line widths (c), iron line flux (d), cyclotron line residual fluxes (e), iron line equivalent widths
(f), and cyclotron line average depth (g) in six 2$-$10 keV, unabsorbed, power law flux bins for the energy and width of the cyclotron lines, and in five flux bins for the residual flux and the depth. 
\label{fig:ecyc_index}}
\end{figure}

\clearpage
\twocolumn
\noindent As shown in \citet{Zheleznyakov96}, 
the width of the resonant layer for the assumed dipole magnetic field is 
$\Delta r_\mathrm{res}\sim \beta_{T_\mathrm{e}} r_\mathrm{res}/3$, where 
$\beta_{T_\mathrm{e}}=v_{T_\mathrm{e}}/c\sim 1/10$ is the thermal velocity of 
post-shock electrons; for typical temperatures $T_\mathrm{e}\sim 10$~keV and 
$\hbar\omega_\mathrm{cyc}\sim 50$~keV, $\Delta  r_\mathrm{res}\sim 6\times 10^4$~cm can be comparable 
with the shock size $H_\mathrm{s}$ and thus can substantially modify the CRSF formation. 
Note that the post-shock electron temperature $T_\mathrm{e}$ does not vary substantially.
The characteristic 
optical depth of the resonant layer in the inhomogeneous dipole magnetic field $B$ is \citep{Zheleznyakov96} 
\begin{multline}\label{taures}
\tau_\mathrm{res}=\frac{16}{3}\frac{\pi^2 e^2}{m_\mathrm{e}c}\frac{n_\mathrm{e}\Delta r_\mathrm{res}}{\omega_\mathrm{cyc}}
\sim \\10^4\left(\frac{n_\mathrm{e}}{10^{20}\mathrm{cm}^{-3}}\right)\myfrac{50\,\mathrm{keV}}{E_\mathrm{cyc}}^\frac{4}{3}\myfrac{R_\mathrm{NS}}{10^6\mathrm{cm}}\myfrac{B_\mathrm{NS}}{10^{12}\mathrm{G}}^\frac{1}{3}\,.
\end{multline} 
It is also known that during the cyclotron resonance scatterings
the number of scatterings of a photon in the resonant layer  
scales as the optical depth, $N_\mathrm{sc}\propto \tau_\mathrm{res}$, in contrast to the 
scaling $N_\mathrm{sc}\propto \tau^2$ for the non-resonance Thomson scattering 
\citep{Wasserman80,Lyutinkov06,Garasev11}. 
This has an important consequence for the CRSF  discussed below.  

\begin{table}
\caption{Fitting formulae for the CRSF centre energy, $E_\mathrm{cyc}$, width $W$, the
residual line flux $r$ and line optical depth $\tau_\ell$
as a function of X-ray flux $F_\mathrm{x}$ in the collisionless shock braking regime, assuming a 
magnetospheric radius $R_\mathrm{m}\sim M^{-x}$}
\begin{tabular}{lcc}
\hline
\hline
Formula & $x_\mathrm{d}=2/7$ & $x_\mathrm{s}=2/11$\\
\hline
$E_\mathrm{cyc}(F_\mathrm{x})=E_0(K_1 F_\mathrm{x}^{-\alpha}+1)^{-3}$ & $\alpha_\mathrm{d}=5/7$ & $\alpha_\mathrm{s}=9/11$\\
$W(F_\mathrm{x})=K_2E_\mathrm{cyc}^{1/3}(F_\mathrm{x})F_\mathrm{x}^\beta$ & $\beta_\mathrm{d}=5/14$ &$\beta_\mathrm{s}=9/22$\\
$r(F_\mathrm{x})=K_3 E_\mathrm{cyc}^{2/3}(F_\mathrm{x}) F_\mathrm{x}^{-\gamma} $ & $\gamma_\mathrm{d}=5/14$ &$\gamma_\mathrm{s}=9/22$\\
$\tau_\ell(F_\mathrm{x})=K_4 + \ln\left(E_\mathrm{cyc}^{-2/3}(F_\mathrm{x}) F_\mathrm{x}^{\gamma}\right) $ & $\gamma_\mathrm{d}=5/14$ &$\gamma_\mathrm{s}=9/22$\\
\hline
\end{tabular}
\label{t:exponents}  
\end{table}

The height of the collisionless shock is $H_\mathrm{s}\sim (v_0/4) t_\mathrm{eq}\propto 1/n_\mathrm{e}$  
\citep[here $t_{eq}$ is the electron-proton equilibration time; see e.g. ][]{ShapiroSalpeter75}. 
The electron number density behind the shock can be estimated from the mass continuity equation
\begin{equation}{}
n_\mathrm{e}= \myfrac{\dot M}{A}\myfrac{4}{v_0 m_\mathrm{p}}\,.
\end{equation}
The accretion area $A$ is determined by the magnetospheric radius $R_\mathrm{m}$ and for the dipole field 
should vary as 
$A\sim (R_\mathrm{NS}\sqrt{R_\mathrm{NS}/R_\mathrm{m}})^2\propto 1/R_\mathrm{m}$.
In the general case, the magnetospheric radius is inversely proportional to the mass accretion rate, 
$R_\mathrm{m}\propto \dot M^{-x}$,
where $x=x_\mathrm{d}=2/7$ for disc accretion or Bondi quasi-spherical accretion, or
$x=x_\mathrm{s}=2/11$ for quasi-spherical settling accretion \citep{Shakura12}, where the latter may be 
realized in the case of GX 304-1 \citep{Postnov15a}. 

With these scalings, we find for the electron number density $n_\mathrm{e}\propto \dot M/A\propto \dot M^{1-x}$.
Therefore, the characteristic shock height scales with accretion rate as  
\begin{equation}\label{l*}
H_\mathrm{s}\propto 1/n_\mathrm{e}\propto A/\dot M \propto \dot M^{-\alpha}\,,
\end{equation}
and $n_\mathrm{e}\propto \dot M^\alpha$, 
where $\alpha=\alpha_\mathrm{d}=1-x_\mathrm{d}=5/7$ for disc or Bondi quasi-spherical accretion
and $\alpha=\alpha_\mathrm{s}=1-x_\mathrm{s}=9/11$ for quasi-spherical settling accretion. 


\subsubsection{Cyclotron line energy scaling with X-ray flux}

Consider the case where the characteristic size of the plasma region, $H_\mathrm{s}\lesssim 10^5$~cm, 
is comparable with the thickness of the resonant layer, $\Delta r_\mathrm{res}\sim 6 \times 10^4$~cm. The optical
depth of the resonant layer is very large (see \Eq{taures}). 
The CRSF is formed at some effective energy corresponding to the magnetic field at some height within the 
resonance layer, which is related to the shock height, $H_\mathrm{CRSF}\lesssim H_\mathrm{s}$, and hence 
should have the same dependence on the mass accretion rate as H$_\mathrm{s}$.
The CRSF energy is $E_\mathrm{cyc}\propto B(R)\propto 1/R^3$, 
and noticing that $R=R_\mathrm{NS}+H_\mathrm{CRSF}$, we find, for the assumed dipole magnetic field, 
\begin{equation}
{E_\mathrm{cyc}(\dot M)}=E_0\myfrac{R_\mathrm{NS}}{H_\mathrm{CRSF}(\dot M)+R_\mathrm{NS}}^3\,,
\end{equation}
where $E_0$ corresponds to the line emitted from the NS surface magnetic field $B_\mathrm{NS}$.
Clearly, the line dependence on the observed X-ray flux is entirely determined by 
how the collisionless shock height $H_\mathrm{s}$ responds to the variable mass accretion rate (see \Eq{l*} above).     
As the observed X-ray flux $F_\mathrm{x}$ is directly proportional to $\dot M$ and 
introducing the relation $H_\mathrm{CRSF}/R_\mathrm{NS}=K_1 F_\mathrm{x}^{-\alpha}$, we arrive at 
\begin{equation}
\label{e:sol1}
E_\mathrm{cyc}(F_\mathrm{x})=E_0(K_1 F_\mathrm{x}^{-\alpha}+1)^{-3}\,.
\end{equation}
The constant $K_1$, which determines the CRSF location height, $H_\mathrm{CRSF}/R_\mathrm{
NS}$, can be found from fitting the observational data, $\alpha_d$=5/7 and $\alpha_s$=9/11. Generally, $K_1$ 
may be a function of $\dot M$ as well, but in view of lack of solid theory of CRSF formation downstream the 
shock we will assume $K_1=const$.

\subsubsection{Cyclotron line width scaling with X-ray flux}

As discussed above, the resonant line is formed by multiple scatterings 
in a resonant layer behind the 
shock. In each single scattering on an electron, moving essentially in one dimension along the magnetic field lines, 
the energy of the resonant photon is Doppler shifted, 
$(\Delta E_\mathrm{cyc}/E_\mathrm{cyc})_1=\pm\beta_{T_\mathrm{e}}$, where 
the post-shock electron temperature $T_\mathrm{e}\sim 10$~keV  does not strongly vary in the scattering region. 
Therefore, after many scatterings the CRSF width will be $W/E_\mathrm{cyc}
\simeq \sqrt{(\Delta E_\mathrm{cyc}/E_\mathrm{cyc})_1^2 N_\mathrm{sc}}\propto \sqrt{T_\mathrm{e} N_\mathrm{sc}}\propto \sqrt{N_\mathrm{sc}} \propto \sqrt{\tau_\mathrm{res}}$. As follows from \Eq{taures}, $\tau_\mathrm{res}\propto n_\mathrm{e}/E_\mathrm{cyc}^{4/3}$, and hence  
the observed CRSF width can be fitted
by the following formula:
\begin{equation}
\label{e:Wsol1}
W(F_\mathrm{x})=K_2E_\mathrm{cyc}^{1/3}(F_\mathrm{x})F_\mathrm{x}^\beta\,,
\end{equation}
where $\beta=\alpha/2$,  $E_\mathrm{cyc}(F_\mathrm{x})$ is determined by formula (\ref{e:sol1}) and $K_2$ is 
a constant. 

\subsubsection{Cyclotron line residual flux and line `depth' scaling with X-ray flux}

Finally, we consider how the residual flux at the line center changes with X-ray luminosity in our model.
Consider the simplest case of an isothermal atmosphere with resonance scattering (the Eddington model),
which can be a good first approximation for the resonant layer behind the collisionless shock front. 
It is easy to check that in our case with $E_\mathrm{cyc}=\hbar\omega_\mathrm{cyc}\sim 50$~keV 
$\gg kT_\mathrm{e}\sim$ 10 keV  
and with typical densities $n_\mathrm{e}\sim 10^{20}$ cm$^{-3}$, the ratio of the absorption
to scattering is very small, i.e. we can neglect absorptions of scattered photons altogether. 
According to the theory of resonance scattering lines in an isothermal atmosphere 
\citep[see, e.g., ][(chapter 7)]{Ivanov69}, and \citep{Ivanov73},
in the limit of high survival probability of scattered photons in the continuum
and neglecting the absorption, the residual flux $r$ of a resonance line (the so-called '$\lambda$-solution') is 
determined solely by the number of scatterings of the line photons and scales as 
\begin{equation}
\label{}
r = \frac{1}{\sqrt{N_\mathrm{sc}}}\propto \frac{1}{\sqrt{\tau_\mathrm{res}}} \propto \frac{E_\mathrm{cyc}^{2/3}}{\sqrt{n_\mathrm{e}}}\,.
\end{equation}
Plugging in the scaling $n_\mathrm{e}\propto \dot M^\alpha$, we can recast this expression
into the convenient form:
\begin{equation}\label{e:r}
r(F_\mathrm{x})=K_3E_\mathrm{cyc}^{2/3}(F_\mathrm{x})F_\mathrm{x}^{-\gamma}\,,
\end{equation}  
where $K_3$ is a constant and $\gamma=\alpha/2$, yielding 
$\gamma_\mathrm{d}=5/14$ and $\gamma_\mathrm{s}=9/22$
for disc and quasi-spherical accretion, respectively.

It is also possible to introduce the line `optical depth' $\tau_\ell$ defined as
$r=e^{-\tau_\ell}$. It is this parameter that is usually inferred from data analysis. The 
application of formula \eqn{e:r} in this case is straightforward:
\begin{equation}
\label{e:taul}
\tau_\ell(F_\mathrm{x})=K_4+\ln(E_\mathrm{cyc}^{-2/3}(F_\mathrm{x})F_\mathrm{x}^{\gamma})\,,
\end{equation}
where $K_4$ is the constant to be found from fitting. (Note that the fitting procedure
of $\tau_\ell(F_\mathrm{x})$ should be done independently of fitting $r(F_\mathrm{x})$, since these 
quantities are derived independently from the data analysis.)

\subsubsection{Fitting the Variance-weighted Data}

\begin{figure}
\includegraphics[width=3.4in]{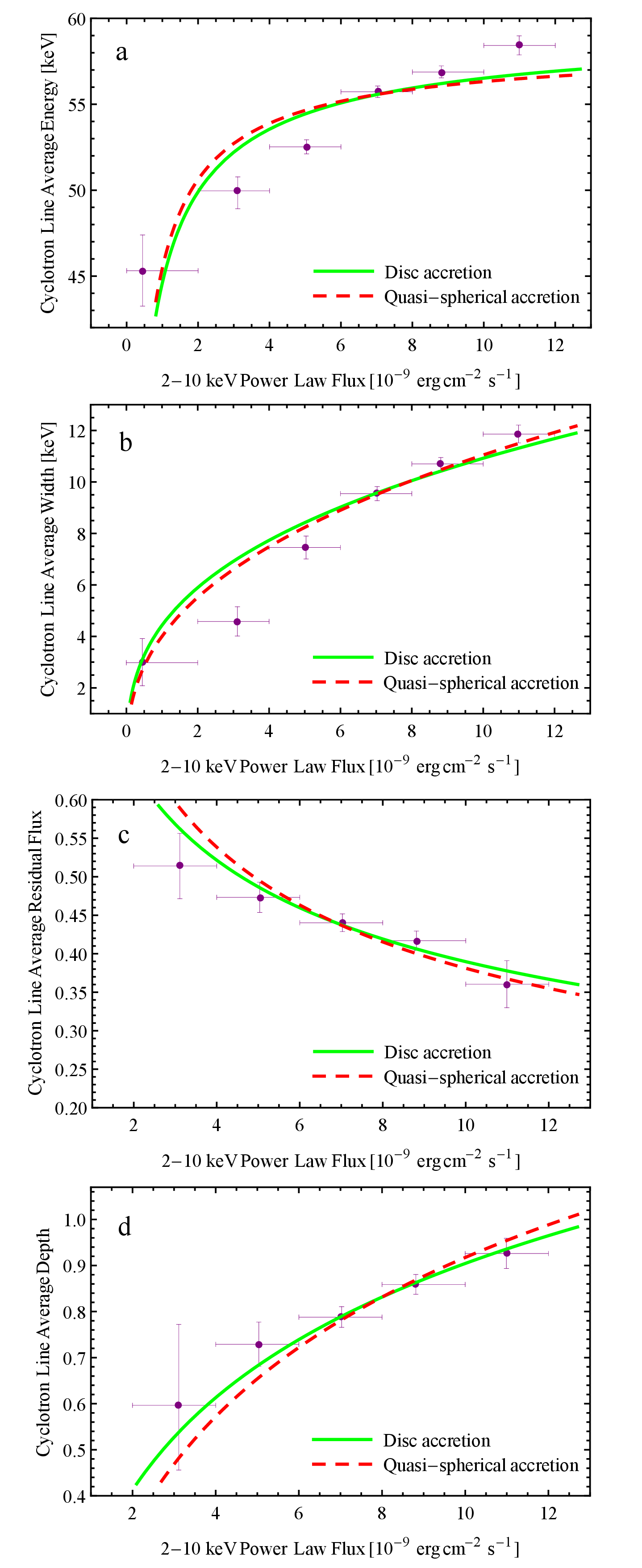}
\caption{Best-fits of the observed cyclotron line parameters versus X-ray flux  for two possible types of accretion in GX 304-1, disc or quasi-spherical,
shown by solid green line and dashed red line, respectively.
\textbf{a)} Cyclotron line energy 
(\ref{e:sol1}); \textbf{b)} cyclotron line 
width 
(\ref{e:Wsol1}); \textbf{c)}
cyclotron line residual flux
(\ref{e:r});  and \textbf{d)}
cyclotron line 'depth' 
(\ref{e:taul}). 
Horizontal bars indicate the width of the flux bins inside which averaging was done.
}
\label{f:plotE}
\end{figure}

The results of fitting the variance-weighted data (described in section \ref{sec:variance} and shown in 
Fig.~\ref{fig:ecyc_index})
by formulae \eqn{e:sol1}, \eqn{e:Wsol1}, \eqn{e:r}, and \eqn{e:taul}
are shown in Fig.~\ref{f:plotE} and
listed in Table~\ref{t:fits}. We do not show formal errors in the fitting coefficients due to roughness of the model 
physical assumptions (constant electron temperature, approximate treatment of the cyclotron resonance scattering, 
etc.). It is also seen that the data do not allow us to distinguish between the two possible
dependences of the magnetospheric radius on $\dot M$ for different types of accretion (disc or quasi-spherical one). 

\begin{table}
\caption{Best-fit parameters of
$E_{\rm cyc}(F_{\rm x})$, $W(F_{\rm x})$, $r(F_{\rm x})$
and $\tau_\ell(F_{\rm x})$ \label{t:fits}}
\begin{tabular}{lcc}
\hline
\hline
Param & Disc accretion & Quasi-spherical accretion \\
\hline
\multicolumn{3}{c}{Fig. \ref{f:plotE}a: fit $E_{\rm cyc}(F_{\rm x})$ by \Eq{e:sol1} }\\
\hline
$E_0$ & $59.99$ &  $58.62$\\
$K_1$ & $0.1$ &  $0.09$ \\
\hline
\multicolumn{3}{c}{Fig. \ref{f:plotE}b: fit $W(F_{\rm x})$ by \Eq{e:Wsol1}}\\
\hline
$K_2$ & $1.25$ &  $1.12$ \\
\hline
\multicolumn{3}{c}{Fig. \ref{f:plotE}c: fit $r(F_{\rm x})$ by \Eq{e:r}}\\
\hline
$K_3$ & $0.06$ &  $0.07$\\
\hline
\multicolumn{3}{c}{Fig. \ref{f:plotE}d: fit $\tau_\ell(F_{\rm x})$ by \Eq{e:taul}}\\
\hline
$K_4$ & $2.77$ &  $2.66$\\
\hline
\hline
\end{tabular}\\
$E_0$ is energy in keV\\
$K_1$ is flux$^{\alpha}$ ($10^{-9\alpha}$ erg$^\alpha$ cm$^{-2\alpha}$ s$^{-\alpha}$)\\
$K_2$ is energy$^{2/3}$ flux$^{-\beta}$ (keV$^{2/3}$ 10$^{9\beta}$ erg$^{-\beta}$ cm$^{2\beta}$ s$^{\beta}$)\\
$K_3$ is energy$^{-2/3}$ flux$^{\gamma}$ (keV$^{-2/3}$ 10$^{-9\gamma}$ erg$^{\gamma}$ cm$^{-2\gamma}$ s$^{-\gamma}$)\\
$K_4$ is dimensionless\\
 \label{t:fitting}
\end{table}

With further increase in accretion rate, the transition to
radiation braking regime and the appearance of an optically thick accretion column
should occur \citep{Basko76}. The critical luminosity for the transition is expected near $10^{37}$erg s$^{-1}$ 
(Becker et al. 2012, see also Mushtukov et al. 2015a for recent  more accurate calculations). While the brightest 
single observation only reached $\sim(7\pm1.4)\times10^{36}$ erg s$^{-1}$ for the 2.4$\pm$0.5 kpc of 
Parkes et al. (1980), it would be interesting to probe the transition between different accretion regimes in 
transient X-ray pulsars with more powerful outbursts.

Note that an alternative explanation of the positive correlations between  $E_\mathrm{cyc}-F_\mathrm{x}$ and
$W-F_\mathrm{x}$ at moderate X-ray luminosities was recently proposed by \citet{Mushtukov15b}.
However, that model predicts the \textit{opposite} sign of the second derivative in the 
$E_\mathrm{cyc}-F_\mathrm{x}$ and 
$W-F_\mathrm{x}$ relations \citep[cf. black solid lines in Fig. 6a and 6b Fig. 7a in ][]{Mushtukov15b}, while the 
simple physical explanation given above is consistent with observations of GX 304$-$1.

\subsection{Outburst Shifts in Orbital Phase}
The shifts in orbital phase applied to the three outbursts can be understood in terms of changes in the size
of the circumstellar disk around V850 Cen. Referring to Fig.~4 in \citet{Postnov15a}, the disk is inclined with 
respect to the orbital plane of the neutron star, and the neutron star passes through the disk at point A, 
accumulating matter that forms a temporary accretion disk \citep{Devasia11}. The lack of a double peak to 
the three outbursts implies that the circumstellar disk does not extend to the recrossing of the line of nodes 
at point B. Changes in the thickness of the circumstellar
disk from one orbit to the next will affect the amount of matter captured in the accretion disk and thus the 
duration of the outburst. The few percent orbital phase shifts imply small variations in the 
circumstellar disk on timescales of a hundred days.

\subsection{Flux Correlation in General}
The strong positive correlations of spectral parameters with source flux, clearly indicate that 
the source flux, or indeed the mass accretion rate, is responsible for the overall continuum shape and 
that of the cyclotron line as well. This is also supported by the nearly identical soft color/intensity curves for the 
three outbursts and the fact that the four early 2010 August observations yield consistency with other observations 
when plotted versus flux as opposed to plotted versus orbital phase. \citet{Kuhnel13} similarly found that the key 
driver for the continuum shape in GRO J1008$-$57 was the power law flux. They found a common spectral model 
based on flux independent parameters and flux correlations for three Type I outbursts 
and a Type II outburst, where the power law flux was the defining variable, when the source was in the subcritical 
state. 

\subsection{Soft Color versus Flux}
We find that the soft color ratio increases with increasing flux along the horizontal branch \citep{Reig13}, 
with excursions from the overall track due to an extra amount of material in the line of sight over about 3 days. 
The hard color ratio shows a similar horizontal branch increase with intensity, but also shows a reversal of 
the trend at the lowest intensities. The changes in the soft and hard color ratios with intensity can be related 
to the overall steepening of the power law index 
with decreasing intensity and its hardening of the falling exponential at the lowest intensities. 
The flattening of spectra with increasing X-ray flux, as is seen in Fig.~\ref{fig:sc_flux}, could be due to 
increase in the optical depth inside the scattering region behind the shock and hence in the
$y$-parameter in the unsaturated Comptonization regime. 

\section{CONCLUSIONS}
This work presents the analysis of \xte observations of the accreting X-ray pulsar GX 304$-$1 that provides 
the finest detail to date on the correlation 
of the cyclotron line parameters (energy, width, depth, and residual flux) with source flux for any accreting 
X-ray binary system. The correlations display, for the first time, a flattening with increasing power law flux.
This is successfully modeled by 
a rather simple one-dimensional physical treatment of both disk accretion and quasi-spherical accretion, 
since in this case no optically thick accretion column is assumed to form above the neutron star polar 
caps, and the 
emergent radiation is thus dynamically unimportant. The neutron star surface magnetic field is measured to be 
$\sim$60 keV in both models. In addition, the correlations of the power law index, break energy, 
and iron line flux with power law flux points strongly to the source flux, and thus the mass accretion rate, 
as the overarching determinant of the spectral behavior. 

\section*{acknowledgements} 
We thank the referee for the careful reading of the paper and the thoughtful comments proferred. We acknowledge the on-going efforts of the Magnet collaboration on accreting X-ray pulsars. Their work over the years has led to a better understanding of emission from the accretion column, and has led to production of physics-based models of both the continuum and cyclotron line shapes. We acknowledge the support of the International Space Science Institute (ISSI) in Bern, Switzerland, for workshops supporting the Magnet collaboration. The work of K. Postnov is supported by RFBF grants 14-02-00657 and 14-02-91345. The work of  M. Gornostaev and N. Shakura (calculations of
the scaling laws) is supported by RSF grant 14-12-00146. The work of D. Klochkov, J. Wilms, and R. Staubert was supported by DFG grants KL2734/2-1 and WI 1860/11-1. M. K\"uhnel acknowledges support by the Bundesministerium f\"ur Wirtschaft und Technologie under Deutsches Zentrum f\"ur Luft- und Raumfahrt grant 50OR1207.




\appendix
\onecolumn
\section{HEXTEBACKEST} \citep[This section is based upon the poster ``Estimating the HEXTE A Background Spectrum'', which was presented at the 9th AAS HEAD meeting, ][]{Pottschmidt06}.
Since the launch of \textsl{RXTE} in 1995, the HEXTE instrument mostly
operated in its standard `rocking' mode where the pointing direction
of each of its two clusters alternated between source and background
measurements in such a way that one cluster was always looking at the
source while the other sampled the background. During the extraction
of source light curves and spectra, each cluster uses its own
background measurements for correction. This allowed HEXTE to achieve
signal to background ratios of <1\% for long observations
($\gtrsim$400\,ks) of weak sources \citep{Rothschild98}. Starting in 2005 December the
rocking mechanism of cluster~A began to display increasingly frequent
interruptions and since 2006 July was permanently fixed in the
on-source staring position. We have developed a FTOOL,
{\texttt{HEXTEBACKEST}}, which for a given observation uses the
background measured by cluster~B to produce an estimated cluster~A
background spectrum. The tool uses a set of channel dependent
parameters to perform a linear transformation of the count rates. We
explain how these parameters were derived, compare estimated and
measured cluster~A backgrounds for archived rocking observations, and
present examples of the application of the method. Cluster~B began experiencing similar rocking interruptions in
2009 December and was permanently fixed in one of the off-source positions at the end of 2010 March.
This enabled cluster~B to collect background data for use with \texttt{HEXTEBACKEST} to estimate cluster~A 
background for the rest of the \xte mission.

\subsection{Introduction}
\begin{figure}
\includegraphics[width=3.2in]{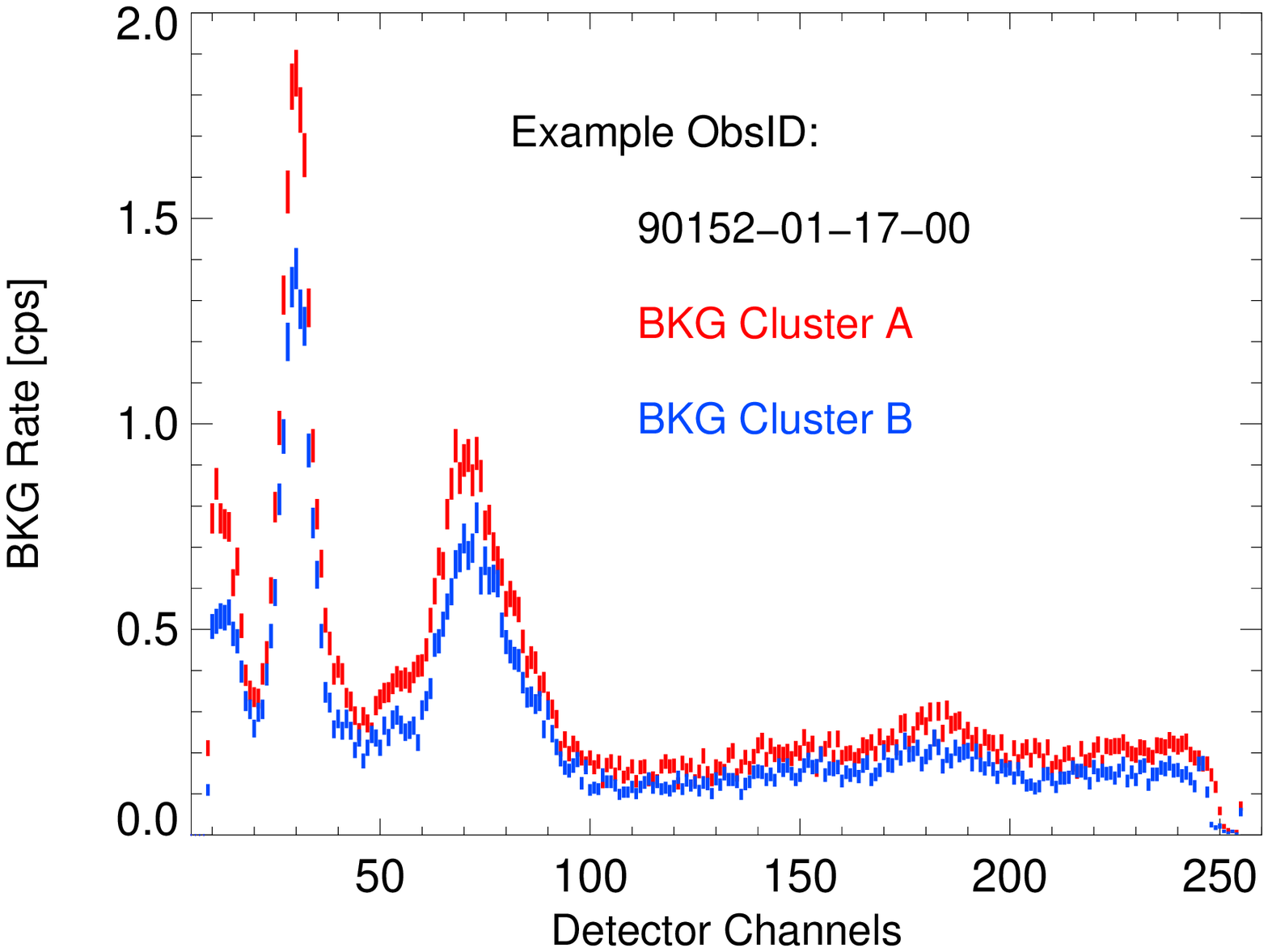}
\includegraphics[width=3.2in]{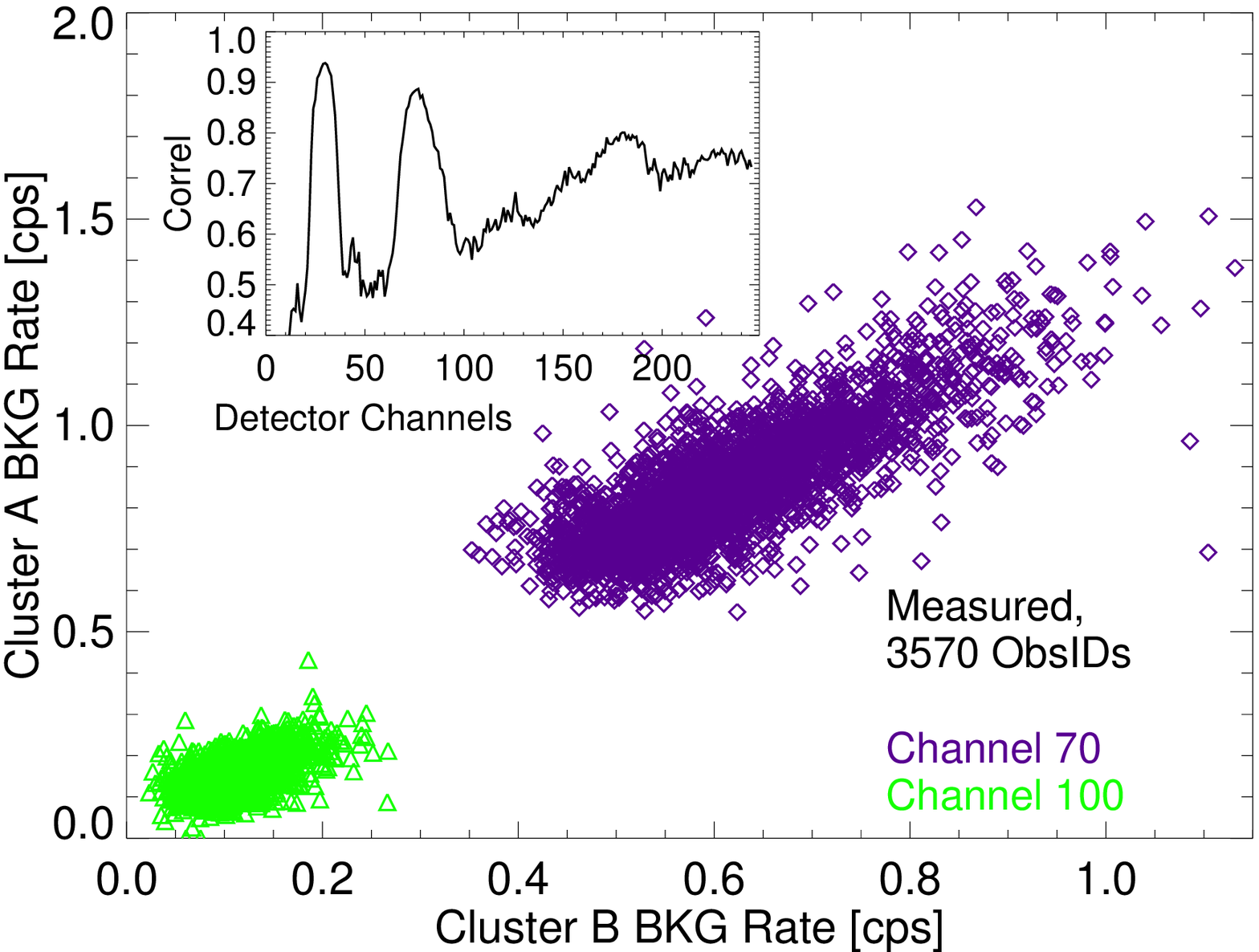}
\caption{\textbf{Left}: Background spectrum measured by HEXTE's clusters A (red) and
  B (blue) for AO9 observation ObsID 90152-01-17. Note that there is
  an instrumental cutoff below channel $\sim10$ and above channel
  $\sim246$ (starting from 0). \textbf{Right}: Cluster A versus cluster B background rates measured in
  channels 70 (purple) and 100 (green) for 3570 ObsIDs of AO9. The
  inset shows the correlation coefficient between the A and B rates
  for all channels based on these observations.}\label{fig:bkgex}
\end{figure}

Both clusters used their off-source observations to measure
their individual backgrounds, which are different from each other
mainly but not only due to the fact that cluster~B had only 3
operating detectors after 1996 March. For an example of the measured background
spectra, see top panel of Fig.~\ref{fig:bkgex}. The cluster~A background
can be estimated based on the measured cluster~B background: their
rates are well correlated for each detector channel (inset of bottom panel of
Fig.~\ref{fig:bkgex}, with varying correlation coefficients which
become especially high in the background lines around 30 and 70\,keV
[detector channels $\sim$ energy channels for HEXTE]). We extracted the
background spectra of several thousand exposures performed during the ninth mission year
(AO9, 2004). Fig.~\ref{fig:bkgex}-bottom panel demonstrates the correlation in two
selected channels, one associated with a peak in the spectrum and one
not.

\subsection{Linear Correction Parameters}
\begin{figure}
\includegraphics[width=3.4in]{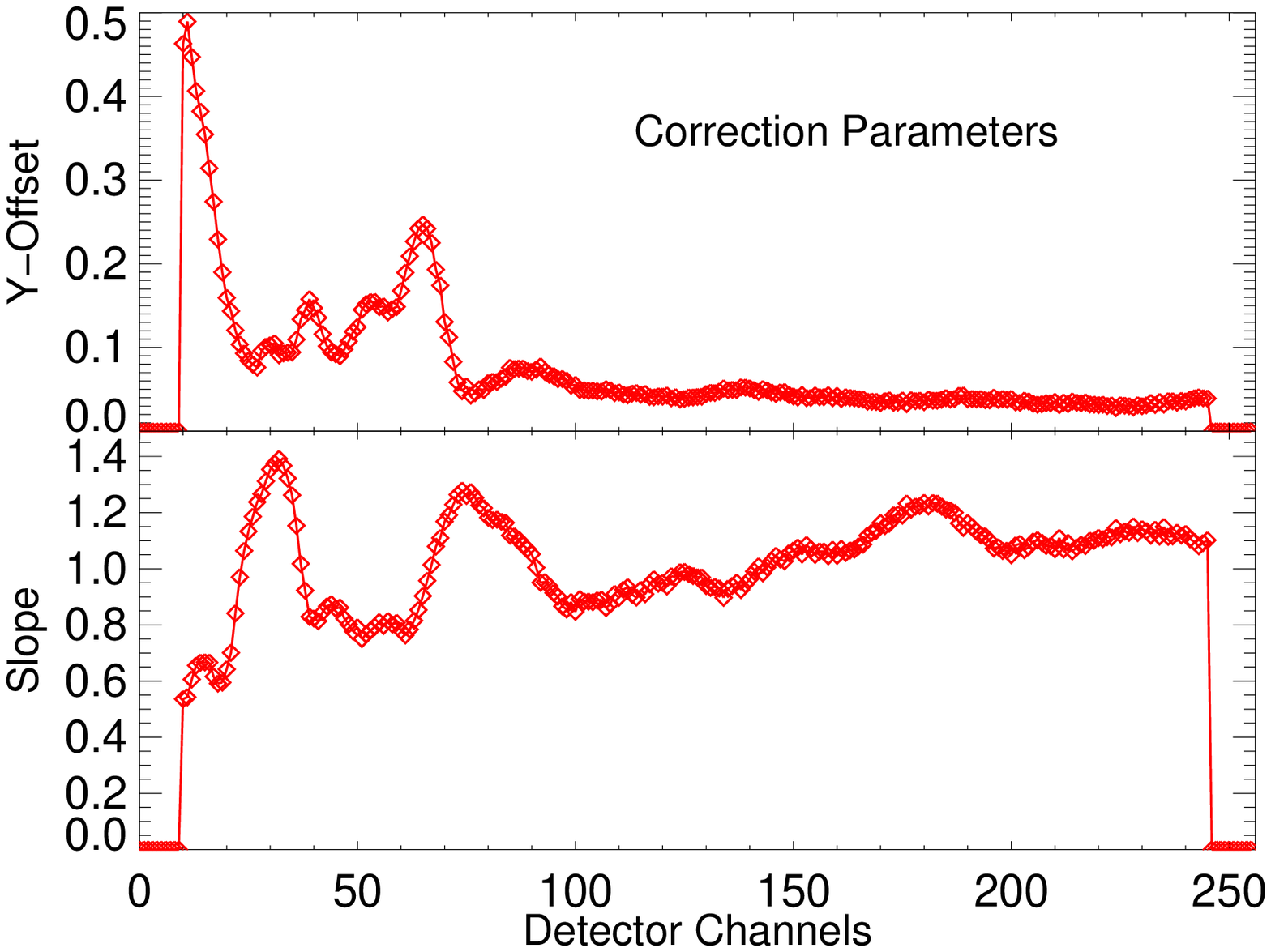}
\includegraphics[width=3.4in]{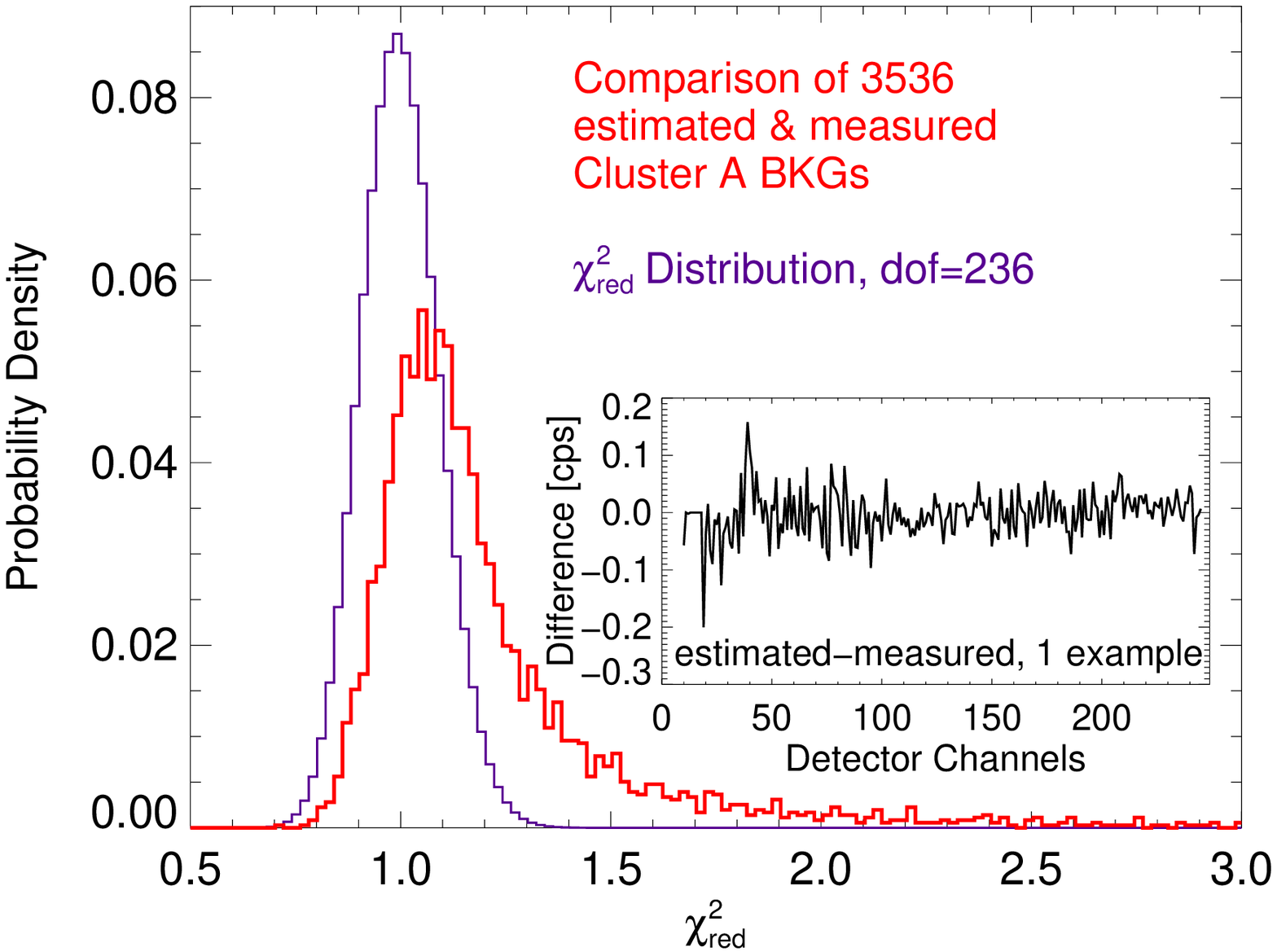}
\caption{\textbf{Left}:Results of linear fits, $rate\_A=m(channel) \times
  rate\_B+\Delta y(channel)$, for 3570 ObsIDs. \textbf{Left, Above:} Offset
  $\Delta y$. \textbf{Left, Below:} Slope $m$. Both parameters have been
  set to $0$ for channels below 10 and above 246. \textbf{Right}: $\chi^2$ 
  comparison of the estimated and measured cluster A
backgrounds for the AO9 ObsIDs (red). The theoretical distribution is
also shown (purple). The inset shows the difference between the
estimated and measured cluster A backgrounds for one typical
observation.}\label{fig:corpar}
\end{figure}

We performed linear fits to the A versus B background rates for each
detector channel based on the AO9 data set using \texttt{poly\_fit} in
IDL and taking A and B uncertainties into account. Note that the 3570
ObsIDs are the result of pre-selection: (1) observations with high A
or B rates in the lower channels have been omitted to screen against
sources in the background field of view, (2) since observations performed far
from the SAA show different background correlations, they have also
been omitted. The top panel in Fig.~\ref{fig:corpar} shows the correction
parameters we obtained. In the bottom panel of Fig.~\ref{fig:corpar}, the estimated and measured 
cluster~A spectra are compared (red) -- the former based on the AO9 cluster B
measurements and on the correlation parameters -- using the statistic
$\chi^2_\text{red}=\sum[d^2/(\sigma_\text{est}^2 +
\sigma_\text{meas}^2)]/$dof for each observation, where $d$ is the
estimated minus the measured cluster A background spectrum, $\sigma_\text{est}$ and
$\sigma_\text{meas}$ are the spectral uncertainties, and the number of
valid channels, dof (degrees of freedom), is 236 (see \citealt[][for comparing two
independent data sets]{bevington}). With respect to the theoretical
distribution (purple) a small shift and a tail of higher $\chi^2$
values can be seen. 

\subsection{Applications}
\begin{figure}
\includegraphics[angle=270,width=7.5in]{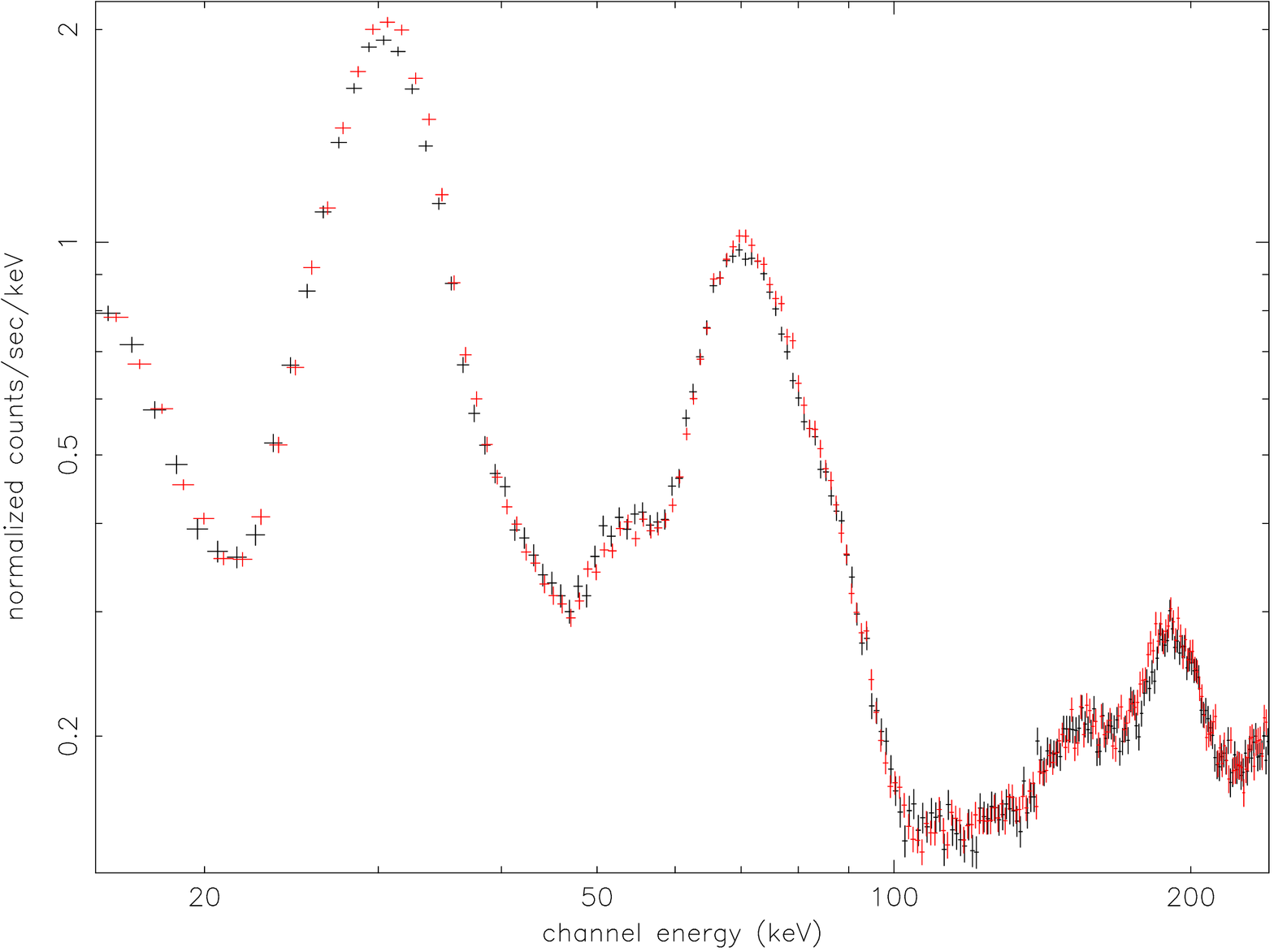}
\caption{Estimated (red) and measured (black) cluster A background for
  the Cyg X-1 observation shown in Fig.~\ref{fig:cyg_hexte}. As
  confirmed by the source fit the estimated background is a good
  match, however, small deviations, especially in the line peaks,
  remain.} \label{fig:cyg_bkg}
\end{figure}

\begin{figure}
\includegraphics[width=3.3in]{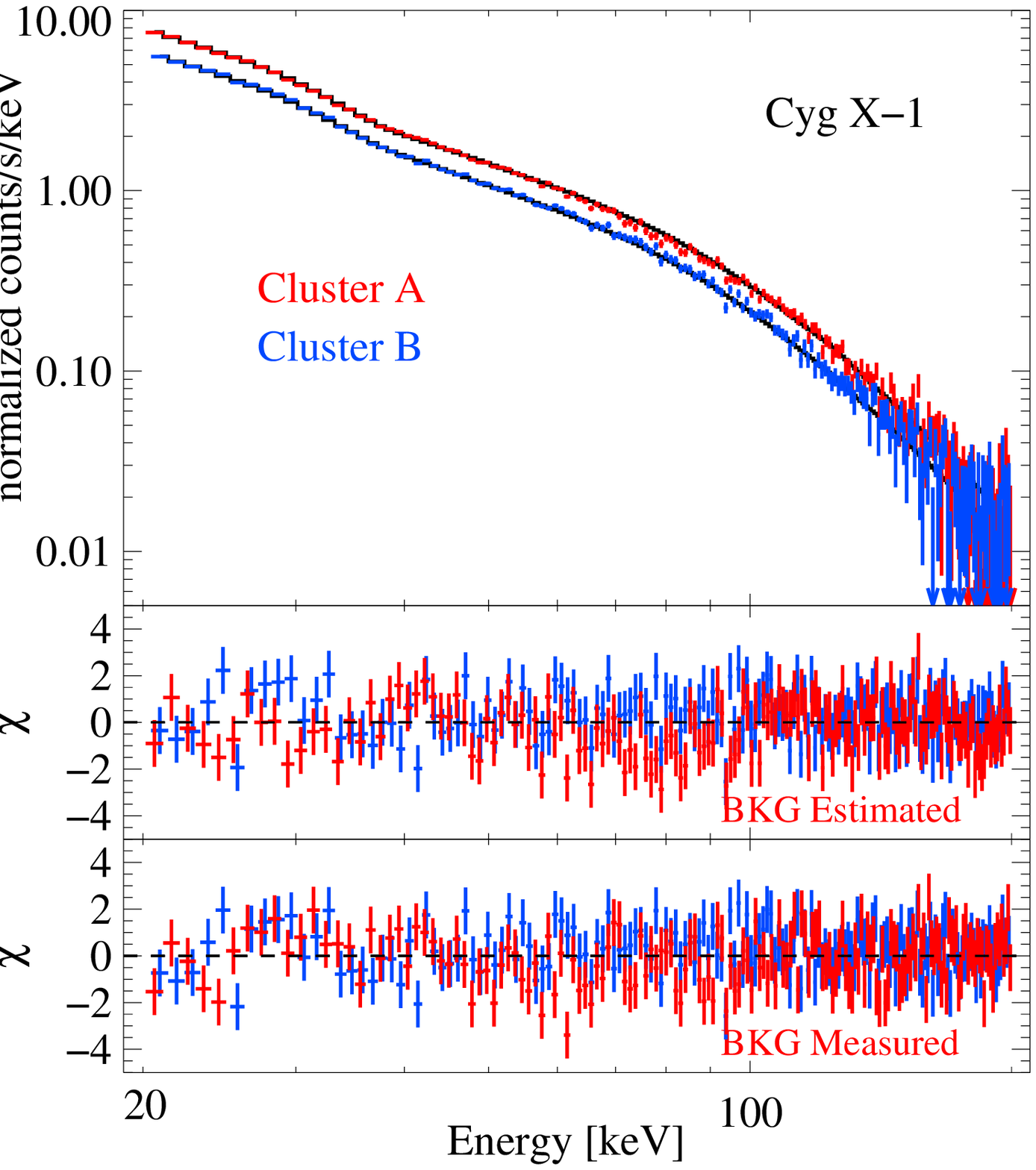}
\includegraphics[width=3.3in]{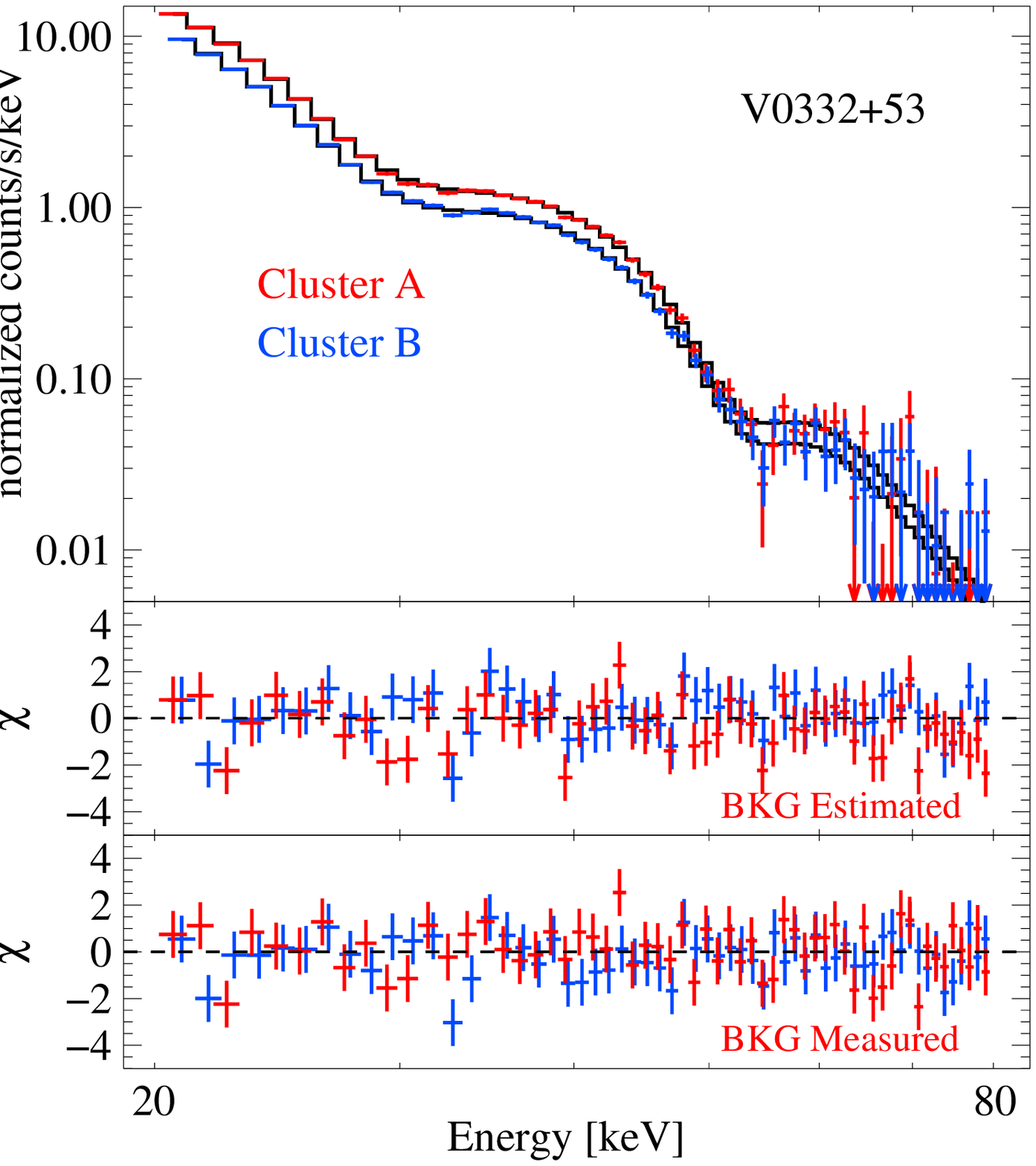}
\caption{
  \textbf{Left, Top:} HEXTE cluster A (red) and B (blue) counts
  spectra with the best fit \texttt{cutoffpl} model (black) for an
  observation of the black hole binary Cyg~X-1 performed on
  2004 Nov.\ 30. The spectrum has been averaged over 5 ObsIDs. The
  spectrum used for the cluster A background subtraction has been
  estimated based on the cluster B background and the correction
  parameters.
 \textbf{Left, Middle:} Residuals using the estimated cluster~A 
  background, best fit parameters: $\Gamma=1.53^{+0.02}_{-0.02}$,
  $E_\text{cut}=132^{+8}_{-7}$\,keV, $K=1.22^{+0.07}_{-0.07}$\,
  photons/keV/cm$^2$/s at 1\,keV.
  \textbf{Left, Bottom:} Residuals using
  the measured cluster~A background, best fit parameters:
  $\Gamma=1.54^{+0.02}_{-0.02}$, $E_\text{cut}=134^{+8}_{-7}$\,keV,
  $K=1.25^{+0.07}_{-0.07}$\, photons/keV/cm$^2$/s at 1\,keV.
  The two fits thus lead to consistent results without applying
  additional systematics in order to take uncertainties in the
  cluster~A background estimate into account. 
 \textbf{Right, Top:} HEXTE cluster~A (red) and B (blue) counts
  spectra with the best fit two-cyclotron-lines model (black) for an
  observation of the transient pulsar V0332$+$53 performed on
  2004 Dec.\ 12. The spectrum used for the cluster~A background
  subtraction has been estimated based on the cluster~B background and
  the correction parameters. 
  \textbf{Right, Middle:} Residuals using the
  estimated cluster A background, best fit parameters:
  $\Gamma=-0.15^{+0.75}_{-0.57}$,
  $E_\text{cycl1}=28.83^{+0.08}_{-0.07}$\,keV,
  $E_\text{cycl2}=51.5^{+0.6}_{-0.6}$\,keV.
  \textbf{Right, Bottom:} Residuals using the measured cluster~A background, best fit
  parameters: $\Gamma=+0.50^{+0.51}_{-0.55}$,
  $E_\text{cycl1}=28.83^{+0.08}_{-0.07}$\,keV,
  $E_\text{cycl2}=51.3^{+0.6}_{-0.6}$\,keV. The two fits thus lead
  to consistent results, in this case, however, systematics of
  2\% had to be applied in order to take uncertainties in the cluster
  A background estimate into account. }
  \label{fig:cyg_hexte}
\end{figure}

The method outlined above is available to derive
HEXTE cluster~A background spectra for post-2006 July observations. Each HEASOFT 
release contains the FTOOL 
\texttt{HEXTEBACKEST}which takes an input .pha file, performs the
linear correction for all channels, and writes a corrected output .pha
file. A FITS file with the correction parameters is part of the
calibration database (CALDB), distributed from NASA's High Energy Science 
Archive and Research Center (HEASARC). As a hidden parameter of \texttt{HEXTEBACKEST} it will
by default be remotely accessed. See `\texttt{fhelp hextebackest}'
for more details (e.g., on spectral binning). Here we show that for recent observations of bright
sources the estimated cluster~A background gives satisfactory results in the sense that the same
source fits as with the measured cluster A backgrounds are obtained
applying systematic uncertainties of 2\% or less. Limited tests with spectra 
from AO4 and earlier show that the correction parameters are not adequate 
for older observations. Fig.~\ref{fig:cyg_bkg} shows the comparison between 
the measured cluster A background and that generated by 
\texttt{hextebackest} for one example observation. Deviations between the two data sets are mostly 
seen at the peaks of the stronger background lines. \texttt{HEXTEBACKEST} was
applied for the observation of a smooth continuum (Cyg X$-$1; Fig.~\ref{fig:cyg_hexte}-left) 
and one with two cyclotron line features imposed on the continuum 
(V0332+653; Fig.~\ref{fig:cyg_hexte}-right). In both cases the residuals to the fit are shown for the
case of estimated and measured backgrounds, and they are comparable in both cases.
This demonstrates that the \texttt{HEXTEBACKEST} does not introduce spurious features in the 
spectra.
\clearpage
\onecolumn


\clearpage
\section{Spectral Fit Tables and Figures}
\label{sec:fits}
This section contains the best fit parameters from the spectral fitting of each observation with both the 
\texttt{highecut} and \texttt{cutoffpl} models. The tables are divided into the continuum and the line parameters. 
After the tables, plots of the various parameters are given. 

\begin{table}
\caption{Best-fit Highecut Continuum Spectral Parameters of GX 304$-$1\label{tab:best_fit_highecut_cont}}
\scriptsize
\begin{tabular}{lrrrrrrrrr}
\# & $N_\mathrm{H}^a$ & Index$^b$ & Flux$^c$ & Ecut$^d$ & Efold$^d$ & Ecyc$^e$ & Width$^e$ & Depth$^e$ & $\chi^2$/dof\\ \hline
1 & $7.30^{+0.50}_{-0.60}$ & $1.700^{+0.050}_{-0.060}$ & $1.377^{+0.028}_{-0.030}$ & $7.90^{+0.50}_{-0.60}$ & $26.90^{+3.30}_{-2.90}$ & $43.30^{+6.60}_{-3.00}$ & $2.40^{+3.70}_{-2.20}$ & $\ge 0.28$ & 1.22/152 \\
2 & $6.70^{+0.60}_{-0.80}$ & $1.640^{+0.060}_{-0.090}$ & $1.410^{+0.040}_{-0.050}$ & $7.50^{+0.70}_{-0.70}$ & $25.00^{+4.00}_{-4.00}$ & $--$ & $--$ & $--$ & 1.04/155 \\
3 & $7.70^{+0.40}_{-0.40}$ & $1.533^{+0.028}_{-0.031}$ & $2.380^{+0.040}_{-0.040}$ & $8.01^{+0.28}_{-0.30}$ & $24.60^{+1.60}_{-1.50}$ & $46.00^{+5.00}_{-5.00}$ & $4.00^{+4.00}_{-4.00}$ & $0.37^{+6.48}_{-0.19}$ & 1.20/152 \\
\hline
4 & $3.10^{+0.23}_{-0.23}$ & $0.848^{+0.030}_{-0.029}$ & $9.280^{+0.080}_{-0.080}$ & $4.96^{+0.24}_{-0.26}$ & $20.30^{+1.20}_{-1.00}$ & $58.80^{+2.20}_{-1.70}$ & $12.90^{+1.20}_{-1.00}$ & $1.18^{+0.15}_{-0.12}$ & 1.18/151 \\
5 & $2.70^{+1.90}_{-0.40}$ & $0.760^{+0.040}_{-0.040}$ & $9.720^{+0.680}_{-0.130}$ & $4.70^{+0.40}_{-2.80}$ & $17.50^{+0.90}_{-0.80}$ & $57.40^{+1.40}_{-1.20}$ & $11.00^{+1.00}_{-0.90}$ & $1.00^{+0.10}_{-0.09}$ & 0.89/151 \\
6 & $2.94^{+0.21}_{-0.16}$ & $0.811^{+0.031}_{-0.022}$ & $10.400^{+0.090}_{-0.060}$ & $5.07^{+0.15}_{-0.15}$ & $19.00^{+1.30}_{-0.70}$ & $60.30^{+8.00}_{-2.10}$ & $\ge 12.2$ & $1.16^{+0.69}_{-0.13}$ & 1.24/151 \\
7 & $2.80^{+0.40}_{-0.40}$ & $0.770^{+0.040}_{-0.040}$ & $11.080^{+0.130}_{-0.120}$ & $5.14^{+0.27}_{-0.26}$ & $18.00^{+0.90}_{-0.80}$ & $58.90^{+1.60}_{-1.30}$ & $11.90^{+0.90}_{-0.80}$ & $1.09^{+0.11}_{-0.09}$ & 0.87/152 \\
8 & $2.60^{+0.60}_{-0.40}$ & $0.820^{+0.090}_{-0.060}$ & $11.070^{+0.260}_{-0.170}$ & $5.40^{+0.60}_{-0.40}$ & $17.30^{+2.00}_{-1.10}$ & $59.00^{+4.10}_{-2.80}$ & $10.80^{+2.00}_{-1.50}$ & $1.09^{+0.35}_{-0.21}$ & 1.01/153 \\
9 & $2.71^{+0.27}_{-0.29}$ & $0.880^{+0.040}_{-0.040}$ & $11.990^{+0.120}_{-0.140}$ & $6.08^{+0.16}_{-0.19}$ & $18.90^{+1.00}_{-1.00}$ & $57.50^{+1.70}_{-1.40}$ & $12.40^{+1.10}_{-1.00}$ & $0.84^{+0.08}_{-0.07}$ & 1.09/151 \\
11 & $3.36^{+0.27}_{-0.30}$ & $0.940^{+0.040}_{-0.040}$ & $11.260^{+0.120}_{-0.130}$ & $6.20^{+0.17}_{-0.19}$ & $19.40^{+1.10}_{-1.10}$ & $58.10^{+2.40}_{-1.90}$ & $12.30^{+1.60}_{-1.50}$ & $0.77^{+0.11}_{-0.09}$ & 1.25/152 \\
12 & $4.10^{+0.60}_{-0.60}$ & $1.080^{+0.070}_{-0.070}$ & $8.300^{+0.170}_{-0.180}$ & $6.30^{+0.40}_{-0.40}$ & $20.40^{+2.20}_{-1.90}$ & $55.20^{+3.40}_{-2.50}$ & $10.10^{+2.40}_{-2.00}$ & $0.69^{+0.19}_{-0.16}$ & 0.75/152 \\
13 & $4.30^{+0.40}_{-0.40}$ & $1.180^{+0.050}_{-0.050}$ & $7.190^{+0.120}_{-0.110}$ & $6.40^{+0.31}_{-0.26}$ & $21.40^{+1.80}_{-1.50}$ & $52.00^{+1.90}_{-1.30}$ & $7.40^{+1.90}_{-1.70}$ & $0.58^{+0.14}_{-0.08}$ & 0.88/152 \\
14 & $5.49^{+0.21}_{-0.23}$ & $1.371^{+0.021}_{-0.027}$ & $6.400^{+0.060}_{-0.060}$ & $7.80^{+0.24}_{-0.30}$ & $26.10^{+1.10}_{-1.20}$ & $54.70^{+2.40}_{-1.70}$ & $8.60^{+1.80}_{-1.60}$ & $0.69^{+0.16}_{-0.13}$ & 1.01/152 \\
15 & $5.70^{+0.50}_{-0.70}$ & $1.320^{+0.060}_{-0.090}$ & $4.930^{+0.100}_{-0.140}$ & $7.50^{+0.60}_{-0.70}$ & $21.50^{+1.90}_{-2.60}$ & $50.90^{+2.70}_{-3.50}$ & $6.10^{+2.50}_{-2.00}$ & $0.62^{+0.57}_{-0.26}$ & 0.93/152 \\
16 & $6.90^{+0.40}_{-0.40}$ & $1.410^{+0.040}_{-0.040}$ & $4.600^{+0.070}_{-0.070}$ & $8.00^{+0.40}_{-0.50}$ & $23.10^{+1.50}_{-1.50}$ & $53.00^{+6.00}_{-4.00}$ & $4.90^{+4.50}_{-2.80}$ & $0.60^{+5.40}_{-0.40}$ & 0.78/152 \\
17 & $7.40^{+0.50}_{-0.50}$ & $1.510^{+0.040}_{-0.050}$ & $3.040^{+0.050}_{-0.060}$ & $7.90^{+0.40}_{-0.40}$ & $23.30^{+1.80}_{-1.70}$ & $52.00^{+5.00}_{-12.00}$ & $4.00^{+5.00}_{-4.00}$ & $\ge 0.10$ & 1.01/152 \\
18 & $7.70^{+0.50}_{-0.60}$ & $1.630^{+0.050}_{-0.060}$ & $2.370^{+0.050}_{-0.060}$ & $7.70^{+0.50}_{-0.60}$ & $24.70^{+2.40}_{-2.40}$ & $--$ & $--$ & $--$ & 1.19/155 \\
19 & $7.00^{+0.80}_{-0.90}$ & $1.880^{+0.070}_{-0.100}$ & $1.150^{+0.040}_{-0.050}$ & $7.40^{+0.80}_{-0.90}$ & $37.00^{+14.00}_{-10.00}$ & $48.90^{+2.50}_{-3.50}$ & $1.20^{+2.10}_{-1.10}$ & $\ge 0.18$ & 1.23/151 \\
20 & $7.50^{+0.70}_{-0.90}$ & $2.010^{+0.060}_{-0.100}$ & $0.801^{+0.026}_{-0.032}$ & $7.80^{+1.10}_{-1.30}$ & $48.00^{+23.00}_{-14.00}$ & $--$ & $--$ & $--$ & 0.96/155 \\
21 & $7.66^{+0.00}_{-1.11}$ & $2.160^{+0.060}_{-0.130}$ & $0.447^{+0.000}_{-0.024}$ & $7.30^{+1.80}_{-5.40}$ & $\ge 44.8$ & $41.00^{+7.00}_{-4.00}$ & $3.30^{+2.90}_{-2.30}$ & $\ge 0.67$ & 1.08/153 \\
22 & $7.70^{+1.00}_{-1.40}$ & $2.160^{+0.060}_{-0.200}$ & $0.504^{+0.026}_{-0.037}$ & $8.00^{+13.00}_{-6.00}$ & $\ge 38.0$ & $--$ & $--$ & $--$ & 0.84/155 \\
23 & $6.80^{+1.70}_{-1.90}$ & $2.160^{+0.100}_{-0.210}$ & $0.328^{+0.027}_{-0.029}$ & $7.00^{+13.00}_{-6.00}$ & $\ge 32.5$ & $40.00^{+8.00}_{-4.00}$ & $4.00^{+4.00}_{-4.00}$ & $\ge 0.21$ & 0.97/153 \\
\hline
24 & $7.47^{+0.17}_{-0.18}$ & $1.672^{+0.016}_{-0.018}$ & $2.060^{+0.015}_{-0.016}$ & $7.85^{+0.20}_{-0.21}$ & $27.30^{+1.00}_{-1.00}$ & $--$ & $--$ & $--$ & 1.85/154 \\
25 & $8.40^{+0.30}_{-0.32}$ & $1.388^{+0.027}_{-0.033}$ & $4.050^{+0.050}_{-0.050}$ & $7.89^{+0.26}_{-0.31}$ & $21.70^{+1.00}_{-1.10}$ & $51.20^{+1.20}_{-5.30}$ & $2.60^{+3.60}_{-1.80}$ & $\ge 0.15$ & 0.67/152 \\
26 & $7.30^{+0.40}_{-0.40}$ & $1.450^{+0.026}_{-0.028}$ & $3.740^{+0.050}_{-0.050}$ & $7.94^{+0.25}_{-0.27}$ & $24.60^{+1.20}_{-1.20}$ & $50.80^{+1.10}_{-1.70}$ & $5.00^{+1.20}_{-1.00}$ & $0.53^{+0.30}_{-0.18}$ & 0.98/152 \\
27 & $6.72^{+0.17}_{-0.18}$ & $1.371^{+0.016}_{-0.019}$ & $5.170^{+0.040}_{-0.040}$ & $7.83^{+0.20}_{-0.23}$ & $27.60^{+0.90}_{-1.00}$ & $54.30^{+1.90}_{-1.30}$ & $8.40^{+1.60}_{-1.40}$ & $0.61^{+0.11}_{-0.09}$ & 1.06/152 \\
28 & $4.10^{+0.40}_{-0.40}$ & $1.070^{+0.040}_{-0.050}$ & $7.000^{+0.090}_{-0.100}$ & $5.79^{+0.24}_{-0.30}$ & $22.30^{+1.40}_{-1.50}$ & $57.90^{+2.60}_{-2.10}$ & $9.60^{+1.50}_{-1.40}$ & $1.00^{+0.19}_{-0.15}$ & 1.13/153 \\
29 & $4.00^{+0.40}_{-0.40}$ & $0.970^{+0.050}_{-0.040}$ & $8.010^{+0.120}_{-0.110}$ & $5.40^{+0.40}_{-0.40}$ & $20.40^{+1.30}_{-1.10}$ & $56.90^{+1.20}_{-1.00}$ & $10.00^{+0.90}_{-0.90}$ & $0.86^{+0.08}_{-0.07}$ & 0.98/152 \\
30 & $4.44^{+0.28}_{-0.30}$ & $1.100^{+0.040}_{-0.040}$ & $7.670^{+0.090}_{-0.100}$ & $5.97^{+0.21}_{-0.24}$ & $23.50^{+1.50}_{-1.50}$ & $56.10^{+1.60}_{-1.40}$ & $10.10^{+1.10}_{-1.10}$ & $0.89^{+0.11}_{-0.10}$ & 1.11/152 \\
31 & $4.60^{+0.40}_{-0.50}$ & $1.090^{+0.050}_{-0.070}$ & $8.110^{+0.120}_{-0.150}$ & $5.70^{+0.40}_{-0.60}$ & $24.10^{+2.00}_{-2.60}$ & $56.70^{+1.80}_{-1.50}$ & $10.80^{+1.10}_{-1.20}$ & $0.86^{+0.11}_{-0.10}$ & 1.11/152 \\
32 & $6.90^{+0.40}_{-0.40}$ & $1.000^{+0.040}_{-0.050}$ & $8.600^{+0.110}_{-0.120}$ & $5.77^{+0.23}_{-0.30}$ & $21.70^{+1.40}_{-1.50}$ & $57.00^{+1.80}_{-1.50}$ & $11.00^{+1.20}_{-1.10}$ & $0.87^{+0.11}_{-0.08}$ & 1.01/152 \\
33 & $5.90^{+0.40}_{-0.50}$ & $0.990^{+0.050}_{-0.070}$ & $9.550^{+0.150}_{-0.170}$ & $5.81^{+0.26}_{-0.33}$ & $21.80^{+2.20}_{-2.10}$ & $54.60^{+4.00}_{-2.90}$ & $11.10^{+2.10}_{-1.90}$ & $0.76^{+0.24}_{-0.17}$ & 1.19/151 \\
34 & $7.10^{+0.30}_{-0.32}$ & $1.100^{+0.040}_{-0.050}$ & $8.290^{+0.100}_{-0.110}$ & $5.83^{+0.25}_{-0.30}$ & $24.20^{+1.70}_{-1.70}$ & $55.90^{+1.40}_{-1.20}$ & $10.10^{+1.00}_{-1.00}$ & $0.91^{+0.10}_{-0.09}$ & 0.90/152 \\
35 & $5.60^{+0.50}_{-1.30}$ & $1.300^{+0.050}_{-0.170}$ & $9.180^{+0.140}_{-0.440}$ & $7.60^{+0.60}_{-1.40}$ & $30.00^{+4.00}_{-8.00}$ & $55.50^{+4.40}_{-2.70}$ & $11.70^{+2.60}_{-3.70}$ & $0.85^{+0.20}_{-0.22}$ & 0.85/152 \\
36 & $4.40^{+0.40}_{-0.40}$ & $1.150^{+0.040}_{-0.050}$ & $7.040^{+0.100}_{-0.100}$ & $6.12^{+0.25}_{-0.27}$ & $23.40^{+1.70}_{-1.60}$ & $54.10^{+1.50}_{-1.20}$ & $8.70^{+1.30}_{-1.20}$ & $0.85^{+0.13}_{-0.12}$ & 0.95/152 \\
37 & $4.91^{+0.23}_{-0.23}$ & $1.197^{+0.028}_{-0.028}$ & $7.280^{+0.070}_{-0.070}$ & $6.43^{+0.20}_{-0.18}$ & $24.40^{+1.40}_{-1.20}$ & $55.70^{+1.70}_{-1.40}$ & $10.20^{+1.20}_{-1.10}$ & $0.76^{+0.09}_{-0.08}$ & 1.06/152 \\
38 & $6.11^{+0.18}_{-0.19}$ & $1.362^{+0.017}_{-0.020}$ & $5.830^{+0.040}_{-0.050}$ & $7.89^{+0.22}_{-0.27}$ & $30.40^{+1.20}_{-1.20}$ & $54.20^{+1.40}_{-1.10}$ & $8.70^{+1.10}_{-1.00}$ & $0.83^{+0.11}_{-0.10}$ & 1.19/152 \\
39 & $6.79^{+0.26}_{-0.29}$ & $1.395^{+0.024}_{-0.032}$ & $4.730^{+0.050}_{-0.060}$ & $7.82^{+0.29}_{-0.36}$ & $26.60^{+1.30}_{-1.50}$ & $50.90^{+1.60}_{-1.80}$ & $6.80^{+1.60}_{-1.40}$ & $0.57^{+0.21}_{-0.14}$ & 0.78/152 \\
40 & $7.02^{+0.17}_{-0.18}$ & $1.512^{+0.016}_{-0.017}$ & $3.260^{+0.022}_{-0.023}$ & $7.83^{+0.17}_{-0.18}$ & $25.60^{+0.90}_{-0.90}$ & $49.80^{+1.30}_{-4.40}$ & $3.90^{+1.10}_{-3.60}$ & $0.66^{+0.56}_{-0.28}$ & 1.15/152 \\
41 & $7.00^{+1.00}_{-1.10}$ & $2.090^{+0.080}_{-0.110}$ & $0.403^{+0.019}_{-0.020}$ & $7.40^{+1.30}_{-1.40}$ & $\ge 33.8$ & $43.00^{+8.00}_{-6.00}$ & $3.60^{+4.00}_{-3.00}$ & $\ge 0.56$ & 1.13/153 \\
42 & $7.80^{+1.80}_{-3.10}$ & $2.300^{+0.140}_{-0.590}$ & $0.196^{+0.019}_{-0.033}$ & $8.00^{+4.00}_{-6.00}$ & $\ge 19.6$ & $--$ & $--$ & $--$ & 1.20/156 \\
43 & $7.30^{+1.40}_{-1.90}$ & $2.160^{+0.110}_{-0.220}$ & $0.234^{+0.016}_{-0.021}$ & $7.80^{+2.20}_{-2.50}$ & $\ge 24.8$ & $--$ & $--$ & $--$ & 0.88/156 \\
44 & $7.60^{+1.10}_{-1.60}$ & $2.140^{+0.080}_{-0.190}$ & $0.304^{+0.016}_{-0.023}$ & $7.70^{+2.50}_{-5.80}$ & $\ge 35.4$ & $--$ & $--$ & $--$ & 0.90/156 \\
\hline
45 & $3.70^{+0.60}_{-0.40}$ & $1.000^{+0.100}_{-0.050}$ & $6.300^{+0.160}_{-0.090}$ & $5.10^{+0.70}_{-0.50}$ & $19.20^{+2.90}_{-1.30}$ & $54.80^{+3.30}_{-1.90}$ & $7.60^{+2.60}_{-2.20}$ & $0.72^{+0.15}_{-0.11}$ & 0.96/152 \\
46 & $3.50^{+0.60}_{-0.50}$ & $1.000^{+0.080}_{-0.060}$ & $6.440^{+0.160}_{-0.110}$ & $5.30^{+0.50}_{-0.40}$ & $20.00^{+2.70}_{-1.70}$ & $52.80^{+3.70}_{-1.60}$ & $7.10^{+4.00}_{-2.00}$ & $0.79^{+0.36}_{-0.18}$ & 0.87/153 \\
47 & $4.60^{+0.60}_{-0.70}$ & $1.140^{+0.060}_{-0.080}$ & $6.260^{+0.130}_{-0.150}$ & $6.00^{+0.40}_{-0.60}$ & $22.60^{+2.50}_{-2.50}$ & $56.40^{+5.20}_{-2.70}$ & $9.30^{+3.30}_{-2.40}$ & $0.82^{+0.25}_{-0.14}$ & 1.02/153 \\
48 & $5.60^{+0.50}_{-0.50}$ & $1.250^{+0.050}_{-0.060}$ & $6.430^{+0.120}_{-0.120}$ & $6.00^{+0.40}_{-0.50}$ & $27.00^{+3.00}_{-2.70}$ & $55.10^{+2.30}_{-1.70}$ & $9.60^{+1.80}_{-1.80}$ & $0.76^{+0.14}_{-0.08}$ & 0.95/152 \\
49 & $12.70^{+0.40}_{-0.40}$ & $1.110^{+0.050}_{-0.050}$ & $6.720^{+0.110}_{-0.100}$ & $5.60^{+0.40}_{-0.40}$ & $23.00^{+2.00}_{-1.80}$ & $56.10^{+1.90}_{-1.50}$ & $9.60^{+1.40}_{-1.40}$ & $0.78^{+0.12}_{-0.11}$ & 1.16/152 \\
50 & $9.40^{+0.40}_{-0.40}$ & $1.070^{+0.050}_{-0.050}$ & $7.220^{+0.120}_{-0.120}$ & $5.70^{+0.40}_{-0.40}$ & $22.20^{+1.60}_{-1.50}$ & $55.60^{+1.10}_{-1.00}$ & $9.50^{+0.90}_{-0.90}$ & $0.79^{+0.08}_{-0.08}$ & 0.87/152 \\
51 & $10.70^{+0.60}_{-0.60}$ & $1.200^{+0.060}_{-0.060}$ & $7.310^{+0.150}_{-0.150}$ & $6.20^{+0.40}_{-0.40}$ & $25.30^{+2.80}_{-2.40}$ & $56.00^{+3.50}_{-2.20}$ & $8.90^{+2.10}_{-1.90}$ & $0.78^{+0.21}_{-0.11}$ & 0.95/152 \\
52 & $11.40^{+0.40}_{-0.40}$ & $1.040^{+0.040}_{-0.040}$ & $7.580^{+0.100}_{-0.100}$ & $5.72^{+0.26}_{-0.30}$ & $21.10^{+1.10}_{-1.00}$ & $56.60^{+0.90}_{-0.80}$ & $10.00^{+0.70}_{-0.70}$ & $0.78^{+0.06}_{-0.06}$ & 1.19/152 \\
54 & $15.10^{+0.60}_{-0.60}$ & $0.860^{+0.060}_{-0.060}$ & $11.640^{+0.220}_{-0.220}$ & $5.80^{+0.27}_{-0.31}$ & $18.90^{+1.80}_{-1.50}$ & $61.00^{+5.00}_{-4.00}$ & $12.60^{+2.70}_{-2.00}$ & $0.97^{+0.30}_{-0.16}$ & 1.12/152 \\
55 & $13.30^{+1.60}_{-0.80}$ & $0.870^{+0.070}_{-0.050}$ & $8.420^{+0.510}_{-0.220}$ & $4.30^{+0.80}_{-2.40}$ & $17.60^{+1.40}_{-1.10}$ & $59.00^{+6.00}_{-4.00}$ & $10.50^{+3.30}_{-2.50}$ & $0.80^{+0.36}_{-0.22}$ & 1.03/152 \\
56 & $14.60^{+0.70}_{-0.70}$ & $1.080^{+0.060}_{-0.070}$ & $6.940^{+0.160}_{-0.150}$ & $5.80^{+0.40}_{-0.40}$ & $21.90^{+2.40}_{-2.10}$ & $60.00^{+10.00}_{-4.00}$ & $11.20^{+4.30}_{-2.60}$ & $1.04^{+0.79}_{-0.24}$ & 0.88/152 \\
57 & $12.60^{+0.40}_{-0.40}$ & $0.940^{+0.040}_{-0.040}$ & $9.010^{+0.120}_{-0.110}$ & $5.66^{+0.24}_{-0.27}$ & $19.30^{+0.90}_{-0.80}$ & $57.00^{+0.90}_{-0.80}$ & $10.40^{+0.70}_{-0.70}$ & $0.81^{+0.05}_{-0.05}$ & 1.11/152 \\
58 & $8.60^{+0.40}_{-0.40}$ & $0.930^{+0.050}_{-0.050}$ & $10.410^{+0.140}_{-0.140}$ & $5.77^{+0.23}_{-0.27}$ & $19.90^{+1.30}_{-1.20}$ & $56.50^{+1.70}_{-1.40}$ & $10.10^{+1.10}_{-1.10}$ & $0.93^{+0.13}_{-0.11}$ & 1.34/152 \\
59 & $8.90^{+0.50}_{-0.50}$ & $1.030^{+0.050}_{-0.070}$ & $8.770^{+0.150}_{-0.170}$ & $6.06^{+0.29}_{-0.36}$ & $22.10^{+2.00}_{-2.00}$ & $56.20^{+2.90}_{-2.00}$ & $11.20^{+1.80}_{-1.60}$ & $0.84^{+0.13}_{-0.10}$ & 1.00/152 \\
60 & $3.10^{+0.50}_{-0.40}$ & $0.830^{+0.050}_{-0.050}$ & $10.620^{+0.170}_{-0.160}$ & $5.58^{+0.30}_{-0.31}$ & $17.60^{+1.10}_{-1.00}$ & $58.60^{+2.40}_{-1.80}$ & $10.70^{+1.50}_{-1.40}$ & $0.84^{+0.13}_{-0.11}$ & 0.91/153 \\
61 & $3.50^{+0.50}_{-0.50}$ & $0.940^{+0.050}_{-0.060}$ & $9.450^{+0.160}_{-0.170}$ & $5.85^{+0.29}_{-0.34}$ & $18.70^{+1.30}_{-1.30}$ & $56.20^{+1.90}_{-1.50}$ & $10.00^{+1.40}_{-1.30}$ & $0.73^{+0.10}_{-0.07}$ & 0.88/153 \\
63 & $1.90^{+0.70}_{-0.60}$ & $0.740^{+0.080}_{-0.080}$ & $11.620^{+0.260}_{-0.240}$ & $5.60^{+0.40}_{-0.40}$ & $16.50^{+1.90}_{-1.50}$ & $57.90^{+4.20}_{-2.90}$ & $11.80^{+2.80}_{-2.60}$ & $0.72^{+0.18}_{-0.14}$ & 0.93/152 \\
64 & $3.80^{+0.40}_{-0.40}$ & $1.000^{+0.050}_{-0.050}$ & $9.320^{+0.130}_{-0.140}$ & $6.25^{+0.25}_{-0.26}$ & $20.10^{+1.40}_{-1.30}$ & $55.30^{+2.30}_{-1.70}$ & $10.00^{+1.60}_{-1.40}$ & $0.80^{+0.14}_{-0.12}$ & 1.03/153 \\
\end{tabular}
\end{table}

\begin{table}
\contcaption{Best-fit Highecut Continuum Spectral Parameters of GX 304$-$1\label{tab:cont_best_fit_highecut_cont}}
\scriptsize
\begin{tabular}{lrrrrrrrrr}
\# & $N_\mathrm{H}^a$ & Index$^b$ & Flux$^c$ & Ecut$^d$ & Efold$^d$ & Ecyc$^e$ & Width$^e$ & Depth$^e$ & $\chi^2$/dof\\ \hline
65 & $4.20^{+0.40}_{-0.40}$ & $1.070^{+0.050}_{-0.050}$ & $8.770^{+0.130}_{-0.130}$ & $6.50^{+0.33}_{-0.27}$ & $22.90^{+2.10}_{-1.70}$ & $56.40^{+3.10}_{-2.20}$ & $11.10^{+2.10}_{-1.70}$ & $0.79^{+0.15}_{-0.11}$ & 1.21/152 \\
66 & $3.79^{+0.24}_{-0.25}$ & $1.056^{+0.028}_{-0.029}$ & $7.980^{+0.080}_{-0.080}$ & $6.30^{+0.16}_{-0.16}$ & $19.80^{+0.90}_{-0.80}$ & $55.50^{+1.90}_{-1.40}$ & $8.50^{+1.40}_{-1.30}$ & $0.74^{+0.11}_{-0.10}$ & 1.14/152 \\
67 & $6.16^{+0.17}_{-0.17}$ & $1.367^{+0.016}_{-0.018}$ & $5.140^{+0.040}_{-0.040}$ & $7.86^{+0.17}_{-0.19}$ & $25.50^{+0.80}_{-0.80}$ & $52.20^{+1.00}_{-0.90}$ & $6.90^{+1.00}_{-1.00}$ & $0.75^{+0.16}_{-0.12}$ & 1.23/152 \\
68 & $6.60^{+0.50}_{-0.50}$ & $1.420^{+0.040}_{-0.050}$ & $4.790^{+0.080}_{-0.090}$ & $7.90^{+0.40}_{-0.50}$ & $24.60^{+1.90}_{-1.90}$ & $51.00^{+2.20}_{-3.50}$ & $5.20^{+2.30}_{-2.10}$ & $0.80^{+2.00}_{-0.40}$ & 0.79/152 \\
69 & $6.80^{+0.29}_{-0.32}$ & $1.531^{+0.028}_{-0.037}$ & $3.340^{+0.040}_{-0.050}$ & $7.80^{+0.40}_{-0.40}$ & $25.20^{+1.50}_{-1.60}$ & $50.00^{+4.00}_{-5.00}$ & $5.90^{+3.20}_{-2.70}$ & $0.39^{+0.88}_{-0.18}$ & 0.86/152 \\
70 & $7.60^{+0.70}_{-0.90}$ & $1.750^{+0.060}_{-0.110}$ & $2.160^{+0.070}_{-0.090}$ & $7.90^{+0.60}_{-1.10}$ & $29.00^{+6.00}_{-6.00}$ & $--$ & $--$ & $--$ & 0.98/155 \\
71 & $7.30^{+0.50}_{-0.50}$ & $1.830^{+0.040}_{-0.050}$ & $1.521^{+0.030}_{-0.032}$ & $7.60^{+0.40}_{-0.50}$ & $30.00^{+5.00}_{-4.00}$ & $--$ & $--$ & $--$ & 1.43/155 \\
72 & $7.90^{+1.90}_{-2.00}$ & $2.130^{+0.100}_{-0.270}$ & $0.490^{+0.050}_{-0.050}$ & $8.00^{+13.00}_{-6.00}$ & $\ge 28.5$ & $40.00^{+19.00}_{-4.00}$ & $4.30^{+4.00}_{-2.50}$ & $\ge 0.69$ & 0.90/153 \\
\hline
\end{tabular}
\\
\noindent $^a$ Column density in 10$^{22}$ cm$^{-2}$\\
$^b$ Power law photon index\\
$^c$ Unabsorbed power law 2$-$10 keV flux in 10$^{-9}$ ergs cm$^{-2}$ s$^{-1}$\\
$^d$ Ecut is cutoff energy in keV; Efold is folding energy in keV\\
$^e$ Ecyc is CRSF energy; Width is CRSF width in keV; Depth is CRSF depth\\
\end{table}
%
\normalsize
\begin{table}
\caption{Best-fit Highecut Spectral Lines\label{tab:best_fit_highecut_lines}}
\scriptsize
\begin{tabular}{lrrrrrrrr}
\# & iron$^a$ & iron$^b$ & 10.5\,keV$^c$ & 3.88\,keV$^d$ & 30\,keV$^e$ & 39\,keV$^f$ & 53\,keV$^g$ & 66\,keV$^h$\\ \hline
1 & $5^{+2}_{-2}$ & $36^{+16}_{-17}$ & $--$ & $-0.7^{+0.4}_{-0.4}$ & $0.9^{+0.5}_{-0.5}$ & $1.3^{+0.7}_{-0.6}$ & $0.7^{+0.5}_{-0.3}$ & $1.4^{+0.5}_{-0.5}$ \\
2 & $5^{+3}_{-3}$ & $31^{+20}_{-21}$ & $--$ & $-0.5^{+0.4}_{-0.4}$ & $1.5^{+0.5}_{-0.5}$ & $1.1^{+0.3}_{-0.3}$ & $0.5^{+0.3}_{-0.3}$ & $2.2^{+0.5}_{-0.5}$ \\
3 & $12^{+3}_{-3}$ & $45^{+11}_{-11}$ & $--$ & $-1.0^{+0.4}_{-0.4}$ & $\le 0.1$ & $1.3^{+0.7}_{-0.5}$ & $\le 1.1$ & $\le 0.7$ \\
\hline
4 & $91^{+8}_{-8}$ & $79^{+7}_{-7}$ & $-11^{+6}_{-6}$ & $-1.9^{+1.1}_{-1.2}$ & $4.9^{+1.4}_{-1.3}$ & $2.9^{+0.9}_{-0.9}$ & $\le 1.0$ & $4.1^{+0.7}_{-0.7}$ \\
5 & $90^{+13}_{-13}$ & $74^{+11}_{-11}$ & $-19^{+8}_{-8}$ & $-2.5^{+2.3}_{-4.7}$ & $3.6^{+0.9}_{-0.9}$ & $2.3^{+0.7}_{-0.7}$ & $\le 0.8$ & $2.9^{+0.6}_{-0.6}$ \\
6 & $89^{+6}_{-6}$ & $69^{+5}_{-5}$ & $-12^{+4}_{-4}$ & $-1.7^{+0.7}_{-0.8}$ & $4.0^{+1.1}_{-2.7}$ & $2.5^{+0.8}_{-1.1}$ & $\le 0.9$ & $3.3^{+0.6}_{-0.6}$ \\
7 & $106^{+15}_{-16}$ & $75^{+12}_{-12}$ & $-17^{+8}_{-8}$ & $--$ & $4.3^{+1.0}_{-0.9}$ & $2.8^{+0.7}_{-0.7}$ & $0.8^{+0.5}_{-0.5}$ & $4.0^{+0.6}_{-0.6}$ \\
8 & $94^{+20}_{-25}$ & $66^{+16}_{-19}$ & $--$ & $--$ & $4.0^{+2.4}_{-2.3}$ & $3.1^{+1.8}_{-1.8}$ & $\le 1.7$ & $3.5^{+1.4}_{-1.5}$ \\
9 & $79^{+8}_{-7}$ & $51^{+6}_{-5}$ & $-10^{+5}_{-5}$ & $-1.3^{+1.0}_{-1.0}$ & $4.8^{+1.1}_{-1.1}$ & $2.6^{+0.8}_{-0.8}$ & $\le 0.4$ & $2.7^{+0.6}_{-0.6}$ \\
11 & $65^{+7}_{-7}$ & $45^{+5}_{-5}$ & $--$ & $-3.2^{+1.0}_{-1.0}$ & $4.6^{+1.1}_{-1.0}$ & $2.4^{+0.8}_{-0.8}$ & $\le 0.6$ & $2.9^{+0.6}_{-0.7}$ \\
12 & $14^{+10}_{-9}$ & $13^{+10}_{-9}$ & $--$ & $-1.8^{+1.5}_{-1.5}$ & $4.2^{+1.5}_{-1.3}$ & $2.1^{+1.2}_{-1.2}$ & $\le 1.3$ & $2.8^{+1.0}_{-1.0}$ \\
13 & $30^{+6}_{-6}$ & $35^{+7}_{-7}$ & $--$ & $-1.1^{+1.0}_{-1.0}$ & $3.4^{+0.9}_{-0.8}$ & $1.4^{+0.9}_{-0.9}$ & $\le 0.5$ & $2.0^{+0.7}_{-0.7}$ \\
14 & $27^{+5}_{-6}$ & $37^{+7}_{-8}$ & $--$ & $-3.0^{+0.7}_{-0.7}$ & $2.6^{+0.6}_{-0.6}$ & $1.2^{+0.6}_{-0.6}$ & $0.7^{+0.5}_{-0.5}$ & $2.9^{+0.6}_{-0.6}$ \\
15 & $18^{+10}_{-11}$ & $31^{+17}_{-19}$ & $--$ & $-2.5^{+1.0}_{-1.0}$ & $2.0^{+0.8}_{-0.8}$ & $1.7^{+0.9}_{-0.8}$ & $\le 1.3$ & $1.4^{+0.8}_{-0.7}$ \\
16 & $22^{+7}_{-7}$ & $43^{+13}_{-13}$ & $--$ & $-2.0^{+0.9}_{-0.9}$ & $2.0^{+0.8}_{-0.8}$ & $0.9^{+0.7}_{-0.6}$ & $\le 1.9$ & $1.8^{+0.8}_{-0.7}$ \\
17 & $12^{+5}_{-5}$ & $36^{+14}_{-14}$ & $--$ & $-1.3^{+0.6}_{-0.6}$ & $1.5^{+0.7}_{-0.7}$ & $1.3^{+1.1}_{-0.5}$ & $\le 1.8$ & $1.6^{+0.6}_{-0.6}$ \\
18 & $8^{+4}_{-4}$ & $32^{+15}_{-16}$ & $--$ & $-0.9^{+0.5}_{-0.5}$ & $1.6^{+0.6}_{-0.6}$ & $1.2^{+0.4}_{-0.4}$ & $0.3^{+0.3}_{-0.3}$ & $1.4^{+0.6}_{-0.6}$ \\
19 & $5^{+3}_{-3}$ & $48^{+23}_{-26}$ & $-2.5^{+1.7}_{-1.7}$ & $-1.1^{+0.5}_{-0.5}$ & $\le 0.9$ & $1.3^{+0.4}_{-0.4}$ & $1.2^{+0.4}_{-0.5}$ & $1.8^{+0.6}_{-0.6}$ \\
20 & $4^{+2}_{-2}$ & $58^{+24}_{-26}$ & $--$ & $-0.3^{+0.3}_{-0.3}$ & $0.5^{+0.6}_{-0.6}$ & $0.7^{+0.3}_{-0.3}$ & $0.6^{+0.2}_{-0.2}$ & $1.7^{+0.5}_{-0.5}$ \\
21 & $3^{+0}_{-2}$ & $77^{+0}_{-39}$ & $--$ & $--$ & $1.3^{+0.8}_{-0.7}$ & $1.1^{+0.4}_{-0.5}$ & $0.7^{+0.5}_{-0.4}$ & $1.5^{+0.7}_{-0.7}$ \\
22 & $\le 3$ & $\le 64$ & $--$ & $-0.4^{+0.3}_{-0.3}$ & $0.7^{+0.7}_{-0.7}$ & $0.5^{+0.4}_{-0.4}$ & $0.5^{+0.4}_{-0.4}$ & $1.5^{+0.7}_{-0.7}$ \\
23 & $\le 3$ & $\le 99$ & $--$ & $--$ & $1.2^{+1.1}_{-1.1}$ & $1.0^{+0.5}_{-0.6}$ & $0.9^{+0.5}_{-0.5}$ & $2.2^{+0.9}_{-0.9}$ \\
\hline
24 & $8^{+1}_{-1}$ & $37^{+6}_{-6}$ & $-1.5^{+0.8}_{-0.8}$ & $-1.0^{+0.2}_{-0.2}$ & $1.6^{+0.2}_{-0.2}$ & $1.1^{+0.1}_{-0.1}$ & $0.3^{+0.1}_{-0.1}$ & $1.5^{+0.2}_{-0.2}$ \\
25 & $16^{+5}_{-5}$ & $35^{+10}_{-10}$ & $--$ & $-1.2^{+0.6}_{-0.5}$ & $1.5^{+0.6}_{-0.6}$ & $0.8^{+0.6}_{-0.4}$ & $\le 2.0$ & $1.5^{+0.5}_{-0.5}$ \\
26 & $19^{+5}_{-5}$ & $45^{+11}_{-11}$ & $--$ & $-2.0^{+0.6}_{-0.6}$ & $1.5^{+0.2}_{-0.2}$ & $1.0^{+0.2}_{-0.2}$ & $0.5^{+0.4}_{-0.4}$ & $1.4^{+0.2}_{-0.2}$ \\
27 & $25^{+4}_{-4}$ & $43^{+6}_{-6}$ & $--$ & $-1.8^{+0.4}_{-0.4}$ & $2.3^{+0.4}_{-0.4}$ & $1.0^{+0.4}_{-0.4}$ & $\le 0.6$ & $1.8^{+0.4}_{-0.4}$ \\
28 & $41^{+8}_{-7}$ & $48^{+10}_{-8}$ & $--$ & $--$ & $2.9^{+0.7}_{-0.7}$ & $1.2^{+0.7}_{-0.7}$ & $\le 1.0$ & $3.5^{+0.7}_{-0.8}$ \\
29 & $56^{+13}_{-13}$ & $57^{+13}_{-14}$ & $--$ & $-1.5^{+1.3}_{-1.4}$ & $3.3^{+0.6}_{-0.5}$ & $1.4^{+0.4}_{-0.4}$ & $0.3^{+0.3}_{-0.3}$ & $2.7^{+0.5}_{-0.5}$ \\
30 & $35^{+7}_{-6}$ & $37^{+7}_{-6}$ & $--$ & $-1.5^{+0.7}_{-0.7}$ & $3.8^{+0.8}_{-0.8}$ & $1.6^{+0.6}_{-0.6}$ & $\le 0.8$ & $2.9^{+0.7}_{-0.7}$ \\
31 & $48^{+14}_{-8}$ & $48^{+15}_{-9}$ & $--$ & $-2.4^{+0.8}_{-0.8}$ & $3.9^{+1.0}_{-1.0}$ & $2.1^{+0.7}_{-0.7}$ & $\le 0.8$ & $3.6^{+0.7}_{-0.8}$ \\
32 & $47^{+9}_{-8}$ & $44^{+9}_{-7}$ & $--$ & $-1.2^{+0.7}_{-0.8}$ & $4.4^{+0.9}_{-0.9}$ & $2.1^{+0.7}_{-0.7}$ & $\le 0.5$ & $3.6^{+0.7}_{-0.7}$ \\
33 & $52^{+11}_{-8}$ & $43^{+10}_{-7}$ & $-9^{+5}_{-5}$ & $-1.4^{+0.9}_{-0.9}$ & $6.7^{+2.1}_{-2.1}$ & $4.4^{+1.4}_{-1.4}$ & $\le 1.8$ & $2.4^{+1.6}_{-1.6}$ \\
34 & $46^{+8}_{-7}$ & $46^{+9}_{-7}$ & $--$ & $-1.7^{+0.7}_{-0.7}$ & $4.6^{+0.9}_{-0.9}$ & $1.9^{+0.7}_{-0.7}$ & $0.8^{+0.5}_{-0.5}$ & $3.2^{+0.7}_{-0.7}$ \\
35 & $57^{+15}_{-26}$ & $53^{+14}_{-24}$ & $--$ & $-4.7^{+2.5}_{-1.6}$ & $5.3^{+2.2}_{-3.3}$ & $2.2^{+1.5}_{-1.5}$ & $\le 1.9$ & $4.3^{+1.3}_{-2.4}$ \\
36 & $29^{+6}_{-6}$ & $34^{+7}_{-7}$ & $--$ & $-1.7^{+0.8}_{-0.8}$ & $2.9^{+0.9}_{-0.8}$ & $1.6^{+0.8}_{-0.8}$ & $0.8^{+0.6}_{-0.6}$ & $2.4^{+0.7}_{-0.7}$ \\
37 & $23^{+4}_{-4}$ & $27^{+4}_{-4}$ & $--$ & $-2.1^{+0.6}_{-0.6}$ & $3.8^{+0.6}_{-0.6}$ & $1.6^{+0.5}_{-0.5}$ & $0.4^{+0.4}_{-0.4}$ & $2.3^{+0.5}_{-0.5}$ \\
38 & $30^{+4}_{-5}$ & $44^{+6}_{-7}$ & $--$ & $-2.0^{+0.5}_{-0.5}$ & $3.5^{+0.6}_{-0.5}$ & $1.3^{+0.5}_{-0.5}$ & $0.7^{+0.4}_{-0.4}$ & $2.7^{+0.5}_{-0.5}$ \\
39 & $21^{+5}_{-5}$ & $39^{+9}_{-10}$ & $--$ & $-1.8^{+0.6}_{-0.6}$ & $1.9^{+0.6}_{-0.6}$ & $1.4^{+0.6}_{-0.6}$ & $\le 0.8$ & $1.4^{+0.5}_{-0.5}$ \\
40 & $15^{+2}_{-2}$ & $41^{+6}_{-6}$ & $--$ & $-1.1^{+0.3}_{-0.3}$ & $1.7^{+0.3}_{-0.3}$ & $1.1^{+0.2}_{-0.2}$ & $\le 1.1$ & $1.3^{+0.2}_{-0.2}$ \\
41 & $3^{+1}_{-1}$ & $80^{+40}_{-40}$ & $--$ & $--$ & $1.1^{+0.7}_{-0.5}$ & $1.1^{+0.3}_{-0.6}$ & $0.6^{+0.3}_{-0.3}$ & $1.8^{+0.5}_{-0.5}$ \\
42 & $\le 1$ & $\le 44$ & $--$ & $--$ & $\le 0.7$ & $0.7^{+0.4}_{-0.4}$ & $0.6^{+0.3}_{-0.3}$ & $1.6^{+0.6}_{-0.6}$ \\
43 & $\le 1$ & $\le 69$ & $--$ & $--$ & $\le 0.9$ & $0.8^{+0.3}_{-0.3}$ & $0.6^{+0.2}_{-0.2}$ & $1.6^{+0.5}_{-0.5}$ \\
44 & $2^{+1}_{-1}$ & $70^{+40}_{-50}$ & $--$ & $--$ & $1.2^{+0.5}_{-0.5}$ & $0.5^{+0.3}_{-0.3}$ & $0.4^{+0.2}_{-0.2}$ & $1.5^{+0.5}_{-0.5}$ \\
\hline
45 & $41^{+9}_{-15}$ & $54^{+12}_{-20}$ & $--$ & $-2.1^{+1.2}_{-1.4}$ & $1.8^{+0.9}_{-0.9}$ & $\le 1.9$ & $\le 0.4$ & $1.9^{+1.0}_{-0.9}$ \\
46 & $48^{+11}_{-13}$ & $62^{+15}_{-16}$ & $--$ & $--$ & $\le 2.4$ & $2.0^{+1.4}_{-1.3}$ & $\le 1.5$ & $\le 2.1$ \\
47 & $24^{+11}_{-8}$ & $31^{+14}_{-10}$ & $--$ & $--$ & $2.7^{+1.1}_{-1.1}$ & $1.0^{+1.1}_{-1.0}$ & $\le 0.6$ & $1.7^{+1.0}_{-1.0}$ \\
48 & $26^{+8}_{-7}$ & $33^{+11}_{-9}$ & $--$ & $-1.3^{+0.9}_{-0.9}$ & $3.5^{+1.0}_{-0.9}$ & $1.2^{+0.8}_{-0.8}$ & $\le 0.6$ & $2.9^{+0.8}_{-0.8}$ \\
49 & $34^{+9}_{-8}$ & $42^{+11}_{-10}$ & $--$ & $-1.0^{+0.6}_{-0.6}$ & $2.8^{+0.8}_{-0.7}$ & $1.4^{+0.7}_{-0.6}$ & $\le 0.8$ & $2.6^{+0.7}_{-0.7}$ \\
50 & $47^{+11}_{-10}$ & $54^{+14}_{-12}$ & $--$ & $-1.5^{+1.0}_{-1.0}$ & $3.2^{+0.6}_{-0.6}$ & $1.1^{+0.5}_{-0.5}$ & $\le 0.6$ & $2.7^{+0.5}_{-0.5}$ \\
51 & $29^{+9}_{-9}$ & $33^{+11}_{-10}$ & $--$ & $-1.7^{+1.1}_{-1.1}$ & $3.2^{+1.1}_{-1.1}$ & $1.0^{+1.0}_{-1.0}$ & $\le 0.8$ & $2.7^{+1.1}_{-1.1}$ \\
52 & $46^{+11}_{-10}$ & $50^{+12}_{-11}$ & $--$ & $-1.3^{+0.9}_{-0.9}$ & $2.9^{+0.4}_{-0.4}$ & $1.4^{+0.3}_{-0.3}$ & $0.2^{+0.2}_{-0.2}$ & $2.4^{+0.4}_{-0.4}$ \\
54 & $87^{+17}_{-15}$ & $59^{+12}_{-11}$ & $--$ & $-1.8^{+1.2}_{-1.3}$ & $2.7^{+1.8}_{-1.7}$ & $1.6^{+1.3}_{-1.3}$ & $\le 0.7$ & $2.6^{+0.9}_{-1.0}$ \\
55 & $83^{+11}_{-11}$ & $80^{+11}_{-10}$ & $--$ & $-2.4^{+2.2}_{-1.7}$ & $\le 2.1$ & $\le 2.4$ & $\le 1.5$ & $\le 2.3$ \\
56 & $33^{+11}_{-10}$ & $40^{+14}_{-12}$ & $--$ & $-1.5^{+1.0}_{-1.0}$ & $2.7^{+1.3}_{-1.5}$ & $1.5^{+1.1}_{-1.0}$ & $\le 0.9$ & $3.2^{+0.9}_{-1.0}$ \\
57 & $58^{+13}_{-12}$ & $52^{+12}_{-11}$ & $--$ & $-1.2^{+1.0}_{-1.0}$ & $3.1^{+0.4}_{-0.4}$ & $1.5^{+0.3}_{-0.3}$ & $0.3^{+0.2}_{-0.2}$ & $2.6^{+0.4}_{-0.4}$ \\
58 & $70^{+11}_{-10}$ & $53^{+9}_{-7}$ & $--$ & $-1.9^{+0.8}_{-0.8}$ & $3.4^{+1.2}_{-1.2}$ & $1.0^{+1.0}_{-1.0}$ & $\le 1.1$ & $1.2^{+0.9}_{-0.9}$ \\
59 & $46^{+11}_{-9}$ & $43^{+11}_{-9}$ & $--$ & $-1.5^{+1.1}_{-1.1}$ & $1.9^{+1.5}_{-1.4}$ & $1.7^{+1.1}_{-1.1}$ & $\le 0.6$ & $2.0^{+0.9}_{-0.9}$ \\
60 & $98^{+18}_{-18}$ & $73^{+15}_{-14}$ & $--$ & $--$ & $4.2^{+1.1}_{-1.1}$ & $1.8^{+1.0}_{-1.0}$ & $\le 1.1$ & $2.7^{+0.8}_{-0.9}$ \\
61 & $56^{+16}_{-14}$ & $47^{+14}_{-12}$ & $--$ & $--$ & $3.0^{+0.9}_{-0.8}$ & $1.7^{+0.7}_{-0.8}$ & $\le 0.4$ & $2.0^{+0.7}_{-0.7}$ \\
63 & $98^{+23}_{-20}$ & $65^{+17}_{-14}$ & $-24^{+11}_{-11}$ & $--$ & $5.3^{+2.3}_{-2.0}$ & $2.9^{+1.7}_{-1.7}$ & $\le 0.7$ & $2.4^{+1.2}_{-1.3}$ \\
64 & $44^{+9}_{-8}$ & $38^{+8}_{-7}$ & $--$ & $--$ & $3.6^{+1.2}_{-1.2}$ & $2.5^{+1.0}_{-1.0}$ & $\le 1.2$ & $2.4^{+0.9}_{-0.9}$ \\
\end{tabular}
\end{table}

\begin{table}
\contcaption{Best-fit Highecut Spectral Lines\label{tab:cont_best_fit_highecut_lines}}
\scriptsize
\begin{tabular}{lrrrrrrrr}
\# & iron$^a$ & iron$^b$ & 10.5\,keV$^c$ & 3.88\,keV$^d$ & 30\,keV$^e$ & 39\,keV$^f$ & 53\,keV$^g$ & 66\,keV$^h$\\ \hline
65 & $37^{+7}_{-7}$ & $34^{+7}_{-7}$ & $--$ & $-1.8^{+1.1}_{-1.1}$ & $2.2^{+1.4}_{-1.3}$ & $1.3^{+1.0}_{-1.0}$ & $\le 0.9$ & $2.1^{+1.0}_{-1.0}$ \\
66 & $33^{+5}_{-5}$ & $33^{+5}_{-5}$ & $--$ & $-1.3^{+0.7}_{-0.7}$ & $2.9^{+0.6}_{-0.6}$ & $1.1^{+0.6}_{-0.6}$ & $\le 0.9$ & $2.3^{+0.6}_{-0.6}$ \\
67 & $28^{+4}_{-4}$ & $47^{+6}_{-6}$ & $--$ & $-1.8^{+0.4}_{-0.4}$ & $1.7^{+0.4}_{-0.4}$ & $1.1^{+0.4}_{-0.4}$ & $0.7^{+0.4}_{-0.4}$ & $1.6^{+0.4}_{-0.4}$ \\
68 & $27^{+8}_{-8}$ & $51^{+14}_{-15}$ & $--$ & $-1.2^{+1.0}_{-1.0}$ & $0.9^{+0.9}_{-0.9}$ & $1.3^{+1.0}_{-0.9}$ & $\le 2.1$ & $1.5^{+0.8}_{-0.8}$ \\
69 & $13^{+4}_{-4}$ & $35^{+10}_{-11}$ & $--$ & $-1.3^{+0.5}_{-0.5}$ & $1.4^{+0.5}_{-0.5}$ & $1.1^{+0.5}_{-0.5}$ & $\le 0.9$ & $1.5^{+0.4}_{-0.4}$ \\
70 & $6^{+5}_{-6}$ & $26^{+22}_{-26}$ & $--$ & $-1.7^{+0.8}_{-0.8}$ & $1.5^{+0.9}_{-0.8}$ & $1.1^{+0.5}_{-0.5}$ & $\le 0.5$ & $1.6^{+0.8}_{-0.8}$ \\
71 & $6^{+2}_{-2}$ & $42^{+15}_{-15}$ & $--$ & $-1.2^{+0.4}_{-0.4}$ & $\le 0.6$ & $1.2^{+0.3}_{-0.3}$ & $0.5^{+0.2}_{-0.2}$ & $1.5^{+0.5}_{-0.4}$ \\
72 & $3^{+3}_{-3}$ & $60^{+70}_{-70}$ & $--$ & $--$ & $\le 1.8$ & $1.2^{+0.6}_{-0.6}$ & $\le 1.2$ & $1.5^{+1.0}_{-1.0}$ \\
\hline
\end{tabular}
\\
$^a$ iron line flux in  10$^{-4}$ photons cm$^{-2}$ s$^{-1}$\\
$^b$ iron line equivalent width in eV\\
$^c$ 10.5 keV negative line flux in units of 10$^{-3}$ photons cm$^{-2}$ s$^{-1}$\\
$^d$ 3.88 keV line flux in units of 10$^{-3}$ photons cm$^{-2}$ s$^{-1}$\\
$^e$ 30.17 keV line flux in units of 10$^{-3}$ photons cm$^{-2}$ s$^{-1}$\\
$^f$ 39.04 keV line flux in units of 10$^{-3}$ photons cm$^{-2}$ s$^{-1}$\\
$^g$ 53.00 keV line flux in units of 10$^{-3}$ photons cm$^{-2}$ s$^{-1}$\\
$^h$ 66.64 keV line flux in units of 10$^{-3}$ photons cm$^{-2}$ s$^{-1}$
\end{table}

\begin{table}
\caption{Best-fit Cutoffpl Continuum Spectral Parameters of GX 304$-$1\label{tab:best_fit_cutoffpl_cont}}
\scriptsize
\begin{tabular}{lrrrrrrrrrr}
\# & $N_\mathrm{H}^a$ & Index$^b$ & Flux$^c$ & Efold$^d$ & kT$^e$ & Flux$_\mathrm{BB}$$^f$ & Ecyc$^g$ & Width$^g$ & Depth$^g$ & $\chi^2$/dof\\ 
\hline
1 & $4.80^{+0.80}_{-0.90}$ & $1.580^{+0.080}_{-0.130}$ & $0.920^{+0.080}_{-0.090}$ & $\ge 55.2$ & $1.86^{+0.09}_{-0.08}$ & $0.31^{+0.05}_{-0.04}$ & $47.30^{+3.00}_{-5.30}$ & $5.70^{+2.40}_{-4.10}$ & $0.84^{+0.65}_{-0.30}$ & 0.89/152 \\
2 & $4.50^{+1.00}_{-1.40}$ & $1.550^{+0.100}_{-0.300}$ & $0.970^{+0.100}_{-0.120}$ & $\ge 30.0$ & $1.85^{+0.11}_{-0.16}$ & $0.31^{+0.05}_{-0.05}$ & $51.60^{+2.50}_{-2.80}$ & $5.10^{+4.70}_{-2.90}$ & $\ge 0.6$ & 0.84/152 \\
3 & $6.90^{+0.50}_{-0.60}$ & $1.620^{+0.060}_{-0.110}$ & $1.880^{+0.070}_{-0.080}$ & $\ge 49.3$ & $2.32^{+0.11}_{-0.12}$ & $0.44^{+0.04}_{-0.05}$ & $49.40^{+2.00}_{-2.40}$ & $7.50^{+2.30}_{-1.80}$ & $0.83^{+0.26}_{-0.22}$ & 1.00/150 \\
\hline
4 & $3.10^{+0.70}_{-0.70}$ & $0.540^{+0.170}_{-0.190}$ & $8.200^{+0.700}_{-0.800}$ & $14.80^{+2.80}_{-2.10}$ & $1.34^{+0.07}_{-0.07}$ & $1.00^{+0.50}_{-0.50}$ & $60.90^{+4.00}_{-2.70}$ & $12.50^{+2.50}_{-2.20}$ & $1.07^{+0.27}_{-0.23}$ & 1.19/151 \\
5 & $3.00^{+0.70}_{-0.80}$ & $0.490^{+0.150}_{-0.180}$ & $8.700^{+0.700}_{-0.800}$ & $14.00^{+1.80}_{-1.60}$ & $1.35^{+0.08}_{-0.10}$ & $1.10^{+0.60}_{-0.50}$ & $57.70^{+2.10}_{-1.80}$ & $10.30^{+1.50}_{-1.70}$ & $0.86^{+0.15}_{-0.15}$ & 0.92/151 \\
6 & $2.40^{+0.50}_{-0.40}$ & $0.370^{+0.140}_{-0.120}$ & $8.600^{+0.600}_{-0.500}$ & $12.80^{+1.50}_{-1.00}$ & $1.40^{+0.03}_{-0.03}$ & $1.60^{+0.40}_{-0.40}$ & $59.00^{+6.00}_{-4.00}$ & $10.80^{+3.20}_{-2.50}$ & $0.82^{+0.39}_{-0.18}$ & 1.19/151 \\
7 & $3.20^{+0.70}_{-0.70}$ & $0.560^{+0.120}_{-0.150}$ & $10.100^{+0.700}_{-0.800}$ & $15.00^{+1.70}_{-1.60}$ & $1.43^{+0.09}_{-0.10}$ & $1.10^{+0.50}_{-0.50}$ & $60.30^{+2.60}_{-1.90}$ & $12.10^{+1.40}_{-1.30}$ & $1.03^{+0.16}_{-0.15}$ & 0.89/151 \\
8 & $2.60^{+0.90}_{-0.80}$ & $0.530^{+0.170}_{-0.180}$ & $9.700^{+0.900}_{-1.000}$ & $13.90^{+2.10}_{-1.60}$ & $1.54^{+0.14}_{-0.08}$ & $1.30^{+0.70}_{-0.60}$ & $59.00^{+6.00}_{-4.00}$ & $9.90^{+2.70}_{-2.30}$ & $1.00^{+0.50}_{-0.26}$ & 0.95/151 \\
9 & $1.60^{+0.60}_{-0.50}$ & $0.410^{+0.130}_{-0.130}$ & $9.800^{+0.600}_{-0.600}$ & $12.80^{+1.40}_{-1.10}$ & $1.59^{+0.08}_{-0.05}$ & $1.70^{+0.50}_{-0.40}$ & $57.10^{+3.30}_{-2.50}$ & $10.10^{+2.60}_{-2.40}$ & $0.56^{+0.16}_{-0.12}$ & 0.98/151 \\
11 & $2.20^{+0.50}_{-0.50}$ & $0.460^{+0.100}_{-0.100}$ & $9.200^{+0.500}_{-0.500}$ & $12.90^{+1.00}_{-0.80}$ & $1.59^{+0.06}_{-0.04}$ & $1.50^{+0.40}_{-0.40}$ & $54.10^{+2.40}_{-1.40}$ & $7.10^{+2.20}_{-1.60}$ & $0.45^{+0.11}_{-0.09}$ & 1.20/151 \\
12 & $3.80^{+0.80}_{-1.00}$ & $0.880^{+0.160}_{-0.210}$ & $7.400^{+0.400}_{-0.700}$ & $18.00^{+6.00}_{-4.00}$ & $1.91^{+0.40}_{-0.29}$ & $0.75^{+0.32}_{-0.24}$ & $55.00^{+5.00}_{-4.00}$ & $11.00^{+4.00}_{-4.00}$ & $0.65^{+0.24}_{-0.25}$ & 0.77/151 \\
13 & $3.30^{+0.80}_{-0.70}$ & $0.860^{+0.150}_{-0.140}$ & $6.100^{+0.400}_{-0.500}$ & $16.40^{+3.10}_{-1.90}$ & $1.71^{+0.25}_{-0.12}$ & $0.78^{+0.26}_{-0.21}$ & $51.30^{+2.00}_{-1.40}$ & $5.90^{+3.00}_{-2.20}$ & $0.51^{+0.28}_{-0.12}$ & 0.88/151 \\
14 & $3.00^{+0.60}_{-0.60}$ & $0.870^{+0.120}_{-0.120}$ & $4.870^{+0.270}_{-0.290}$ & $16.50^{+2.40}_{-1.80}$ & $1.75^{+0.13}_{-0.09}$ & $0.89^{+0.17}_{-0.14}$ & $54.10^{+3.60}_{-1.90}$ & $6.90^{+3.40}_{-2.60}$ & $0.53^{+0.35}_{-0.19}$ & 1.03/151 \\
15 & $3.50^{+1.40}_{-1.10}$ & $0.900^{+0.400}_{-0.230}$ & $3.760^{+0.270}_{-0.380}$ & $16.40^{+21.50}_{-1.90}$ & $1.91^{+0.40}_{-0.18}$ & $0.75^{+0.21}_{-0.15}$ & $50.00^{+6.00}_{-6.00}$ & $6.00^{+10.00}_{-4.00}$ & $0.53^{+0.60}_{-0.30}$ & 0.95/151 \\
16 & $5.70^{+0.70}_{-1.00}$ & $1.370^{+0.130}_{-0.290}$ & $3.530^{+0.180}_{-0.210}$ & $42.00^{+22.00}_{-22.00}$ & $2.26^{+0.16}_{-0.22}$ & $0.85^{+0.10}_{-0.17}$ & $54.00^{+8.00}_{-4.00}$ & $\ge 5.4$ & $0.77^{+0.23}_{-0.28}$ & 0.75/151 \\
17 & $5.60^{+0.80}_{-1.10}$ & $1.400^{+0.140}_{-0.250}$ & $2.180^{+0.140}_{-0.150}$ & $41.00^{+30.00}_{-10.00}$ & $2.06^{+0.13}_{-0.18}$ & $0.62^{+0.07}_{-0.09}$ & $52.70^{+3.50}_{-2.50}$ & $8.00^{+4.00}_{-4.00}$ & $0.90^{+0.80}_{-0.40}$ & 0.89/152 \\
18 & $6.30^{+0.80}_{-1.10}$ & $1.620^{+0.100}_{-0.220}$ & $1.760^{+0.120}_{-0.130}$ & $\ge 49.5$ & $2.00^{+0.12}_{-0.16}$ & $0.46^{+0.06}_{-0.06}$ & $49.50^{+2.80}_{-5.10}$ & $5.90^{+3.30}_{-3.00}$ & $0.60^{+0.80}_{-0.40}$ & 1.04/152 \\
19 & $3.00^{+1.40}_{-1.50}$ & $1.550^{+0.150}_{-0.190}$ & $0.660^{+0.130}_{-0.130}$ & $\ge 70.4$ & $1.58^{+0.07}_{-0.06}$ & $0.29^{+0.08}_{-0.07}$ & $49.30^{+1.70}_{-3.10}$ & $1.40^{+1.70}_{-1.00}$ & $\ge 0.9$ & 1.03/151 \\
20 & $4.40^{+1.50}_{-1.30}$ & $1.750^{+0.150}_{-0.200}$ & $0.540^{+0.100}_{-0.100}$ & $\ge 56.2$ & $1.55^{+0.16}_{-0.09}$ & $0.14^{+0.05}_{-0.05}$ & $45.00^{+6.00}_{-6.00}$ & $0.40^{+0.80}_{-0.40}$ & $\ge 0.0$ & 0.87/152 \\
21 & $\le 2.1$ & $1.150^{+0.290}_{-0.280}$ & $0.144^{+0.069}_{-0.023}$ & $\ge 31.7$ & $1.40^{+0.05}_{-0.08}$ & $0.16^{+0.02}_{-0.04}$ & $44.00^{+4.00}_{-6.00}$ & $4.50^{+2.80}_{-2.70}$ & $\ge 0.7$ & 0.95/152 \\
22 & $\le 1.6$ & $1.300^{+0.400}_{-0.400}$ & $0.200^{+0.100}_{-0.060}$ & $\ge 54.0$ & $1.39^{+0.07}_{-0.09}$ & $0.16^{+0.04}_{-0.05}$ & $48.30^{+2.10}_{-4.80}$ & $1.80^{+3.50}_{-1.10}$ & $\ge 0.8$ & 0.76/152 \\
23 & $\le 1.0$ & $0.820^{+0.230}_{-0.820}$ & $0.140^{+0.050}_{-0.060}$ & $\ge 27.8$ & $1.33^{+0.06}_{-0.09}$ & $0.15^{+0.07}_{-0.04}$ & $37.00^{+4.00}_{-0.00}$ & $\ge 11.1$ & $1.50^{+0.80}_{-0.50}$ & 0.86/152 \\
\hline
24 & $6.00^{+0.50}_{-0.40}$ & $1.560^{+0.090}_{-0.080}$ & $1.600^{+0.050}_{-0.050}$ & $44.00^{+19.00}_{-9.00}$ & $1.95^{+0.10}_{-0.08}$ & $0.32^{+0.02}_{-0.02}$ & $50.60^{+1.00}_{-1.80}$ & $4.20^{+2.40}_{-1.30}$ & $0.80^{+0.90}_{-0.40}$ & 1.09/150 \\
25 & $5.60^{+0.60}_{-0.70}$ & $0.840^{+0.110}_{-0.120}$ & $2.940^{+0.170}_{-0.190}$ & $14.80^{+1.80}_{-1.30}$ & $1.82^{+0.11}_{-0.09}$ & $0.66^{+0.11}_{-0.10}$ & $51.20^{+1.00}_{-1.80}$ & $1.00^{+0.80}_{-0.70}$ & $\ge 0.4$ & 0.74/152 \\
26 & $4.90^{+0.70}_{-0.70}$ & $1.060^{+0.110}_{-0.110}$ & $2.780^{+0.150}_{-0.160}$ & $19.60^{+3.10}_{-2.20}$ & $1.92^{+0.12}_{-0.09}$ & $0.59^{+0.08}_{-0.08}$ & $50.50^{+1.30}_{-2.10}$ & $5.40^{+1.60}_{-1.30}$ & $0.49^{+0.24}_{-0.17}$ & 0.92/151 \\
27 & $4.60^{+0.40}_{-0.50}$ & $0.930^{+0.080}_{-0.090}$ & $4.160^{+0.150}_{-0.180}$ & $17.60^{+1.60}_{-1.50}$ & $1.77^{+0.11}_{-0.09}$ & $0.58^{+0.10}_{-0.08}$ & $53.00^{+1.70}_{-1.20}$ & $5.70^{+2.20}_{-1.90}$ & $0.51^{+0.36}_{-0.16}$ & 1.10/151 \\
28 & $4.20^{+0.80}_{-0.50}$ & $0.910^{+0.140}_{-0.090}$ & $6.590^{+0.440}_{-0.280}$ & $19.10^{+3.90}_{-1.70}$ & $1.63^{+0.59}_{-0.14}$ & $0.43^{+0.17}_{-0.25}$ & $58.30^{+3.60}_{-2.70}$ & $9.50^{+2.10}_{-1.90}$ & $0.96^{+0.26}_{-0.17}$ & 1.15/151 \\
29 & $4.60^{+0.70}_{-0.60}$ & $0.860^{+0.100}_{-0.110}$ & $7.700^{+0.500}_{-0.500}$ & $18.40^{+2.00}_{-1.70}$ & $1.59^{+0.35}_{-0.15}$ & $0.44^{+0.28}_{-0.29}$ & $57.10^{+1.40}_{-1.20}$ & $10.00^{+1.10}_{-1.10}$ & $0.81^{+0.09}_{-0.09}$ & 1.00/151 \\
30 & $4.10^{+0.80}_{-0.50}$ & $0.860^{+0.140}_{-0.090}$ & $6.950^{+0.460}_{-0.300}$ & $18.40^{+3.70}_{-1.60}$ & $1.60^{+0.34}_{-0.10}$ & $0.60^{+0.18}_{-0.26}$ & $56.10^{+2.60}_{-1.70}$ & $9.40^{+2.10}_{-1.50}$ & $0.79^{+0.18}_{-0.11}$ & 1.18/151 \\
31 & $3.80^{+0.70}_{-0.60}$ & $0.740^{+0.150}_{-0.120}$ & $6.900^{+0.600}_{-0.500}$ & $16.60^{+2.70}_{-1.70}$ & $1.45^{+0.09}_{-0.05}$ & $0.93^{+0.27}_{-0.33}$ & $57.30^{+3.50}_{-2.40}$ & $9.90^{+2.30}_{-2.10}$ & $0.69^{+0.18}_{-0.12}$ & 1.00/151 \\
32 & $6.60^{+0.70}_{-0.50}$ & $0.720^{+0.140}_{-0.110}$ & $7.700^{+0.600}_{-0.400}$ & $16.40^{+2.50}_{-1.50}$ & $1.53^{+0.17}_{-0.07}$ & $0.79^{+0.25}_{-0.33}$ & $57.10^{+3.10}_{-2.10}$ & $10.00^{+2.20}_{-1.80}$ & $0.74^{+0.16}_{-0.10}$ & 1.01/151 \\
33 & $4.50^{+0.50}_{-0.50}$ & $0.450^{+0.120}_{-0.120}$ & $7.500^{+0.500}_{-0.500}$ & $13.20^{+1.20}_{-1.00}$ & $1.46^{+0.03}_{-0.03}$ & $1.50^{+0.40}_{-0.40}$ & $51.20^{+2.70}_{-1.90}$ & $6.10^{+2.00}_{-1.70}$ & $0.47^{+0.21}_{-0.15}$ & 1.22/151 \\
34 & $6.80^{+0.60}_{-0.70}$ & $0.880^{+0.100}_{-0.130}$ & $7.500^{+0.400}_{-0.500}$ & $19.10^{+2.30}_{-2.30}$ & $1.56^{+0.16}_{-0.10}$ & $0.65^{+0.29}_{-0.22}$ & $56.30^{+2.00}_{-1.60}$ & $9.90^{+1.50}_{-1.60}$ & $0.82^{+0.13}_{-0.13}$ & 0.91/151 \\
35 & $4.30^{+0.80}_{-2.10}$ & $1.020^{+0.140}_{-0.500}$ & $7.800^{+0.500}_{-1.500}$ & $23.00^{+7.00}_{-11.00}$ & $2.00^{+0.40}_{-0.50}$ & $0.90^{+0.40}_{-0.40}$ & $57.00^{+8.00}_{-6.00}$ & $\ge 4.2$ & $0.80^{+0.40}_{-0.50}$ & 0.89/151 \\
36 & $3.70^{+0.70}_{-0.70}$ & $0.860^{+0.120}_{-0.140}$ & $6.200^{+0.400}_{-0.500}$ & $17.50^{+2.60}_{-2.10}$ & $1.59^{+0.20}_{-0.10}$ & $0.64^{+0.26}_{-0.23}$ & $53.80^{+1.80}_{-1.40}$ & $7.70^{+1.80}_{-1.70}$ & $0.75^{+0.17}_{-0.15}$ & 0.95/151 \\
37 & $3.30^{+0.60}_{-0.60}$ & $0.760^{+0.120}_{-0.120}$ & $5.900^{+0.400}_{-0.400}$ & $16.00^{+2.10}_{-1.50}$ & $1.58^{+0.10}_{-0.06}$ & $0.93^{+0.22}_{-0.19}$ & $54.40^{+3.20}_{-1.60}$ & $8.20^{+2.90}_{-2.10}$ & $0.53^{+0.13}_{-0.11}$ & 1.03/151 \\
38 & $4.80^{+0.40}_{-0.50}$ & $1.090^{+0.090}_{-0.100}$ & $5.000^{+0.110}_{-0.170}$ & $23.10^{+4.40}_{-3.00}$ & $2.07^{+0.21}_{-0.21}$ & $0.52^{+0.09}_{-0.06}$ & $54.00^{+1.90}_{-1.30}$ & $8.60^{+1.70}_{-1.40}$ & $0.77^{+0.14}_{-0.13}$ & 1.20/150 \\
39 & $5.00^{+0.70}_{-0.70}$ & $1.040^{+0.150}_{-0.130}$ & $3.830^{+0.170}_{-0.220}$ & $19.60^{+5.60}_{-2.80}$ & $1.91^{+0.24}_{-0.17}$ & $0.54^{+0.11}_{-0.09}$ & $50.20^{+2.00}_{-2.30}$ & $6.30^{+2.60}_{-2.00}$ & $0.48^{+0.24}_{-0.16}$ & 0.85/151 \\
40 & $4.70^{+0.50}_{-0.50}$ & $1.140^{+0.080}_{-0.090}$ & $2.450^{+0.090}_{-0.100}$ & $20.90^{+2.80}_{-2.20}$ & $1.83^{+0.09}_{-0.08}$ & $0.50^{+0.05}_{-0.05}$ & $50.20^{+1.10}_{-2.10}$ & $4.10^{+1.30}_{-1.00}$ & $0.80^{+0.70}_{-0.40}$ & 1.11/151 \\
41 & $\le 1.5$ & $1.230^{+0.270}_{-0.160}$ & $0.146^{+0.048}_{-0.017}$ & $\ge 42.6$ & $1.42^{+0.05}_{-0.06}$ & $0.14^{+0.02}_{-0.03}$ & $49.00^{+2.70}_{-6.90}$ & $2.30^{+6.30}_{-1.50}$ & $\ge 0.8$ & 0.95/152 \\
42 & $\le 1.0$ & $0.690^{+0.230}_{-0.700}$ & $0.068^{+0.029}_{-0.027}$ & $\ge 29.1$ & $1.31^{+0.06}_{-0.07}$ & $0.10^{+0.04}_{-0.01}$ & $37.00^{+1.20}_{-0.00}$ & $\ge 12.1$ & $2.00^{+1.10}_{-0.70}$ & 1.06/152 \\
43 & $\le 7.5$ & $1.800^{+0.400}_{-0.700}$ & $0.150^{+0.080}_{-0.080}$ & $\ge 31.0$ & $1.44^{+0.59}_{-0.14}$ & $0.05^{+0.04}_{-0.04}$ & $--$ & $--$ & $--$ & 0.86/155 \\
44 & $\le 4.3$ & $1.400^{+0.400}_{-0.400}$ & $0.130^{+0.090}_{-0.050}$ & $\ge 43.2$ & $1.38^{+0.08}_{-0.10}$ & $0.09^{+0.04}_{-0.05}$ & $48.00^{+4.00}_{-9.00}$ & $1.60^{+2.60}_{-1.30}$ & $\ge 0.3$ & 0.86/152 \\
\hline
45 & $5.40^{+0.60}_{-0.70}$ & $1.140^{+0.120}_{-0.170}$ & $6.440^{+0.190}_{-0.150}$ & $25.00^{+8.00}_{-7.00}$ & $2.70^{+0.40}_{-1.00}$ & $0.30^{+0.19}_{-0.22}$ & $55.30^{+2.50}_{-1.90}$ & $9.80^{+2.00}_{-2.50}$ & $0.82^{+0.12}_{-0.13}$ & 0.95/151 \\
46 & $5.30^{+0.40}_{-0.40}$ & $1.230^{+0.060}_{-0.090}$ & $6.440^{+0.170}_{-0.150}$ & $34.00^{+7.00}_{-8.00}$ & $3.00^{+0.00}_{-0.21}$ & $0.52^{+0.14}_{-0.19}$ & $54.20^{+3.00}_{-1.70}$ & $9.90^{+2.30}_{-1.80}$ & $1.00^{+0.18}_{-0.17}$ & 0.79/150 \\
47 & $4.00^{+1.10}_{-1.00}$ & $0.830^{+0.200}_{-0.200}$ & $5.500^{+0.600}_{-0.600}$ & $16.50^{+4.20}_{-2.50}$ & $1.54^{+0.43}_{-0.12}$ & $0.60^{+0.40}_{-0.40}$ & $55.20^{+6.50}_{-2.40}$ & $7.30^{+4.10}_{-2.80}$ & $0.66^{+0.36}_{-0.18}$ & 1.06/152 \\
48 & $5.00^{+1.00}_{-1.20}$ & $1.000^{+0.190}_{-0.260}$ & $5.800^{+0.600}_{-0.800}$ & $20.00^{+7.00}_{-5.00}$ & $1.50^{+0.47}_{-0.14}$ & $0.49^{+0.44}_{-0.28}$ & $54.60^{+3.40}_{-2.20}$ & $8.00^{+4.00}_{-4.00}$ & $0.68^{+0.74}_{-0.12}$ & 0.99/151 \\
49 & $12.50^{+0.70}_{-0.70}$ & $0.860^{+0.130}_{-0.140}$ & $6.100^{+0.500}_{-0.500}$ & $17.80^{+2.60}_{-2.10}$ & $1.46^{+0.15}_{-0.08}$ & $0.57^{+0.26}_{-0.25}$ & $55.90^{+2.60}_{-1.90}$ & $8.60^{+2.00}_{-2.30}$ & $0.67^{+0.15}_{-0.14}$ & 1.15/151 \\
50 & $9.90^{+0.60}_{-0.70}$ & $0.970^{+0.080}_{-0.120}$ & $7.000^{+0.400}_{-0.500}$ & $20.50^{+2.30}_{-2.40}$ & $1.78^{+0.46}_{-0.24}$ & $0.33^{+0.26}_{-0.18}$ & $55.70^{+1.20}_{-1.00}$ & $9.60^{+1.00}_{-1.10}$ & $0.77^{+0.09}_{-0.10}$ & 0.90/151 \\
51 & $9.30^{+1.10}_{-1.00}$ & $0.820^{+0.200}_{-0.200}$ & $5.900^{+0.700}_{-0.600}$ & $17.60^{+4.30}_{-2.80}$ & $1.58^{+0.19}_{-0.09}$ & $1.00^{+0.40}_{-0.40}$ & $55.90^{+6.00}_{-2.60}$ & $7.90^{+3.70}_{-2.70}$ & $0.68^{+0.26}_{-0.16}$ & 0.96/151 \\
52 & $11.40^{+0.70}_{-0.60}$ & $0.830^{+0.100}_{-0.090}$ & $6.900^{+0.500}_{-0.400}$ & $17.70^{+1.80}_{-1.30}$ & $1.63^{+0.16}_{-0.09}$ & $0.67^{+0.21}_{-0.26}$ & $57.00^{+1.30}_{-1.10}$ & $10.10^{+1.00}_{-0.90}$ & $0.70^{+0.08}_{-0.07}$ & 1.15/151 \\
54 & $15.30^{+0.90}_{-0.80}$ & $0.660^{+0.120}_{-0.160}$ & $10.800^{+0.800}_{-0.900}$ & $16.00^{+3.30}_{-2.10}$ & $1.70^{+0.46}_{-0.17}$ & $0.90^{+0.50}_{-0.50}$ & $62.00^{+6.00}_{-5.00}$ & $12.00^{+4.00}_{-4.00}$ & $0.90^{+0.35}_{-0.24}$ & 1.20/151 \\
55 & $14.40^{+0.80}_{-1.00}$ & $0.920^{+0.150}_{-0.190}$ & $8.500^{+0.400}_{-0.800}$ & $19.00^{+6.00}_{-4.00}$ & $2.00^{+1.00}_{-1.40}$ & $0.24^{+0.41}_{-0.25}$ & $59.00^{+7.00}_{-4.00}$ & $11.70^{+3.80}_{-2.80}$ & $0.84^{+0.36}_{-0.22}$ & 1.04/151 \\
56 & $15.00^{+1.00}_{-1.00}$ & $0.970^{+0.160}_{-0.180}$ & $6.600^{+0.500}_{-0.600}$ & $20.00^{+7.00}_{-4.00}$ & $1.78^{+0.76}_{-0.30}$ & $0.39^{+0.32}_{-0.23}$ & $61.00^{+9.00}_{-5.00}$ & $\ge 8.4$ & $1.05^{+0.95}_{-0.28}$ & 0.92/151 \\
57 & $12.70^{+0.60}_{-0.60}$ & $0.720^{+0.100}_{-0.100}$ & $8.100^{+0.500}_{-0.500}$ & $16.40^{+1.40}_{-1.20}$ & $1.64^{+0.12}_{-0.09}$ & $0.86^{+0.28}_{-0.29}$ & $57.60^{+1.30}_{-1.10}$ & $10.50^{+0.90}_{-1.00}$ & $0.73^{+0.08}_{-0.08}$ & 1.08/151 \\
58 & $7.70^{+0.50}_{-0.50}$ & $0.480^{+0.110}_{-0.110}$ & $8.500^{+0.500}_{-0.500}$ & $13.60^{+1.10}_{-1.00}$ & $1.51^{+0.04}_{-0.04}$ & $1.50^{+0.40}_{-0.40}$ & $55.90^{+2.00}_{-1.30}$ & $7.70^{+1.50}_{-1.30}$ & $0.77^{+0.15}_{-0.14}$ & 1.21/151 \\
59 & $9.10^{+0.60}_{-0.70}$ & $0.900^{+0.100}_{-0.130}$ & $8.200^{+0.400}_{-0.500}$ & $20.00^{+4.00}_{-4.00}$ & $1.96^{+0.39}_{-0.29}$ & $0.60^{+0.23}_{-0.24}$ & $56.70^{+3.60}_{-2.30}$ & $11.80^{+2.20}_{-2.10}$ & $0.85^{+0.15}_{-0.16}$ & 0.99/151 \\
60 & $2.10^{+0.80}_{-0.70}$ & $0.310^{+0.180}_{-0.170}$ & $8.200^{+0.900}_{-0.800}$ & $12.20^{+1.50}_{-1.10}$ & $1.54^{+0.06}_{-0.06}$ & $2.00^{+0.60}_{-0.60}$ & $57.00^{+5.00}_{-4.00}$ & $8.00^{+4.00}_{-4.00}$ & $0.58^{+0.19}_{-0.14}$ & 0.81/151 \\
61 & $3.40^{+0.90}_{-0.80}$ & $0.690^{+0.160}_{-0.150}$ & $8.500^{+0.800}_{-0.700}$ & $15.30^{+2.50}_{-1.70}$ & $1.65^{+0.30}_{-0.12}$ & $0.90^{+0.50}_{-0.50}$ & $56.10^{+2.60}_{-1.90}$ & $9.20^{+2.10}_{-1.90}$ & $0.64^{+0.13}_{-0.11}$ & 0.90/151 \\
63 & $3.30^{+0.60}_{-1.30}$ & $0.740^{+0.130}_{-0.300}$ & $11.600^{+0.500}_{-1.200}$ & $17.30^{+2.70}_{-4.40}$ & $2.50^{+0.60}_{-1.00}$ & $0.60^{+0.70}_{-0.40}$ & $57.60^{+4.20}_{-2.60}$ & $12.20^{+2.70}_{-2.90}$ & $0.78^{+0.18}_{-0.28}$ & 0.97/150 \\
64 & $3.40^{+0.80}_{-0.70}$ & $0.750^{+0.150}_{-0.150}$ & $8.300^{+0.500}_{-0.600}$ & $16.50^{+3.50}_{-2.20}$ & $1.83^{+0.36}_{-0.18}$ & $0.88^{+0.31}_{-0.28}$ & $55.30^{+3.20}_{-2.20}$ & $9.90^{+2.50}_{-2.20}$ & $0.71^{+0.21}_{-0.17}$ & 0.99/151 \\
\end{tabular}
\end{table}

\begin{table}
\contcaption{Best-fit Cutoffpl Continuum Spectral Parameters of GX 304$-$1\label{tab:cont_best_fit_cutoffpl_cont}}
\scriptsize
\begin{tabular}{lrrrrrrrrrr}
\# & $N_\mathrm{H}^a$ & Index$^b$ & Flux$^c$ & Efold$^d$ & kT$^e$ & Flux$_\mathrm{BB}$$^f$ & Ecyc$^g$ & Width$^g$ & Depth$^g$ & $\chi^2$/dof\\ \hline
\hline
65 & $3.30^{+0.60}_{-0.90}$ & $0.810^{+0.110}_{-0.230}$ & $7.500^{+0.400}_{-0.600}$ & $19.00^{+4.00}_{-5.00}$ & $1.86^{+0.21}_{-0.24}$ & $0.98^{+0.25}_{-0.23}$ & $58.00^{+6.00}_{-6.00}$ & $\ge 7.6$ & $0.79^{+0.29}_{-0.33}$ & 1.27/151 \\
66 & $2.90^{+0.50}_{-0.50}$ & $0.730^{+0.100}_{-0.090}$ & $6.770^{+0.300}_{-0.310}$ & $15.40^{+1.50}_{-1.20}$ & $1.74^{+0.13}_{-0.09}$ & $0.91^{+0.19}_{-0.18}$ & $54.90^{+2.40}_{-1.40}$ & $7.30^{+2.20}_{-1.90}$ & $0.64^{+0.17}_{-0.13}$ & 1.13/151 \\
67 & $3.90^{+0.50}_{-0.50}$ & $0.920^{+0.100}_{-0.090}$ & $3.970^{+0.150}_{-0.160}$ & $17.50^{+2.20}_{-1.60}$ & $1.84^{+0.12}_{-0.09}$ & $0.70^{+0.09}_{-0.08}$ & $51.80^{+1.10}_{-1.10}$ & $6.00^{+1.60}_{-1.40}$ & $0.67^{+0.25}_{-0.16}$ & 1.32/151 \\
68 & $5.50^{+0.70}_{-0.70}$ & $1.400^{+0.100}_{-0.160}$ & $3.710^{+0.170}_{-0.190}$ & $49.00^{+25.00}_{-19.00}$ & $2.26^{+0.13}_{-0.15}$ & $0.87^{+0.10}_{-0.12}$ & $51.70^{+2.50}_{-1.90}$ & $9.00^{+2.20}_{-2.10}$ & $0.94^{+0.23}_{-0.21}$ & 0.71/152 \\
69 & $4.80^{+1.10}_{-0.80}$ & $1.170^{+0.310}_{-0.170}$ & $2.640^{+0.140}_{-0.180}$ & $19.30^{+23.80}_{-2.60}$ & $1.85^{+0.31}_{-0.18}$ & $0.42^{+0.09}_{-0.07}$ & $49.00^{+8.00}_{-7.00}$ & $\ge 1.5$ & $0.33^{+5.69}_{-0.20}$ & 0.87/151 \\
70 & $5.40^{+1.30}_{-1.70}$ & $1.650^{+0.120}_{-0.330}$ & $1.510^{+0.200}_{-0.240}$ & $\ge 29.8$ & $1.83^{+0.15}_{-0.17}$ & $0.43^{+0.11}_{-0.10}$ & $52.00^{+6.00}_{-7.00}$ & $6.00^{+6.00}_{-4.00}$ & $0.90^{+2.70}_{-0.60}$ & 0.88/151 \\
71 & $3.90^{+0.90}_{-1.50}$ & $1.600^{+0.090}_{-0.250}$ & $0.930^{+0.110}_{-0.150}$ & $\ge 80.1$ & $1.70^{+0.06}_{-0.09}$ & $0.37^{+0.08}_{-0.06}$ & $43.10^{+3.60}_{-2.60}$ & $6.50^{+8.70}_{-1.90}$ & $0.51^{+0.28}_{-0.18}$ & 0.86/151 \\
72 & $\le 2.2$ & $0.800^{+1.100}_{-0.800}$ & $0.140^{+0.110}_{-0.060}$ & $\ge 22.7$ & $1.38^{+0.10}_{-0.14}$ & $0.20^{+0.11}_{-0.14}$ & $42.00^{+12.00}_{-6.00}$ & $\ge 2.8$ & $\ge 0.9$ & 0.87/152 \\
\end{tabular}
\\
$^a$ Column density in 10$^{22}$ cm$^{-2}$\\
$^b$ Power law photon index\\
$^c$ Unabsorbed power law 2$-$10 keV flux in 10$^{-9}$ ergs cm$^{-2}$ s$^{-1}$ \\
$^d$ Efold is folding energy in keV\\
$^e$ Blackbody temperature in keV\\
$^f$ Blackbody flux in 10$^{-9}$ ergs cm$^{-2}$ s$^{-1}$ \\
$^g$ Ecyc is CRSF energy; Width is CRSF width in keV; Depth is CRSF depth\\
\end{table}

\clearpage
\begin{table}
\caption{Best-fit Cutoffpl Spectral Lines of GX 304$-$1\label{tab:best_fit_cutoffpl_lines}}
\scriptsize
\begin{tabular}{lrrrrrrrr}
\# & iron$^a$ & iron$^b$ & 10.5\,keV$^c$ & 3.88\,keV$^d$ & 30\,keV$^e$ & 39\,keV$^f$ & 53\,keV$^g$ & 66\,keV$^h$\\ \hline
1 & $3^{+2}_{-2}$ & $19^{+15}_{-14}$ & $--$ & $--$ & $1.5^{+0.6}_{-0.7}$ & $1.0^{+0.6}_{-0.6}$ & $\le 1.2$ & $1.9^{+0.6}_{-0.6}$ \\
2 & $3^{+2}_{-2}$ & $23^{+16}_{-16}$ & $--$ & $--$ & $2.0^{+0.7}_{-1.0}$ & $0.7^{+0.7}_{-0.5}$ & $1.1^{+0.7}_{-0.7}$ & $2.8^{+0.7}_{-0.9}$ \\
3 & $9^{+2}_{-2}$ & $36^{+10}_{-10}$ & $-5.9^{+2.0}_{-2.0}$ & $-0.7^{+0.4}_{-0.4}$ & $\le 1.6$ & $1.2^{+0.6}_{-0.6}$ & $\le 0.8$ & $\le 1.1$ \\
\hline
4 & $88^{+7}_{-7}$ & $76^{+6}_{-6}$ & $--$ & $-3.8^{+1.0}_{-1.0}$ & $2.7^{+1.5}_{-1.1}$ & $2.3^{+1.0}_{-0.9}$ & $\le 0.8$ & $2.7^{+0.8}_{-0.8}$ \\
5 & $83^{+12}_{-12}$ & $67^{+10}_{-10}$ & $--$ & $-4.3^{+1.8}_{-1.8}$ & $2.5^{+0.9}_{-0.8}$ & $2.1^{+0.7}_{-0.7}$ & $\le 0.7$ & $2.0^{+0.7}_{-0.8}$ \\
6 & $85^{+5}_{-5}$ & $66^{+4}_{-4}$ & $--$ & $-3.7^{+0.7}_{-0.7}$ & $2.1^{+0.9}_{-0.9}$ & $2.0^{+0.8}_{-0.7}$ & $\le 0.8$ & $1.7^{+0.8}_{-0.9}$ \\
7 & $102^{+13}_{-13}$ & $72^{+10}_{-10}$ & $--$ & $-4.0^{+1.9}_{-1.9}$ & $3.1^{+1.0}_{-1.0}$ & $2.6^{+0.7}_{-0.7}$ & $0.6^{+0.5}_{-0.5}$ & $3.2^{+0.7}_{-0.7}$ \\
8 & $92^{+14}_{-14}$ & $66^{+10}_{-10}$ & $--$ & $-3.9^{+2.1}_{-2.1}$ & $3.0^{+2.2}_{-2.2}$ & $2.8^{+1.7}_{-1.7}$ & $\le 1.7$ & $2.6^{+1.5}_{-1.7}$ \\
9 & $109^{+6}_{-6}$ & $73^{+4}_{-4}$ & $--$ & $-2.9^{+1.0}_{-1.0}$ & $2.4^{+0.9}_{-0.9}$ & $2.1^{+0.8}_{-0.8}$ & $\le 0.3$ & $1.1^{+0.7}_{-0.7}$ \\
11 & $94^{+6}_{-6}$ & $68^{+5}_{-5}$ & $--$ & $-4.4^{+0.9}_{-0.9}$ & $2.5^{+0.8}_{-0.8}$ & $1.6^{+0.8}_{-0.7}$ & $\le 0.8$ & $1.0^{+0.7}_{-0.6}$ \\
12 & $35^{+9}_{-9}$ & $35^{+9}_{-9}$ & $--$ & $-2.7^{+1.4}_{-1.4}$ & $3.8^{+2.1}_{-1.5}$ & $2.1^{+1.2}_{-1.2}$ & $\le 1.2$ & $2.6^{+1.3}_{-1.1}$ \\
13 & $47^{+6}_{-6}$ & $56^{+7}_{-7}$ & $--$ & $-1.3^{+1.0}_{-1.0}$ & $2.7^{+0.8}_{-0.8}$ & $1.1^{+1.0}_{-0.8}$ & $\le 0.7$ & $1.7^{+0.7}_{-0.6}$ \\
14 & $24^{+4}_{-4}$ & $34^{+6}_{-6}$ & $--$ & $-1.7^{+0.7}_{-0.7}$ & $1.7^{+0.6}_{-0.6}$ & $1.0^{+0.6}_{-0.6}$ & $0.6^{+0.7}_{-0.6}$ & $2.2^{+0.6}_{-0.6}$ \\
15 & $20^{+6}_{-6}$ & $36^{+11}_{-11}$ & $--$ & $-1.7^{+1.0}_{-1.0}$ & $1.4^{+4.2}_{-0.8}$ & $1.6^{+1.6}_{-0.9}$ & $\le 1.2$ & $1.2^{+2.7}_{-0.7}$ \\
16 & $19^{+6}_{-6}$ & $37^{+11}_{-11}$ & $--$ & $-1.2^{+0.9}_{-0.9}$ & $3.7^{+2.0}_{-2.6}$ & $1.3^{+1.0}_{-1.3}$ & $\le 1.1$ & $3.5^{+1.3}_{-2.0}$ \\
17 & $9^{+4}_{-4}$ & $29^{+12}_{-12}$ & $--$ & $--$ & $1.7^{+1.4}_{-1.2}$ & $0.9^{+0.8}_{-0.8}$ & $0.8^{+0.6}_{-0.6}$ & $2.2^{+1.1}_{-1.0}$ \\
18 & $7^{+4}_{-4}$ & $26^{+13}_{-13}$ & $--$ & $--$ & $2.1^{+1.0}_{-1.1}$ & $0.8^{+0.7}_{-0.6}$ & $\le 1.1$ & $1.9^{+0.9}_{-0.9}$ \\
19 & $4^{+2}_{-2}$ & $32^{+22}_{-22}$ & $--$ & $-0.6^{+0.5}_{-0.5}$ & $\le 1.3$ & $0.6^{+0.4}_{-0.5}$ & $1.0^{+0.6}_{-0.9}$ & $1.8^{+0.6}_{-0.6}$ \\
20 & $3^{+2}_{-2}$ & $38^{+23}_{-23}$ & $--$ & $--$ & $0.6^{+0.6}_{-0.6}$ & $0.5^{+1.1}_{-0.3}$ & $0.4^{+0.5}_{-0.3}$ & $1.8^{+0.5}_{-0.5}$ \\
21 & $\le 3$ & $\le 75$ & $--$ & $--$ & $1.6^{+0.8}_{-0.9}$ & $1.3^{+0.4}_{-1.3}$ & $1.0^{+0.5}_{-1.0}$ & $1.7^{+0.7}_{-0.7}$ \\
22 & $\le 2$ & $\le 38$ & $--$ & $--$ & $0.8^{+0.8}_{-0.7}$ & $\le 1.0$ & $0.7^{+0.5}_{-0.5}$ & $1.7^{+0.7}_{-0.7}$ \\
23 & $\le 3$ & $\le 74$ & $--$ & $--$ & $5.0^{+2.4}_{-2.1}$ & $1.3^{+0.9}_{-0.7}$ & $0.9^{+0.6}_{-0.6}$ & $3.9^{+1.2}_{-1.3}$ \\
\hline
24 & $6^{+1}_{-1}$ & $28^{+5}_{-5}$ & $-2.0^{+0.9}_{-0.9}$ & $-0.5^{+0.2}_{-0.2}$ & $1.5^{+0.4}_{-0.2}$ & $0.6^{+0.3}_{-0.2}$ & $0.5^{+0.4}_{-0.4}$ & $1.6^{+0.3}_{-0.2}$ \\
25 & $14^{+4}_{-4}$ & $31^{+8}_{-8}$ & $--$ & $--$ & $1.1^{+0.6}_{-0.6}$ & $1.0^{+0.5}_{-0.5}$ & $1.3^{+0.7}_{-0.9}$ & $1.6^{+0.5}_{-0.5}$ \\
26 & $15^{+4}_{-4}$ & $37^{+10}_{-10}$ & $--$ & $-1.0^{+0.6}_{-0.6}$ & $1.0^{+0.3}_{-0.3}$ & $0.8^{+0.3}_{-0.3}$ & $0.4^{+0.4}_{-0.4}$ & $1.4^{+0.2}_{-0.2}$ \\
27 & $24^{+2}_{-2}$ & $41^{+5}_{-5}$ & $--$ & $-1.0^{+0.4}_{-0.4}$ & $1.4^{+0.4}_{-0.4}$ & $0.8^{+0.4}_{-0.4}$ & $\le 0.8$ & $1.1^{+0.4}_{-0.4}$ \\
28 & $50^{+6}_{-5}$ & $59^{+7}_{-6}$ & $--$ & $-1.9^{+0.8}_{-1.0}$ & $2.3^{+0.9}_{-0.7}$ & $1.1^{+0.7}_{-0.7}$ & $\le 1.0$ & $3.0^{+1.0}_{-0.8}$ \\
29 & $59^{+10}_{-9}$ & $60^{+10}_{-10}$ & $--$ & $-3.4^{+1.4}_{-1.5}$ & $2.9^{+0.7}_{-0.6}$ & $1.4^{+0.4}_{-0.4}$ & $\le 0.5$ & $2.4^{+0.6}_{-0.6}$ \\
30 & $47^{+5}_{-5}$ & $52^{+6}_{-5}$ & $--$ & $-2.5^{+0.7}_{-0.9}$ & $2.6^{+1.1}_{-0.7}$ & $1.3^{+0.8}_{-0.6}$ & $\le 0.7$ & $2.0^{+1.1}_{-0.7}$ \\
31 & $54^{+5}_{-5}$ & $56^{+5}_{-6}$ & $--$ & $-3.4^{+0.8}_{-0.8}$ & $2.2^{+0.9}_{-0.7}$ & $1.8^{+0.7}_{-0.7}$ & $\le 0.6$ & $2.4^{+0.8}_{-0.8}$ \\
32 & $58^{+6}_{-5}$ & $55^{+5}_{-5}$ & $--$ & $-2.4^{+0.7}_{-0.8}$ & $3.0^{+1.0}_{-0.7}$ & $1.9^{+0.7}_{-0.7}$ & $\le 0.4$ & $2.6^{+0.8}_{-0.7}$ \\
33 & $65^{+5}_{-6}$ & $56^{+5}_{-5}$ & $--$ & $-2.4^{+0.8}_{-0.8}$ & $3.6^{+1.8}_{-1.8}$ & $3.4^{+1.3}_{-1.3}$ & $\le 1.9$ & $\le 1.7$ \\
34 & $56^{+5}_{-5}$ & $57^{+5}_{-6}$ & $--$ & $-2.7^{+0.8}_{-0.8}$ & $3.3^{+1.0}_{-0.9}$ & $1.7^{+0.7}_{-0.7}$ & $0.5^{+0.5}_{-0.5}$ & $2.3^{+0.8}_{-0.9}$ \\
35 & $59^{+11}_{-11}$ & $56^{+10}_{-11}$ & $--$ & $-3.9^{+1.7}_{-1.6}$ & $4.0^{+4.0}_{-5.0}$ & $\le 3.6$ & $\le 1.6$ & $3.7^{+1.9}_{-3.2}$ \\
36 & $42^{+5}_{-5}$ & $50^{+6}_{-6}$ & $--$ & $-2.4^{+0.8}_{-0.8}$ & $1.9^{+0.9}_{-0.8}$ & $1.3^{+0.8}_{-0.7}$ & $0.6^{+0.6}_{-0.6}$ & $1.6^{+0.8}_{-0.7}$ \\
37 & $37^{+4}_{-4}$ & $43^{+5}_{-4}$ & $--$ & $-2.0^{+0.6}_{-0.6}$ & $2.3^{+0.7}_{-0.6}$ & $1.3^{+0.5}_{-0.5}$ & $\le 0.6$ & $1.3^{+0.7}_{-0.5}$ \\
38 & $28^{+4}_{-4}$ & $42^{+5}_{-5}$ & $-6.1^{+2.9}_{-2.9}$ & $-1.4^{+0.6}_{-0.5}$ & $2.5^{+1.0}_{-0.7}$ & $1.2^{+0.5}_{-0.5}$ & $0.6^{+0.4}_{-0.4}$ & $2.2^{+0.8}_{-0.6}$ \\
39 & $20^{+4}_{-4}$ & $38^{+7}_{-7}$ & $--$ & $-1.1^{+0.6}_{-0.6}$ & $1.2^{+0.9}_{-0.6}$ & $1.3^{+0.6}_{-0.6}$ & $\le 0.7$ & $1.2^{+0.6}_{-0.5}$ \\
40 & $12^{+2}_{-2}$ & $35^{+5}_{-5}$ & $--$ & $-0.3^{+0.3}_{-0.3}$ & $1.3^{+0.3}_{-0.3}$ & $0.8^{+0.3}_{-0.2}$ & $0.7^{+0.5}_{-0.5}$ & $1.3^{+0.2}_{-0.2}$ \\
41 & $1^{+1}_{-1}$ & $40^{+40}_{-40}$ & $--$ & $--$ & $1.3^{+0.5}_{-0.6}$ & $0.5^{+0.9}_{-0.3}$ & $0.7^{+0.4}_{-0.6}$ & $2.0^{+0.5}_{-0.5}$ \\
42 & $\le 1$ & $\le 30$ & $--$ & $--$ & $3.5^{+1.6}_{-1.4}$ & $1.3^{+0.5}_{-0.5}$ & $0.8^{+0.4}_{-0.4}$ & $2.9^{+0.8}_{-0.9}$ \\
43 & $\le 1$ & $\le 47$ & $--$ & $--$ & $\le 0.9$ & $0.7^{+0.3}_{-0.3}$ & $0.5^{+0.2}_{-0.3}$ & $1.6^{+0.5}_{-0.5}$ \\
44 & $\le 2$ & $\le 78$ & $--$ & $--$ & $1.3^{+0.5}_{-0.5}$ & $0.4^{+1.0}_{-0.3}$ & $0.5^{+0.4}_{-0.4}$ & $1.6^{+0.5}_{-0.5}$ \\
\hline
45 & $40^{+7}_{-7}$ & $51^{+9}_{-9}$ & $--$ & $-3.8^{+1.1}_{-1.1}$ & $2.9^{+2.1}_{-1.6}$ & $1.3^{+1.1}_{-1.0}$ & $\le 0.4$ & $3.0^{+1.4}_{-1.3}$ \\
46 & $49^{+8}_{-8}$ & $61^{+10}_{-10}$ & $-13^{+7}_{-7}$ & $-3.2^{+1.2}_{-1.2}$ & $4.3^{+2.1}_{-2.3}$ & $2.9^{+1.3}_{-1.3}$ & $\le 1.3$ & $2.6^{+1.3}_{-1.5}$ \\
47 & $36^{+7}_{-7}$ & $49^{+10}_{-10}$ & $--$ & $--$ & $2.1^{+1.0}_{-1.0}$ & $1.0^{+1.0}_{-1.0}$ & $\le 0.8$ & $0.9^{+1.2}_{-0.9}$ \\
48 & $34^{+6}_{-6}$ & $46^{+8}_{-8}$ & $--$ & $-1.9^{+0.9}_{-1.0}$ & $2.3^{+1.5}_{-0.9}$ & $\le 1.8$ & $\le 1.4$ & $2.0^{+1.3}_{-1.1}$ \\
49 & $39^{+5}_{-5}$ & $48^{+6}_{-6}$ & $--$ & $-1.8^{+0.6}_{-0.6}$ & $2.0^{+0.7}_{-0.7}$ & $1.2^{+0.6}_{-0.6}$ & $\le 0.8$ & $1.9^{+0.7}_{-0.7}$ \\
50 & $55^{+9}_{-9}$ & $63^{+10}_{-10}$ & $--$ & $-2.7^{+1.1}_{-1.0}$ & $2.9^{+0.7}_{-0.7}$ & $1.0^{+0.5}_{-0.5}$ & $\le 0.5$ & $2.5^{+0.6}_{-0.6}$ \\
51 & $39^{+8}_{-8}$ & $46^{+10}_{-10}$ & $--$ & $-1.7^{+1.1}_{-1.1}$ & $1.9^{+1.2}_{-1.1}$ & $\le 1.8$ & $\le 0.7$ & $2.0^{+1.3}_{-1.1}$ \\
52 & $52^{+9}_{-8}$ & $57^{+10}_{-9}$ & $--$ & $-2.1^{+0.9}_{-1.0}$ & $2.3^{+0.5}_{-0.4}$ & $1.4^{+0.3}_{-0.3}$ & $\le 0.3$ & $1.9^{+0.4}_{-0.4}$ \\
54 & $105^{+13}_{-11}$ & $72^{+8}_{-8}$ & $--$ & $-3.2^{+1.2}_{-1.3}$ & $1.7^{+1.9}_{-1.6}$ & $1.4^{+1.4}_{-1.3}$ & $\le 0.7$ & $2.3^{+1.0}_{-1.1}$ \\
55 & $73^{+12}_{-12}$ & $69^{+11}_{-11}$ & $--$ & $-2.1^{+1.4}_{-1.3}$ & $\le 2.5$ & $\le 2.7$ & $\le 1.4$ & $1.3^{+1.5}_{-1.2}$ \\
56 & $43^{+9}_{-9}$ & $51^{+11}_{-11}$ & $--$ & $-2.3^{+1.0}_{-1.0}$ & $2.3^{+1.9}_{-1.7}$ & $1.5^{+1.1}_{-1.0}$ & $\le 0.9$ & $3.0^{+1.2}_{-1.0}$ \\
57 & $65^{+10}_{-10}$ & $58^{+10}_{-9}$ & $--$ & $-2.3^{+1.0}_{-1.0}$ & $2.4^{+0.5}_{-0.4}$ & $1.4^{+0.3}_{-0.3}$ & $\le 0.4$ & $2.2^{+0.4}_{-0.4}$ \\
58 & $82^{+6}_{-6}$ & $64^{+5}_{-5}$ & $--$ & $-3.1^{+0.8}_{-0.8}$ & $1.9^{+1.0}_{-1.0}$ & $\le 1.4$ & $\le 1.4$ & $\le 0.6$ \\
59 & $64^{+9}_{-8}$ & $60^{+8}_{-8}$ & $--$ & $-2.6^{+1.1}_{-1.2}$ & $\le 3.3$ & $1.8^{+1.0}_{-1.1}$ & $\le 0.5$ & $2.0^{+0.9}_{-1.0}$ \\
60 & $100^{+14}_{-14}$ & $75^{+11}_{-11}$ & $--$ & $-3.0^{+2.0}_{-2.1}$ & $3.4^{+1.1}_{-1.1}$ & $1.7^{+1.0}_{-1.0}$ & $\le 1.4$ & $1.2^{+1.0}_{-1.0}$ \\
61 & $70^{+12}_{-12}$ & $60^{+10}_{-10}$ & $--$ & $-2.6^{+1.7}_{-1.9}$ & $2.2^{+0.9}_{-0.8}$ & $1.6^{+0.8}_{-0.8}$ & $\le 0.3$ & $1.4^{+0.9}_{-0.8}$ \\
63 & $116^{+14}_{-14}$ & $77^{+9}_{-10}$ & $-31^{+14}_{-13}$ & $-5.1^{+2.1}_{-2.0}$ & $5.7^{+3.2}_{-2.4}$ & $3.0^{+1.7}_{-1.7}$ & $\le 0.8$ & $2.8^{+1.9}_{-2.0}$ \\
64 & $67^{+9}_{-8}$ & $59^{+8}_{-7}$ & $--$ & $-1.7^{+1.3}_{-1.4}$ & $2.6^{+1.7}_{-1.2}$ & $2.3^{+1.1}_{-1.0}$ & $\le 1.0$ & $1.8^{+1.3}_{-1.0}$ \\
\end{tabular}
\end{table}

\begin{table}
\contcaption{Best-fit Cutoffpl Spectral Lines of GX 304$-$1\label{cont_best_fit_cutoffpl_lines}}
\scriptsize
\begin{tabular}{lrrrrrrrr}
\# & iron$^a$ & iron$^b$ & 10.5\,keV$^c$ & 3.88\,keV$^d$ & 30\,keV$^e$ & 39\,keV$^f$ & 53\,keV$^g$ & 66\,keV$^h$\\ \hline
65 & $56^{+7}_{-7}$ & $54^{+7}_{-7}$ & $--$ & $-2.1^{+1.1}_{-1.1}$ & $\le 3.2$ & $1.5^{+1.0}_{-1.1}$ & $\le 0.6$ & $2.0^{+0.9}_{-1.4}$ \\
66 & $53^{+5}_{-5}$ & $55^{+5}_{-5}$ & $--$ & $-1.9^{+0.7}_{-0.7}$ & $2.2^{+0.6}_{-0.6}$ & $1.0^{+0.6}_{-0.6}$ & $\le 1.0$ & $1.7^{+0.7}_{-0.6}$ \\
67 & $25^{+3}_{-2}$ & $43^{+5}_{-5}$ & $--$ & $-0.9^{+0.4}_{-0.5}$ & $1.0^{+0.4}_{-0.4}$ & $1.0^{+0.4}_{-0.4}$ & $0.6^{+0.5}_{-0.4}$ & $1.3^{+0.4}_{-0.4}$ \\
68 & $26^{+6}_{-6}$ & $48^{+12}_{-12}$ & $--$ & $--$ & $3.0^{+2.1}_{-2.3}$ & $1.9^{+1.1}_{-1.2}$ & $1.0^{+0.9}_{-0.9}$ & $3.5^{+1.6}_{-1.7}$ \\
69 & $12^{+3}_{-3}$ & $33^{+8}_{-8}$ & $--$ & $-0.7^{+0.5}_{-0.5}$ & $1.1^{+2.4}_{-0.5}$ & $1.1^{+0.9}_{-0.6}$ & $\le 1.2$ & $1.5^{+1.1}_{-0.4}$ \\
70 & $\le 6$ & $\le 30$ & $--$ & $-1.0^{+0.8}_{-0.8}$ & $2.1^{+0.9}_{-1.4}$ & $\le 1.5$ & $\le 1.5$ & $2.4^{+1.1}_{-1.3}$ \\
71 & $3^{+2}_{-2}$ & $23^{+14}_{-13}$ & $--$ & $-0.5^{+0.4}_{-0.4}$ & $0.9^{+1.9}_{-0.7}$ & $1.3^{+0.6}_{-0.6}$ & $\le 0.3$ & $1.5^{+0.4}_{-0.4}$ \\
72 & $\le 5$ & $\le 98$ & $--$ & $--$ & $\le 7.0$ & $1.6^{+1.0}_{-1.0}$ & $1.0^{+0.9}_{-0.9}$ & $2.8^{+1.6}_{-2.1}$ \\
\hline
\end{tabular}
\\
$^a$ iron line flux in  10$^{-4}$ photons cm$^{-2}$ s$^{-1}$\\
$^b$ iron line equivalent width in eV\\
$^c$ 10.5 keV negative line flux in units of 10$^{-3}$ photons cm$^{-2}$ s$^{-1}$\\
$^d$ 3.88 keV line flux in units of 10$^{-3}$ photons cm$^{-2}$ s$^{-1}$\\
$^e$ 30.17 keV line flux in units of 10$^{-3}$ photons cm$^{-2}$ s$^{-1}$\\
$^f$ 39.04 keV line flux in units of 10$^{-3}$ photons cm$^{-2}$ s$^{-1}$\\
$^g$ 53.00 keV line flux in units of 10$^{-3}$ photons cm$^{-2}$ s$^{-1}$\\
$^h$ 66.64 keV line flux in units of 10$^{-3}$ photons cm$^{-2}$ s$^{-1}$
\end{table}

\normalsize

\clearpage
\begin{figure}
\includegraphics[width=7.0in]{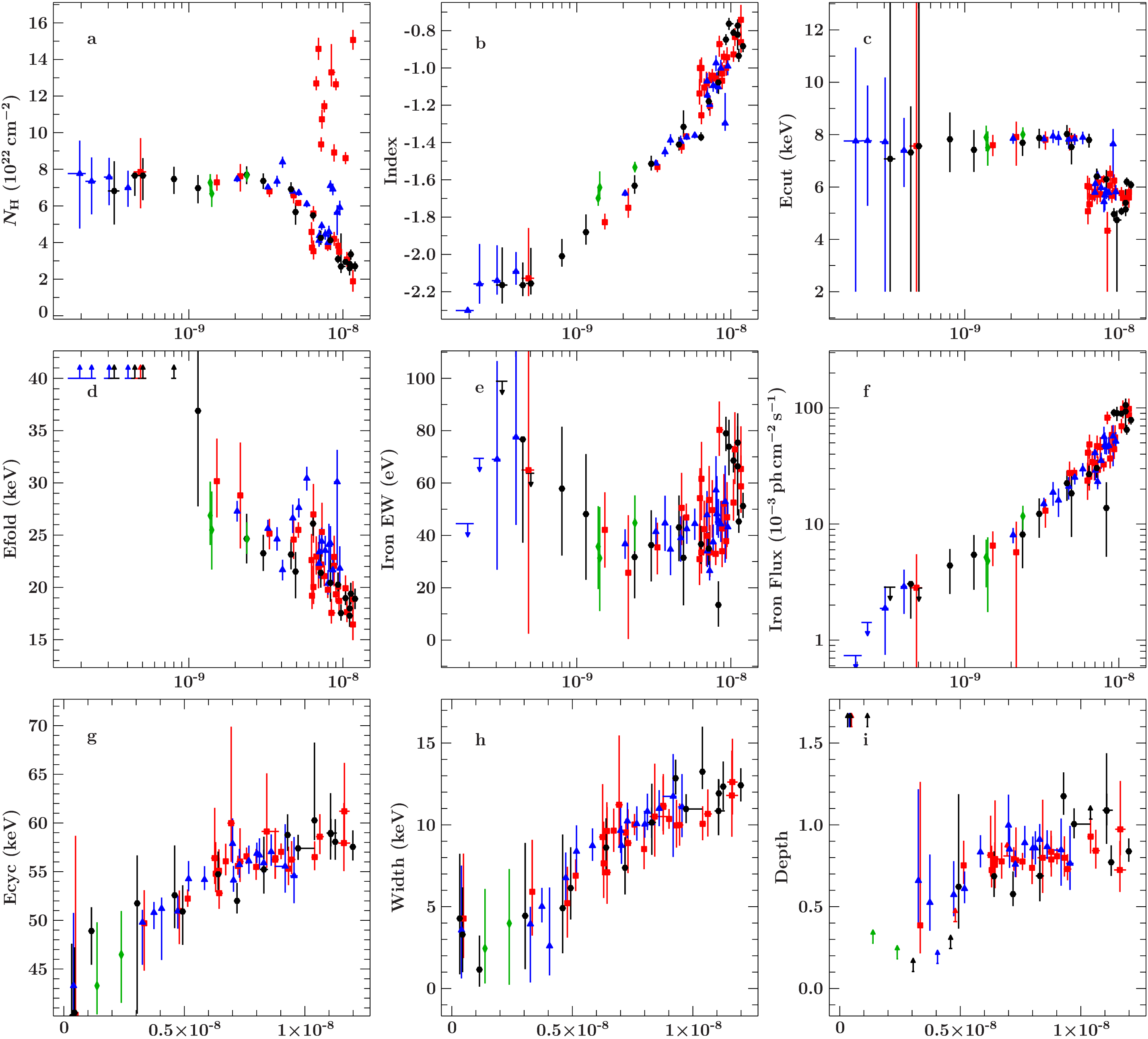}
\caption{The continuum parameters for the \texttt{highecut} model 
plotted versus 2$-$10 keV unabsorbed flux in units of ergs cm$^{-2}$ s$^{-1}$. Data from 2010 March/April are in green, 2010 August are in 
black, 2010 December are in blue, and 2011 May are in red.\label{fig:highecut_continuum_flux}}
\end{figure}

\begin{figure}
\includegraphics[width=7.0in]{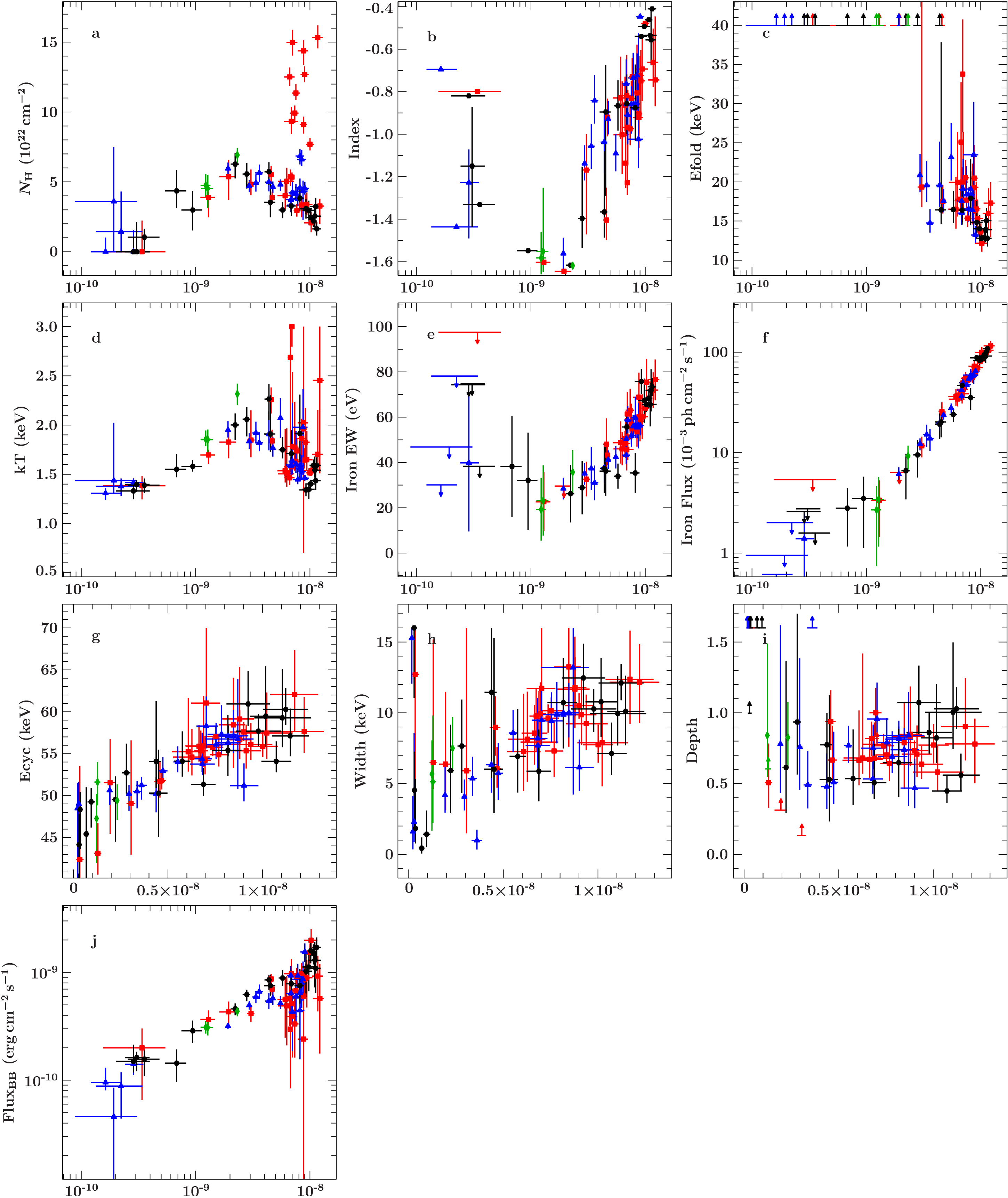}
\caption{The continuum parameters for the \texttt{cutoffpl} model plotted versus 2$-$10 keV unabsorbed  
flux in units of ergs cm$^{-2}$ s$^{-1}$. The blackbody flux (Flux$_{BB}$) is in units of L$_{39}$/D$^2$, where L$_{39}$ is the flux in units of 10$^{39}$ ergs s$^{-1}$, and D is the distance to the source in units of 10 kpc. Data from 2010 March/April are in green, 2010 August are in 
black, 2010 December are in blue, and 2011 May are in red.\label{fig:cutoffpl_continuum_flux}}
\end{figure}

\begin{figure}
\includegraphics[width=6.6in]{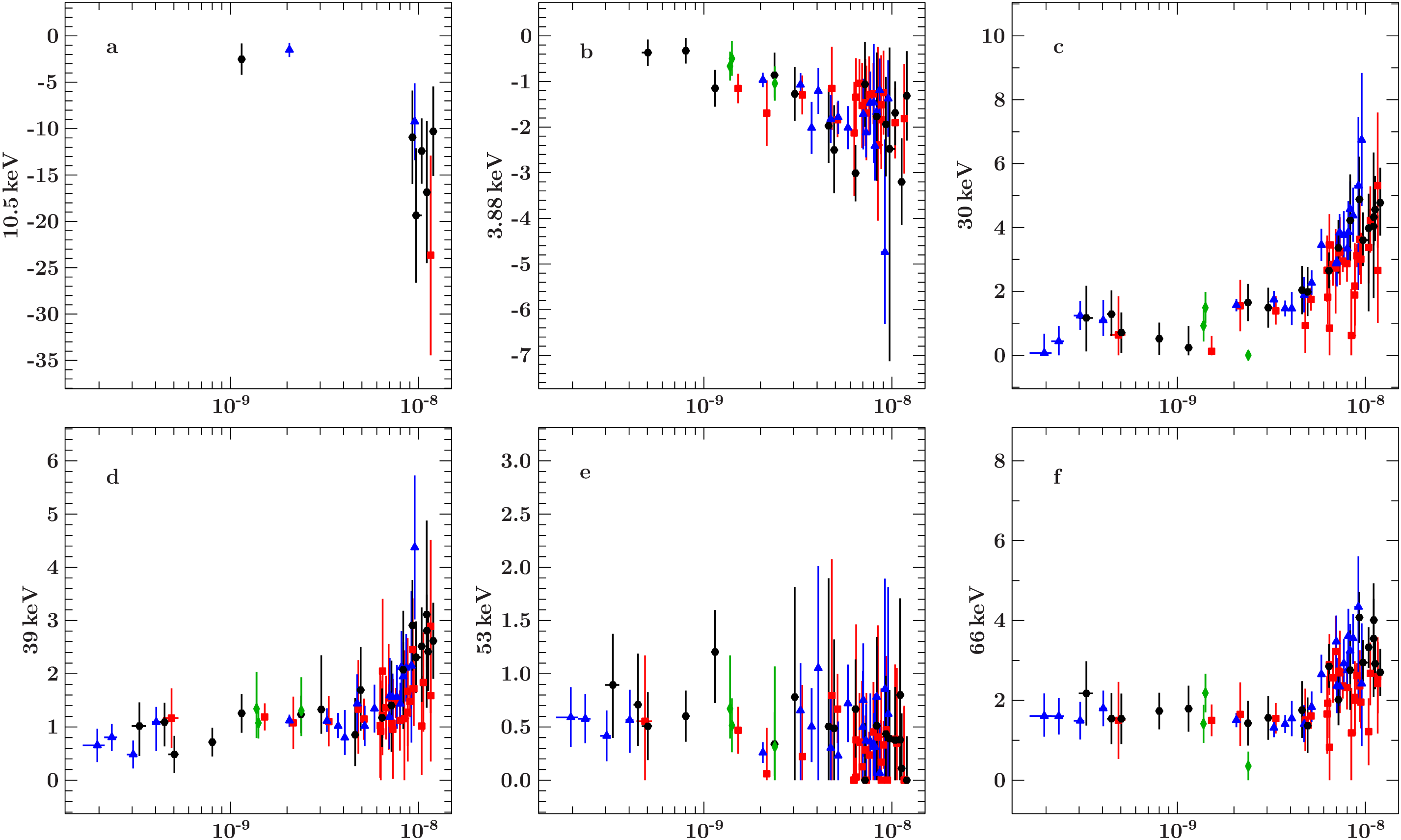}
\caption{The various line fluxes in units of 10$^{-3}$ cm$^{-2}$ s$^{-1}$ for the \texttt{highecut} model plotted 
versus 2$-$10 keV unabsorbed 
flux in units of ergs cm$^{-2}$ s$^{-1}$. Data from 2010 March/April are in green, 2010 August are in 
black, 2010 December are in blue, and 2011 May are in red.\label{fig:highecut_line_flux}}
\end{figure}

\begin{figure}
\includegraphics[width=6.6in]{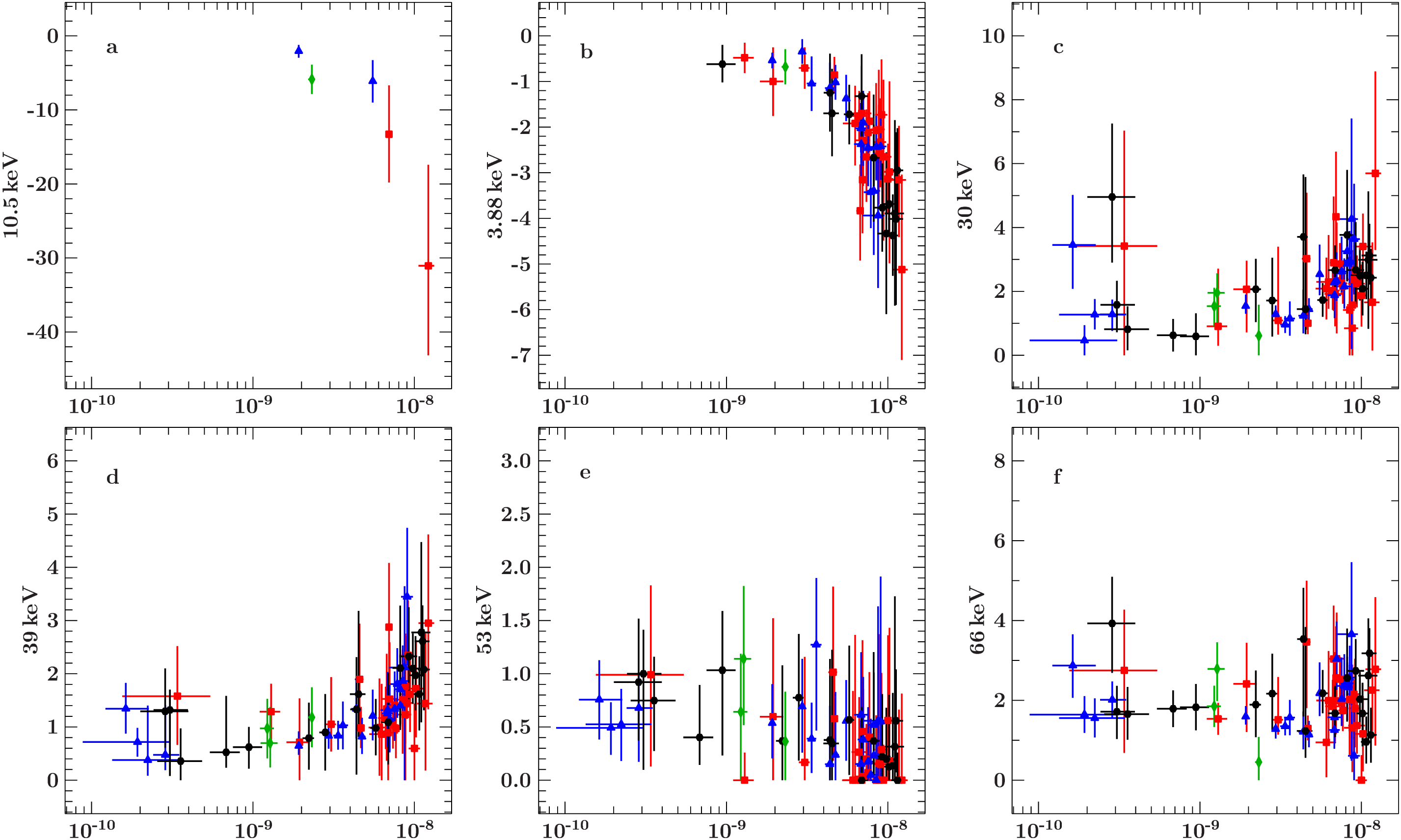}
\caption{The various line fluxes in units of 10$^{-3}$ cm$^{-2}$ s$^{-1}$ for the \texttt{cutoffpl} model plotted 
versus 2$-$10 keV unabsorbed  
flux in units of ergs cm$^{-2}$ s$^{-1}$. Data from 2010 March/April are in green, 2010 August are in 
black, 2010 December are in blue, and 2011 May are in red.\label{fig:cutoffpl_line_flux}}
\end{figure}

\begin{figure}
\includegraphics[width=6.in]{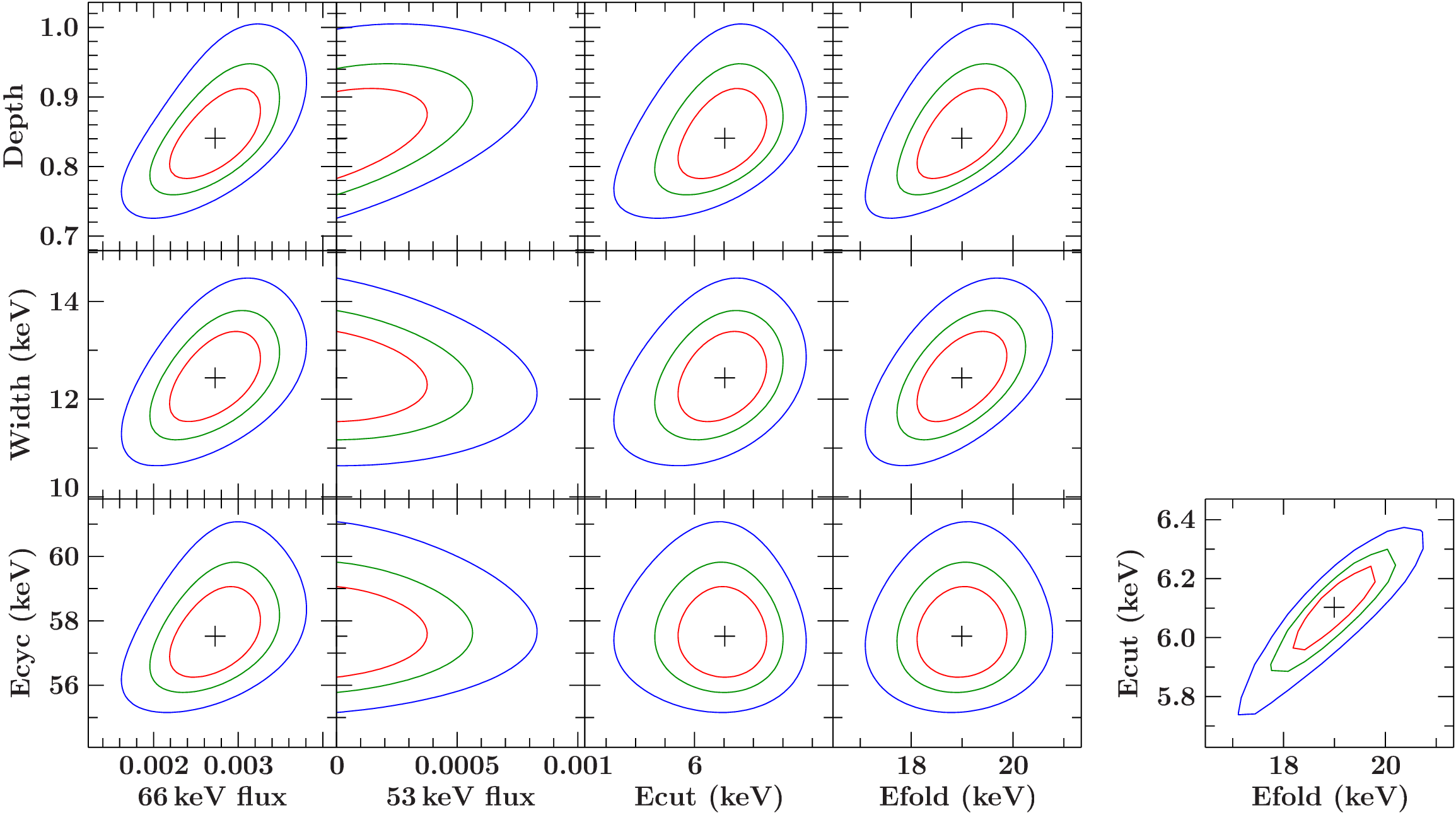}
\caption{Contours of the cyclotron line fitted parameters versus the background lines at 66 keV, 53 keV, and the continuum parameters Ecut and Efold, plus the contours for Ecut versus Efold for observation 9. The red, green, and blue contours represent the 68\%, 90\%, and 99\% significance levels.\label{obs9}}
\end{figure}

\begin{figure}
\includegraphics[width=6.in]{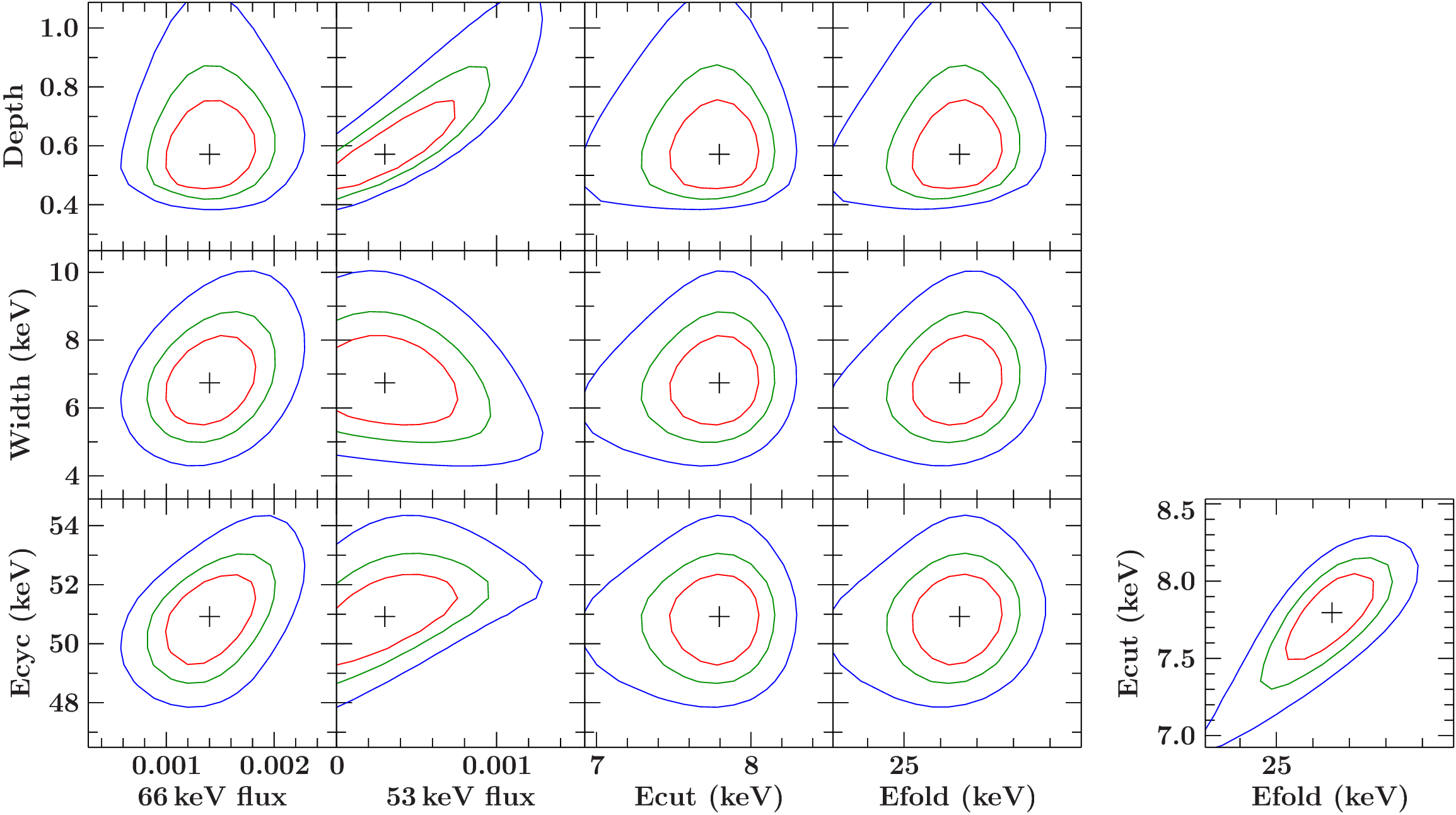}
\caption{Contours of the cyclotron line fitted parameters versus the background lines at 66 keV, 53 keV, and the continuum parameters Ecut and Efold, plus the contours for Ecut versus Efold for observation 39. The red, green, and blue contours represent the 68\%, 90\%, and 99\% significance levels.\label{obs39}}
\end{figure}

Correlations between the fitted cyclotron line parameters and background lines at 53 keV and 66 keV, 
as well as versus the cutoff energy and folding energy of the continuum, are displayed in Fig.~\ref{obs9} and 
\ref{obs39} for high and low flux observations \#9 and \#39, respectively. In addition the correlation between the 
folding and cutoff energies is shown for those observations. 

\clearpage
\normalsize
\section{Test of HEXTE Background Estimation for GX 304$-$1}
\subsection{Counts and Rates}
\begin{figure}
\includegraphics[width=3.3in]{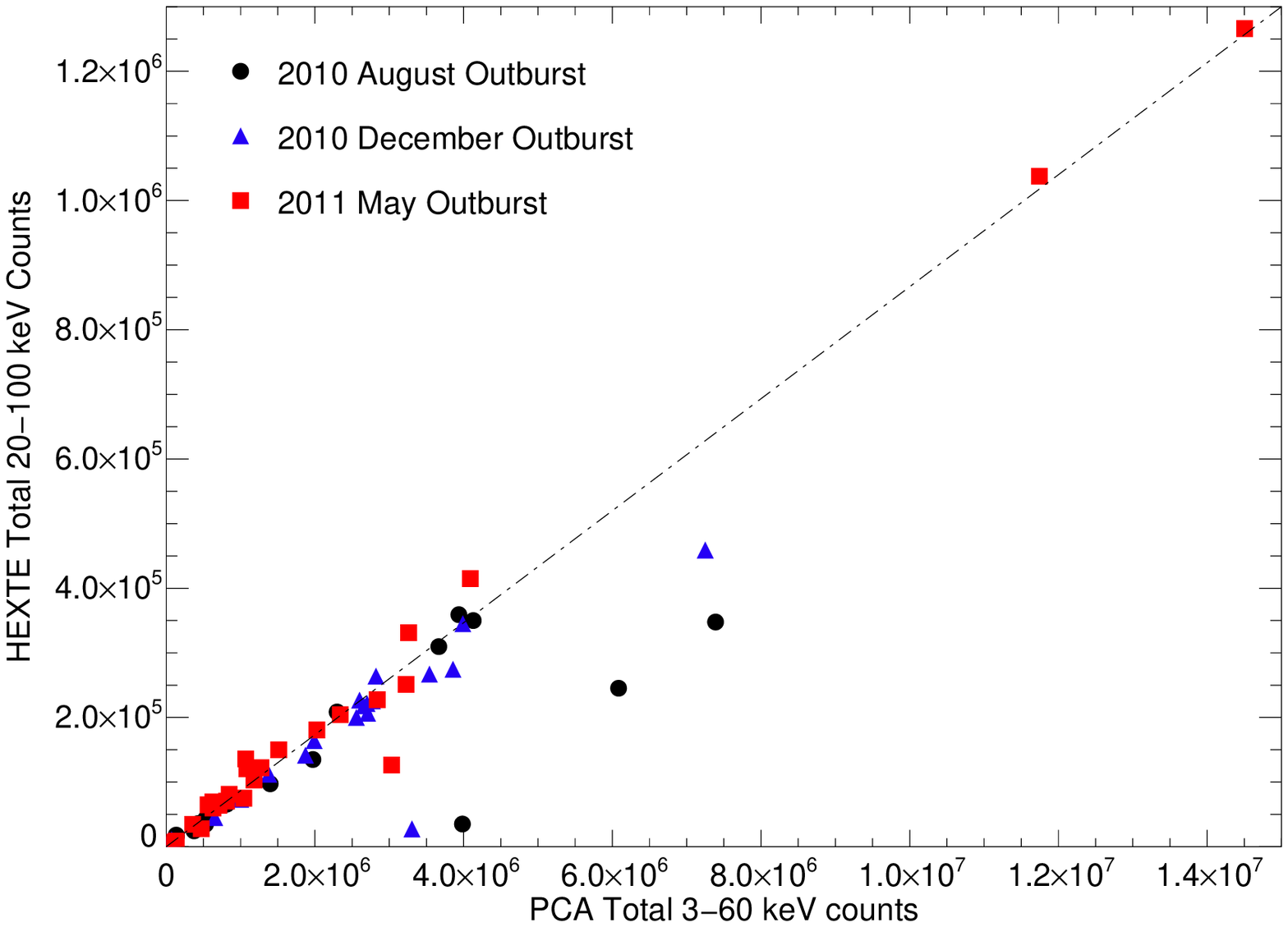}
\includegraphics[width=3.3in]{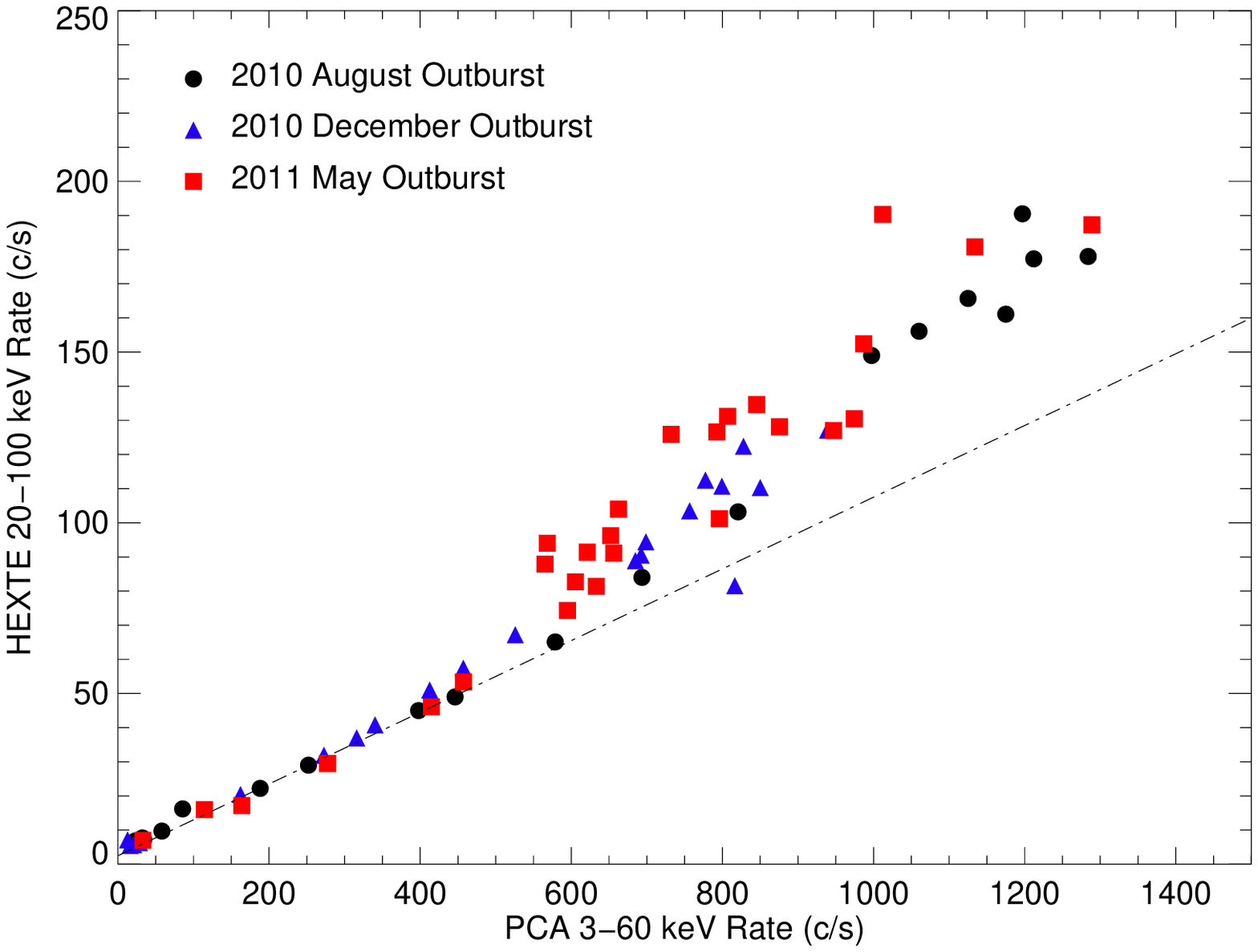}
\caption{\textbf{Left}: Total PCU2 counts 3$-$60 keV versus total HEXTE-A counts 20$-$100 keV. Total 
counts are calculated as lifetime times count rate. The six outliers result from significantly lower HEXTE~A lifetime 
than expected. 
\textbf{Right}: Comparison of PCU2 and HEXTE~A counting rates when the HEXTE background estimation method is used. The deviations from the linear relation are due to the variation in the column depth at the higher flux levels 
and the column density enhancement events 
(see Fig.~\ref{fig:highecut_continuum_flux}).\label{fig:total_counts}}
\end{figure}

One test of the HEXTE background estimation method described above is whether or not the total number of 
counts in the background-subtracted HEXTE spectrum was linearly proportional to that in the 
background-subtracted PCU2 spectrum. 
Fig.~\ref{fig:total_counts}-Left shows the product of the counting rate in the spectral band (3$-$60 keV PCU2; 
20$-$100 keV HEXTE) times the lifetime per observation. The linear relationship is clearly followed with 
the exception of 6 observations where the HEXTE total counts are low. Since the 6 outliers are 
not evident in the rate plot (Fig.~\ref{fig:total_counts}-Right), the HEXTE spectral data, from which the rates 
were extracted using the \texttt{SHOW RATE} command in XSPEC, are not suspect, and the  
outliers appear to be due to 
abnormally low lifetimes in the spectral extraction (as compared to that expected from the value of the 
PCU2 livetime) that resulted from missing HEXTE data. This can also be seen when one calculates the ratio 
of PCU2 to HEXTE livetimes.

Fig.~\ref{fig:total_counts}-Right shows the HEXTE 20$-$100 keV counting rate versus the PCU2 3-60 keV rate. 
The HEXTE and PCU2 counting rates again are linearly correlated until about 500 counts/s in the PCU2. 
The deviation from linearity at higher rates is due to the change in the column density above a flux of 
$\sim 4 \times 10^{-9}$ ergs cm$^{-2}$ s$^{-1}$, and the added column density during the 2011 May column 
density enhancement. The column density variations will affect the 3$-$60 keV PCU2 rate while leaving the 
20$-$100 keV HEXTE rate unaffected.

\subsection{Lines and Normalizations}
The various line fluxes' variations with power law flux from the \texttt{highecut} and \texttt{cutoffpl} model 
fittings are shown in Figs.~\ref{fig:highecut_line_flux} and \ref{fig:cutoffpl_line_flux}. The 10.5 keV feature 
(panel a) appears at the very highest fluxes in the \texttt{highecut} model, while being rare in the \texttt{cutoffpl} 
modeling. The 3.88 PCU2 systematic feature (panel b) shows a stronger correlation with the power law flux for 
the \texttt{cutoffpl} model than that for \texttt{highecut}. This could be due to the more curved shape of the 
\texttt{cutoffpl} model as compared to the straight power law in \texttt{highecut}. The 30 keV HEXTE background 
line (panel c) appears stronger at high power law flux levels in the \texttt{highecut} model. The other HEXTE 
background lines (panels d, e, \& f) show similar behaviors with increasing flux in both models.

\begin{figure}
\includegraphics[width=7.0in]{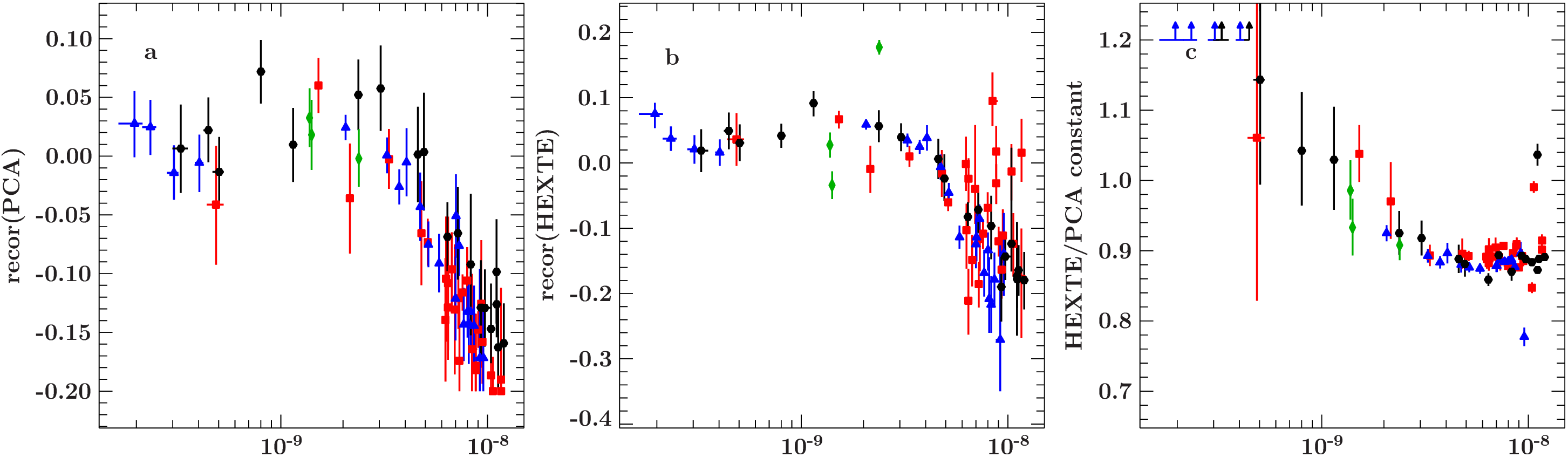}\\
\includegraphics[width=7.0in]{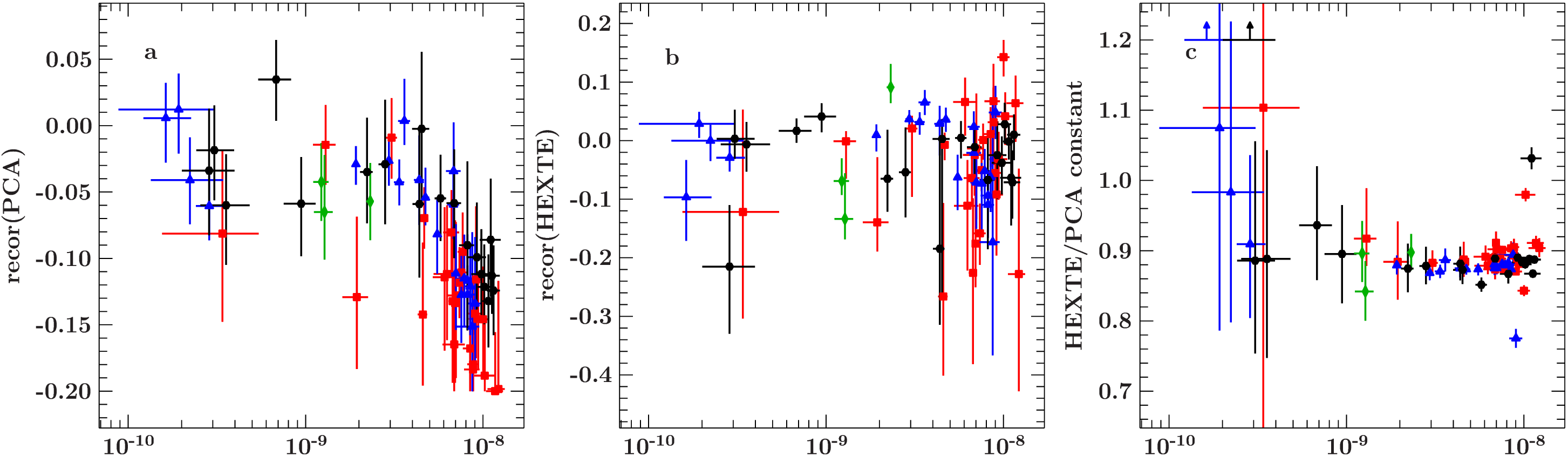}\\
\caption{The variation of the recor normalization versus power law continuum flux in units of ergs cm$^{-2}$ s$^{-1}$ 
is shown for the PCU2 (a) and 
HEXTE (b), plus the relative normalization constant for the HEXTE cluster A with respect to the PCU2 
normalization (c). The values resulting from \texttt{highecut} are plotted above those from \texttt{cutoffpl}.\label{fig:recor}}
\end{figure}

Fig.~\ref{fig:recor} gives the values of the \texttt{recor} parameter for PCU2 and HEXTE as well as the HEXTE 
constant with respect to the PCU2 flux. The top panels are for the \texttt{highecut} model and the bottom panels 
are for the \texttt{cutoffpl} model. In the fitting process, the \texttt{corback} function found in ISIS and \texttt{recor} 
function in XSPEC are used to optimize the background subtraction by adjusting the background live time as part 
of the fitting process. 
PCU2 background estimates are based upon the observed background as a function of certain instrument charged 
particle average counting rates, and as such, may not reflect the exact background experienced during a given 
observation. The spectral shape of the background is assumed to remain the same and just its intensity is 
adjusted via the live time. 
A similar estimation is done for both the effects of the averages associated with HEXTE deadtime and HEXTEBACKEST. The \texttt{recor/corback} free 
parameter is the fraction of the estimated background to be added or subtracted. , 
The originally determined background normalizations have to be reduced by increasing amounts for increasing source fluxes, since true X-rays can 
contaminate the average charged particle counting rates for sources at high fluxes in the PCU2 and HEXTE. These counting rates are the basis for the background estimates.
The effect is larger in the \texttt{highecut} models as 
compared to that in the \texttt{cutoffpl} models.

The relative normalization between the PCU2 and HEXTE instruments is plotted in 
Fig.~\ref{fig:recor}-Top/Bottom panel c. The HEXTE normalization constant is around 0.88 except at the lower 
power law fluxes where it increases with ever larger uncertainties, and at higher fluxes when the column density
enhancements affect the 2$-$10 keV PCU2 fluxes. 

\bsp	
\label{lastpage}
\end{document}